\documentclass[12pt,english,floatfix,nofootinbib,superscriptaddress,aps,prd,preprint]{revtex4}
\usepackage[utf8]{inputenc}
\usepackage{float}
\usepackage{array}
\usepackage{bbold}
\usepackage{lipsum}
\usepackage{dsfont}
\usepackage{graphicx}
\usepackage{amsmath,amsthm,amsfonts,amssymb}
\usepackage{graphicx}
\usepackage[english]{babel} 
\usepackage{color}
\usepackage{tensor}
\usepackage{esint}
\usepackage[dvips]{epsfig}
\usepackage[dvips]{graphicx}
\usepackage{float}
\usepackage{units}
\usepackage{textcomp}
\usepackage{mathrsfs}
\usepackage{amsmath}
\usepackage[makeroom]{cancel}
\usepackage{amssymb}
\usepackage{amsbsy}
\usepackage{amsfonts}
\usepackage{amssymb,mathrsfs,xcolor}
\usepackage{esint}
\usepackage{braket}
\usepackage{array}
\usepackage{graphicx}

\usepackage{wasysym}
\usepackage{multirow}
\usepackage{wrapfig}
\usepackage{subfig}

\usepackage{stmaryrd}
\usepackage{upgreek}

\makeatletter
\makeatletter\usepackage{babel}
\usepackage{hyperref}
\hypersetup{
    colorlinks,
    citecolor=blue,
    filecolor=green,
    linkcolor=purple,
    urlcolor=red,
}

\usepackage{slashed}

\newcommand{\ie}{\begin{equation}}
\newcommand{\fe}{\end{equation}}
\newcommand{\se}{\begin{eqnarray}}
\newcommand{\ff}{\end{eqnarray}}

\begin{document}

\title{Lorentz--violating modifications to particle dynamics, thermodynamics and vacuum energy in bumblebee gravity}

\author{A. A. Ara\'{u}jo Filho}
\email{dilto@fisica.ufc.br}

\affiliation{Departamento de Física, Universidade Federal da Paraíba, Caixa Postal 5008, 58051-970, João Pessoa, Paraíba,  Brazil.}
\affiliation{Departamento de Física, Universidade Federal de Campina Grande Caixa Postal 10071, 58429-900 Campina Grande, Paraíba, Brazil.}
\affiliation{Center for Theoretical Physics, Khazar University, 41 Mehseti Street, Baku, AZ-1096, Azerbaijan.}

\author{K. E. L. de Farias }
\email{klecio.lima@academico.ufpb.br}

\affiliation{Departamento de Física, Universidade Federal de Campina Grande Caixa Postal 10071, 58429-900 Campina Grande, Paraíba, Brazil.}
\author{E. Passos}
\email{passos@df.ufcg.edu.br}

\affiliation{Departamento de Física, Universidade Federal de Campina Grande Caixa Postal 10071, 58429-900 Campina Grande, Paraíba, Brazil.}

\author{F. A. Brito}
\email{fabrito@df.ufcg.edu.br}

\affiliation{Departamento de Física, Universidade Federal de Campina Grande Caixa Postal 10071, 58429-900 Campina Grande, Paraíba, Brazil.}


\author{Ali \"Ovg\"un}
\email{ali.ovgun@emu.edu.tr}
\affiliation{Physics Department, Eastern Mediterranean
University, Famagusta, 99628 North Cyprus, via Mersin 10, Turkiye.}


\author{Hassan  Hassanabadi}
\email{hha1349@gmail.com}
\affiliation{Department   of   Physics, Faculty of Science,   University   of   Hradec   Kr\'{a}lov\'{e},  Rokitansk\'{e}ho 62, 500   03   Hradec   Kr\'{a}lov\'{e},   Czechia}
\affiliation{Khazar University, Department of Physics and Electronics, 41 Mahsati Str, AZ1096, Baku, Azerbaijan}
\affiliation{Al-Farabi Kazakh National University, Al-Farabi av. 71, 050040 Almaty, Kazakhstan}

\author{V. B. Bezerra}
\email{valdir@fisica.ufpb.br}
\affiliation{Departamento de Física, Universidade Federal da Paraíba, Caixa Postal 5008, 58051-970, João Pessoa, Paraíba,  Brazil.}

\author{Amilcar R. Queiroz}
\email{amilcarq@df.ufcg.edu.br}

\affiliation{Departamento de Física, Universidade Federal de Campina Grande Caixa Postal 10071, 58429-900 Campina Grande, Paraíba, Brazil.}

\date{\today}

\begin{abstract}

We investigate how spontaneous Lorentz symmetry breaking in bumblebee gravity modifies particle dynamics, thermodynamics, and vacuum energy around a static black hole background. Starting from the optical--mechanical correspondence, we derive a modified dispersion relation that encodes the influence of the Lorentz--violating parameter $\lambda$ on the propagation of massive and massless modes. We analyze the resulting optical properties, including the effective refractive index, group velocity, and energy--dependent time delay, and show how the non--asymptotically flat geometry reshapes signal propagation. From the same dispersion relation, we construct the interparticle potential for massive and massless excitations and evaluate the electron scattering cross section within the Born approximation, identifying characteristic Lorentz--violating corrections. We then develop a statistical--ensemble description based on the deformed energy--momentum relation and obtain analytic expressions for the thermodynamic observables of a massless bosonic gas. The pressure, mean energy, entropy, and heat capacity are examined in three representative regimes—extremely close to the horizon, near the photon sphere, and in the asymptotic region—where Lorentz violation systematically increase the magnitude of these quantities and leads to finite asymptotic plateaus. Finally, we analyze the vacuum state in the curved background and compute the regularized Casimir energy at zero and finite temperature.

\end{abstract}

\maketitle
     
\tableofcontents


\section{Introduction }

Although modern experiments have repeatedly confirmed that the laws of physics remain unchanged for observers in uniform motion, several proposals inspired by physics at very high energies have challenged this expectation. Instead of assuming exact Lorentz invariance, these approaches allow for departures that could emerge in extreme regimes, a possibility investigated within a variety of theoretical constructions \cite{1,4,2,6,3,7,8,5}. From this perspective, Lorentz violation has been organized according to how the symmetry is lost. One route introduces preferred directions or structures directly at the level of the fundamental equations, so that the symmetry is no longer present from the outset and observable anisotropies naturally arise. A different route keeps the dynamical laws formally invariant, but lets the ground state of the theory select a noninvariant configuration. In this latter case, symmetry breaking is not imposed by hand but develops through the vacuum itself, giving rise to novel and sometimes counterintuitive physical consequences \cite{bluhm2006overview,araujo2025impact,Liu:2025bpp}.

A broad class of studies has treated possible violations of Lorentz symmetry within the effective field theory language provided by the Standard Model Extension, where special emphasis has been placed on realizations driven by vacuum dynamics rather than by explicit symmetry loss \cite{liu2024shadow,filho2023vacuum,12,AraujoFilho:2024ykw,13,11,10,9}. Within this setting, particular interest has converged on scenarios in which an additional vector degree of freedom permeates spacetime. Known as the bumblebee field, this vector does not remain trivial in the ground state; instead, it settles into a configuration with a finite expectation value, thereby endowing spacetime with a preferred orientation \cite{amarilo2024gravitational,schreck2014quantum,schreck2014quantum2}. Once such a direction is selected, local isotropy ceases to hold, and both the effective geometry and the motion of particles reflect this built-in anisotropy. Motivated by these geometric modifications, a growing body of work has shifted attention toward gravitational and thermal manifestations of the mechanism, showing that horizon physics and bulk thermodynamic responses depart from their relativistic counterparts, with measurable changes in entropy, temperature behavior, and heat capacities across distinct regions of the spacetime \cite{paperrainbow,aa2021lorentz,araujo2021higher,araujo2021thermodynamic,araujo2022thermal,araujo2022does,Anacleto:2018wlj,reis2021thermal,Capozziello:2024ucm,DeBianchi:2025bgn,Capozziello:2025wwl}.

A sustained effort to construct black hole spacetimes beyond the standard predictions of general relativity has led to several extensions of Schwarzschild and (A)dS backgrounds, particularly in settings where the vacuum structure itself is allowed to deviate from the usual relativistic constraints \cite{20}. Within this broader program, models governed by a Lorentz--violating vector sector have emerged as a natural arena. In these scenarios, the bumblebee field develops a nontrivial temporal profile, and the resulting ground state reshapes the gravitational field, producing black hole geometries with modified metric functions \cite{23,24,22,21}.

The earliest realization of such ideas for static and spherically symmetric spacetimes appeared in Ref. \cite{14}, where the basic framework of bumblebee gravity was adapted to black hole configurations. Later studies reformulated these solutions into Schwarzschild--like forms and pursued their physical consequences in diverse directions. The modified geometries were shown to reshape the spectrum of quasinormal ringing modes \cite{19,Liu:2022dcn}, and to influence the behavior of infalling matter and accretion flows \cite{18,17}. Using EHT-inspired shadow measurements together with weak-field lensing and complementary probes (accretion, greybody bounds, neutrino propagation, and topological photon-sphere diagnostics), recent works place quantitative constraints on Lorentz-violating bumblebee gravity and Kalb–Ramond black-hole spacetimes and explore related phase effects such as a gravitational Aharonov–Bohm signature \cite{Lambiase:2024uzy,Lambiase:2023zeo,Pantig:2025paj,Ovgun:2025pwy,Pantig:2024kqy}.

Recent progress on the optical--mechanical analogy in curved backgrounds has opened new routes to probe how geometry reshapes both motion and thermodynamic behavior \cite{Nandi2016,Filho:2023ydb,araujo2023thermodynamical,filho2025modified,filho2025particlemotion}. Inspired by these advances, this work focused on a static black hole spacetime emerging from bumblebee gravity and examined how its Lorentz--violating vacuum structure affects physical observables. Rather than beginning with dynamics, the analysis first established an effective dispersion law obtained from the optical--mechanical correspondence, in which the symmetry--breaking parameter $\lambda$ enters as a deformation controlling the propagation of massive and massless excitations. This relation was then used to characterize signal transmission through the spacetime by means of an effective refractive index, modified group velocities, and frequency--dependent delays, highlighting the impact of the non–asymptotically flat geometry on wave travel.

{The novelty of the present work lies in treating particle propagation, scattering, thermodynamics, and vacuum energy within a single dispersion relation based framework for the bumblebee black hole background. Previous studies have mainly focused on specific observables, such as quasinormal modes, shadows, lensing, accretion, or black hole thermodynamics. Here, instead, the optical--mechanical correspondence is used to derive a Lorentz--violating modified dispersion relation, which is then employed as the common starting point for the refractive index, group velocity, time delay, interparticle potential, Born scattering cross section, statistical ensemble thermodynamics, and Casimir energy. In other words, such a construction allows the role of the parameter $\lambda$ to be followed consistently across classical propagation, quantum scattering, thermal observables, and vacuum fluctuations.}

This paper is organized as follows. In Sec. \ref{sec1}, we derive the modified dispersion relation arising in bumblebee gravity. Section \ref{sec2} is devoted to its general implications, including the effective refractive index, the group velocity, and the associated time delay. The interparticle potential in the presence of Lorentz invariance violation is obtained in Sec. \ref{sec3}, while Sec. \ref{sec4} presents the calculation of the electron scattering cross section. In Sec. \ref{sec5}, we develop the thermodynamic description, analyze the asymptotic behavior of the observables, and examine how each quantity depends on the Lorentz--violating parameter. The Casimir energy in the bumblebee black hole background is investigated in Sec. \ref{sec6}, where both zero-- and finite--temperature contributions are discussed together with the corresponding thermodynamic quantities. Finally, Sec. \ref{sec7} contains the conclusions and closing remarks. Throughout this work, we adopt natural units $\hbar=c=G=1$.


\section{Deriving the modified dispersion relation }
\label{sec1}

The discussion begins with a reexamination of the geometry of a general static, spherically symmetric spacetime, which provides the foundation for subsequent developments. After setting this stage, the connection between the Hamiltonian formalism and the canonical momentum of massive particles is reformulated through the optical--mechanical analogy, allowing the spacetime geometry to manifest itself directly in the particle dynamics. Among the scenarios considered, attention is given to a Lorentz--violating background—the bumblebee black hole introduced in Ref.~\cite{14}.

Once the geometric and dynamical aspects are in place, the analysis is redirected toward the thermodynamics of the system. This investigation is carried out by distinguishing three different regimes: the near--horizon domain, the neighborhood of the photon sphere, and the asymptotic region.

In this manner, let us consider the following metric to proceed our forthcoming analysis:
\ie
\mathrm{d}\mathrm{s}^{2} = {-}\mathrm{A}(r) \,\mathrm{d}t^{2} + \frac{1}{\mathrm{B}(r)}\mathrm{d}r^{2} + r^{2}(\mathrm{d}\theta^{2} + \sin^{2}\theta ,\mathrm{d}\varphi^{2}). \label{metric}
\fe

A notable feature of the Bumblebee black hole solution lies in the fact that the geometry is not asymptotically flat. Now, our attention turns to the motion of test particles subjected to such a background. The dynamics of a massive particle, in this context, is obtained through an action principle that dictates its trajectory in the curved spacetime:
\ie
\label{action}
\mathcal{S} = - m \int \mathrm{d}{s},
\fe
in which the integration is carried out along the worldline followed by the particle. When the motion is confined exclusively to the radial direction, the general relation reduces to a simplified form given by:
\ie
\mathrm{d}s = {\sqrt{\mathrm{A}(r) - \frac{1}{\mathrm{B}(r)}\,\mathrm{v}^{2}}}\,\mathrm{d}t.
\fe
Here, the radial velocity is defined as $\mathrm{v} = \dot{r} = \mathrm{d}r/\mathrm{d}t$. Imposing this relation, the action given in Eq.~(\ref{action}) takes the following reduced form:
\ie
\mathcal{L} \equiv - m {\sqrt{\mathrm{A}(r) - \frac{1}{\mathrm{B}(r)}\,\mathrm{v}^{2}}}.
\fe
Applying the usual variational rule to this relation allows one to extract the canonical momentum. The procedure leads directly to the following result:
\ie
{\vec{\mathrm{p}}} = \frac{\partial \mathcal{L}}{\partial \dot{r}}
= {m \, \frac{{\vec{\mathrm{v}}}}{\mathrm{B}(r)\sqrt{\mathrm{A}(r) - \tfrac{1}{\mathrm{B}(r)}\,{\mathrm{v}^2}}}}.
\fe

Starting from the previous relation and making use of the standard definition of the Hamiltonian, $\mathcal{H} = \vec{p}\cdot \vec{v} - \mathcal{L}$, one arrives at the following expression:
\ie
\mathcal{H} = \frac{m \, \mathrm{A}(r)}{{\sqrt{\mathrm{A}(r) - \frac{1}{\mathrm{B}(r)}\mathrm{v}^{2}}}}.
\fe
The next step is to rewrite $\mathcal{H}$ exclusively as a function of the canonical momentum $\vec{p}$ together with the metric components $A(r)$ and $B(r)$. After carrying out the algebraic rearrangements, the velocity becomes:
\ie
\Vec{v} = \frac{\sqrt{\mathrm{A}(r)} \, \Vec{\mathrm{p}}\, \mathrm{B}(r)}{{\sqrt{m^{2} + \mathrm{p}^{2} \mathrm{B}(r)}}}.
\fe
Consequently, the Hamiltonian $\mathcal{H}$ takes the form
\ie
\label{hamiltonian}
\mathcal{H} = \mathrm{E} = {m \sqrt{\mathrm{A}(r) + \frac{\mathrm{A}(r)\mathrm{B}(r)\,\mathrm{p}^{2}}{m^{2}}}}
= {\sqrt{\mathrm{A}(r)\left[m^{2}+\mathrm{B}(r)\mathrm{p}^{2}\right]}}.
\fe

In summary, the obtained relation corresponds to a relationship between momentum and energy determined by the radial components of the metric\footnote{{Notice that Ref.~\cite{filho2025particlemotion} adopts a metric signature different from the one used in the present work. To avoid ambiguities, the conventions employed here should be read from Eqs.~(\ref{metric}) and (\ref{hamiltonian}). In addition, when reproducing or comparing the results of Ref.~\cite{filho2025particlemotion} with the present analysis, one must account for the difference in signature, since it may introduce relative sign changes in some expressions and consequently affect their physical interpretation.}}
The next stage of the study addresses the main aspects of the black hole background considered here. Attention is first directed to the effective refractive index, the group velocity of particle modes, and the associated time delay. The discussion is then extended to the interaction potential, treating both massless and massive excitations. Finally, a thermodynamic analysis is developed within the framework of ensemble theory, where the focus is restricted to massless bosons, making it possible to compute all thermodynamic quantities in closed analytic form.


\section{General properties }
\label{sec2}

\subsection{The black hole spacetime }

In recent studies of bumblebee gravity, two different classes of black hole solutions have been reported: one constructed in the framework of the \textit{metric--affine} approach \cite{filho2023vacuum} and another developed within the purely \textit{metric} formulation. For the purposes of the present work, attention is confined to the latter case \cite{14}:
\ie
\begin{split}
\label{model2}
\mathrm{d}s^{2} = & - \left( 1 - \frac{2M}{r}   \right) \mathrm{d}t^{2} + \frac{(1 + \lambda)}{1 - \frac{2M}{r} } \, \mathrm{d}r^{2} + r^{2}\mathrm{d}\theta^{2} + r^{2} \sin^{2}{\theta}\,\mathrm{d}\varphi^{2},
\end{split}
\fe
where, in this scenario, Lorentz violation does not alter the position of the event horizon when compared with the Schwarzschild solution. In other words, the horizon remains located at $r_{h} = 2M$.

More recently, the literature has reported charged, spherically symmetric and slowly rotating black hole solutions in bumblebee gravity \cite{liu2025charged}, as well as wormhole configurations and exact black hole solutions supported by lightlike or spacelike vacuum expectation values in the same framework \cite{Liu:2025oho}.


\subsection{Index of refraction}

\begin{figure}
    \centering
     \includegraphics[scale=0.62]{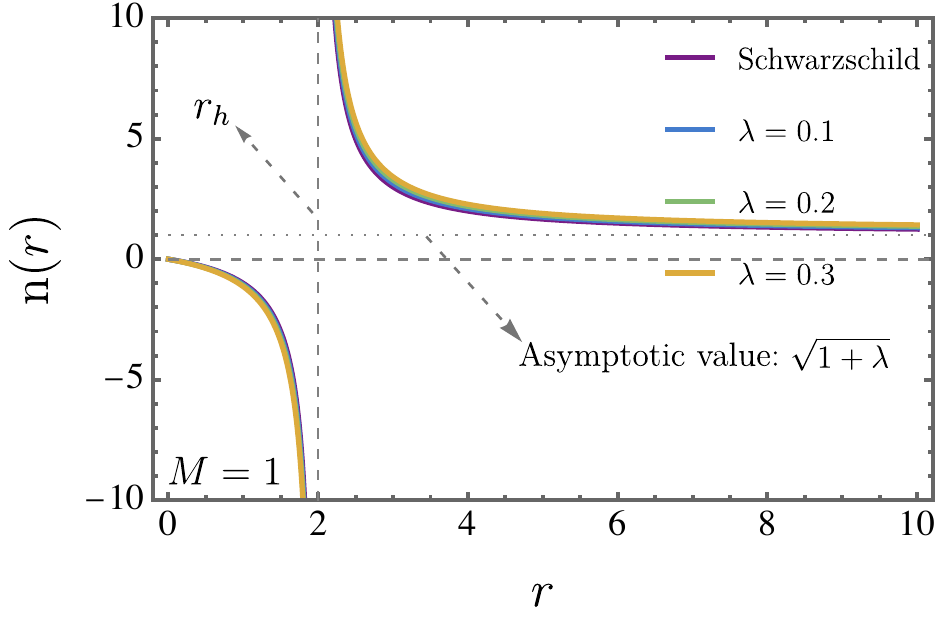}
    \caption{The behavior of the refractive index $\mathrm{n}(r)$ is illustrated as a function of the radial coordinate for different choices of the parameter $\lambda$, while the Schwarzschild limit is displayed for reference. Throughout the analysis the mass has been fixed to $M=1$. In addition, the curves approach the asymptotic constant value $\sqrt{1+\lambda}$ at large distances.}
    \label{nindex}
\end{figure}

We now adopt a definition for the refractive index analogous to the one proposed in Ref.~\cite{Nandi2016}, which leads to the relation
\ie
\mathrm{n}(r) = \frac{1}{\sqrt{\mathrm{A}(r) \mathrm{B}(r)}}.
\fe
Because the underlying geometry arises from a black hole solution in bumblebee gravity, the spacetime configuration naturally reshapes the dynamics of the fields. Under these conditions, the following expression is derived:
\ie
\mathrm{n}(r) = \frac{\sqrt{\lambda +1} \, r}{{r-2M}} \,.
\fe
Figure \ref{nindex} displays the radial dependence of the refractive index. The graph reveals a divergence at the event horizon $r_{h}$. Two additional aspects can be emphasized: outside the horizon ($r>r_{h}$) the function stays {negative}, while in the exterior region it turns {positive}; moreover, in the asymptotic regime the index does not decay to zero but rather tends to the constant limit $\sqrt{1+\lambda}$, determined solely by the parameter $\lambda$.

\subsection{Group velocity}

\begin{figure}
    \centering
     \includegraphics[scale=0.6]{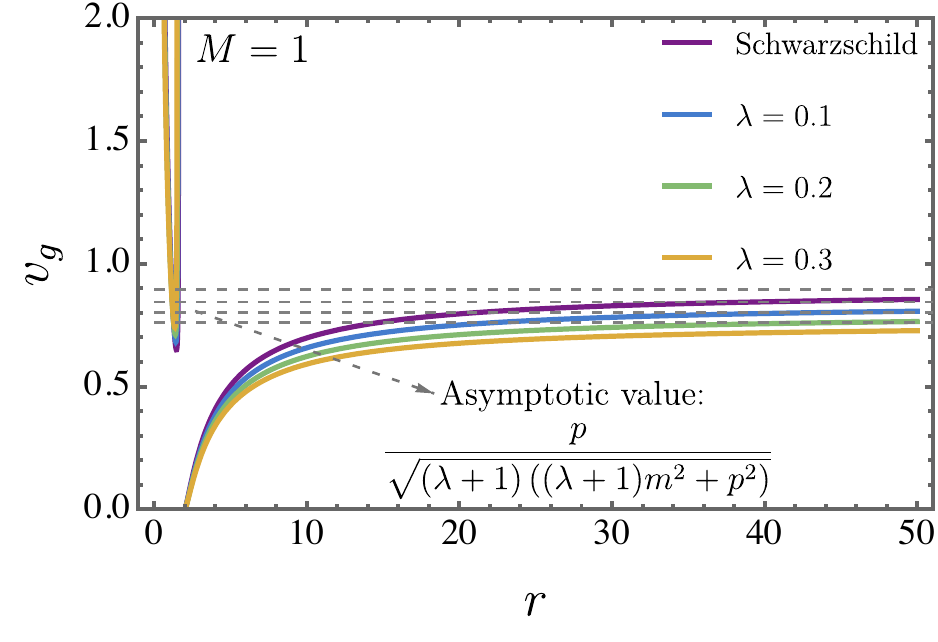}
    \caption{The behavior of the group velocity $v_{g}$ is shown for different choices of the parameter $\lambda$, alongside the Schwarzschild solution, which is plotted for reference.}
    \label{normalgv}
\end{figure}

\begin{figure}
    \centering
     \includegraphics[scale=0.6]{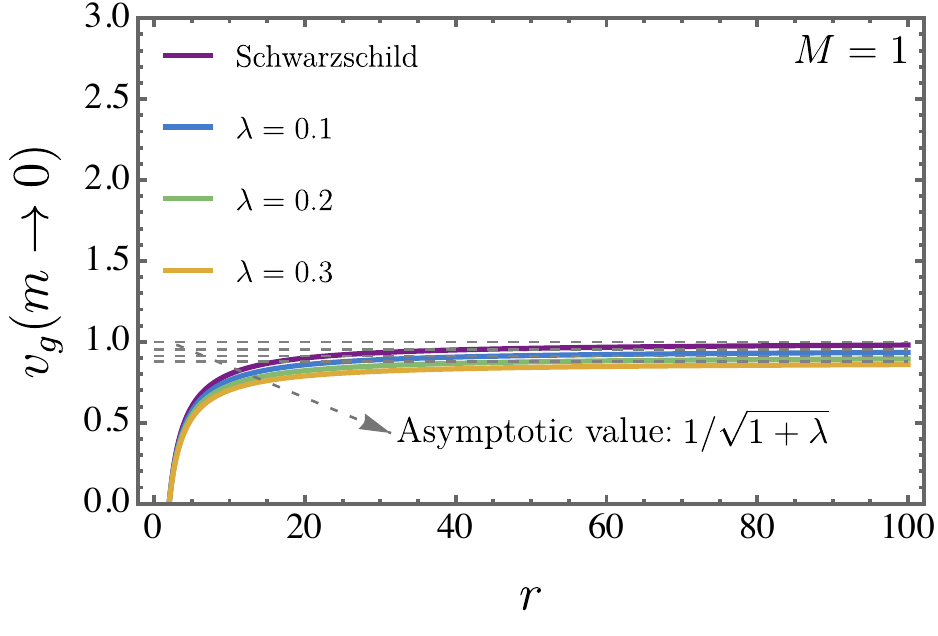}
    \caption{In the massless regime, the group velocity $v_{g}(m \to 0)$ is plotted for various values of $\lambda$, with the Schwarzschild solution included as a benchmark.}
    \label{masslessgv}
\end{figure}

The motion of a particle, or equivalently the propagation of an associated wave packet, is characterized by the group velocity $v_{g}(p,M,\lambda,r)$. This velocity is obtained by evaluating the derivative of the energy with respect to the momentum, thereby reflecting the effect of momentum changes on the dispersion relation. Accordingly, one obtains
\ie
\label{groupvelocity}
v_{g}(\mathrm{p},M,\lambda,r) = \frac{\mathrm{p} \,(r-2 M)^2}{r \sqrt{(\lambda +1) ({-}2 M {+}r) \left((\lambda +1) m^2 r{-}2 M \mathrm{p}^2 {+} \mathrm{p}^2 r\right)}}.
\fe
In other words, notice that the Lorentz--violating parameter introduces subtle corrections that cause the photon velocity to vary with the radial coordinate. To illustrate this effect, Figure \ref{normalgv} shows a configuration in which the system parameters change with position. In addition, applying Eq. (\ref{groupvelocity}) in the asymptotic limit $r \to \infty$ gives
\ie
\lim\limits_{r \to \infty}v_{g}(\mathrm{p},M,\lambda,r) = \frac{\mathrm{p}}{\sqrt{(\lambda +1) \left((\lambda +1) m^2 {+}\mathrm{p}^2\right)}}.
\fe
The above expression highlights that at large radial distances the group velocity approaches a constant value instead of tending to zero, which indicates the lack of asymptotic flatness in this geometry, as mentioned previously. A further point of interest is the limit $m \to 0$, where the analysis of $v_{g}(\mathrm{p},M,\lambda,r)$ leads to
\ie
\label{masslessvelocity}
\lim\limits_{m \to 0}v_{g}(r,M,\lambda) = \frac{r-2 M}{\sqrt{\lambda +1} \, r}.
\fe
Figure \ref{masslessgv} illustrates the behavior in the limit of vanishing mass. In this regime, as the radial coordinate grows ($r \to \infty$), the group velocity tends to the constant $1/\sqrt{1+\lambda}$. Just as in the case of massive particles, the persistence of a nonzero asymptotic value confirms that the geometry does not exhibit asymptotic flatness.

\subsection{Time--delay}

When light travels through an effective medium in which its propagation velocity depends on the radial coordinate, the total propagation time becomes sensitive to the path followed by the signal. {In the present massless limit, however, the resulting expression is not energy--dependent. In other words, the quantity below should be interpreted as the coordinate propagation time difference between two radial positions.} To evaluate this effect, let a photon propagate between two radial positions $d_{1}$ and $d_{2}$, with $d_{2}>d_{1}>2M$. The corresponding propagation-time difference is given by
\ie
\begin{split}
\Delta t(d_{1},d_{2};M,\lambda)
&= t(d_{2})-t(d_{1})
= \int_{d_{1}}^{d_{2}}\frac{\mathrm{d}r}{\lim\limits_{m\to0}v_{g}(r,M,\lambda)}
\\
&= \sqrt{\lambda+1}\left[-2M\ln(d_{1}-2M)-d_{1}
+2M\ln(d_{2}-2M)+d_{2}\right]
\\
&{
= \sqrt{\lambda+1}\left[
d_{2}-d_{1}
+2M\ln\left(\frac{d_{2}-2M}{d_{1}-2M}\right)
\right].
}
\end{split}
\fe
{The last form makes explicit that the result is controlled by three contributions: the coordinate separation $d_{2}-d_{1}$, the logarithmic correction associated with the black-hole mass, and the global factor $\sqrt{1+\lambda}$. The expression is real and well-defined only outside the horizon, namely for $d_{1},d_{2}>2M$.}

\begin{figure}
    \centering
     \includegraphics[scale=0.5]{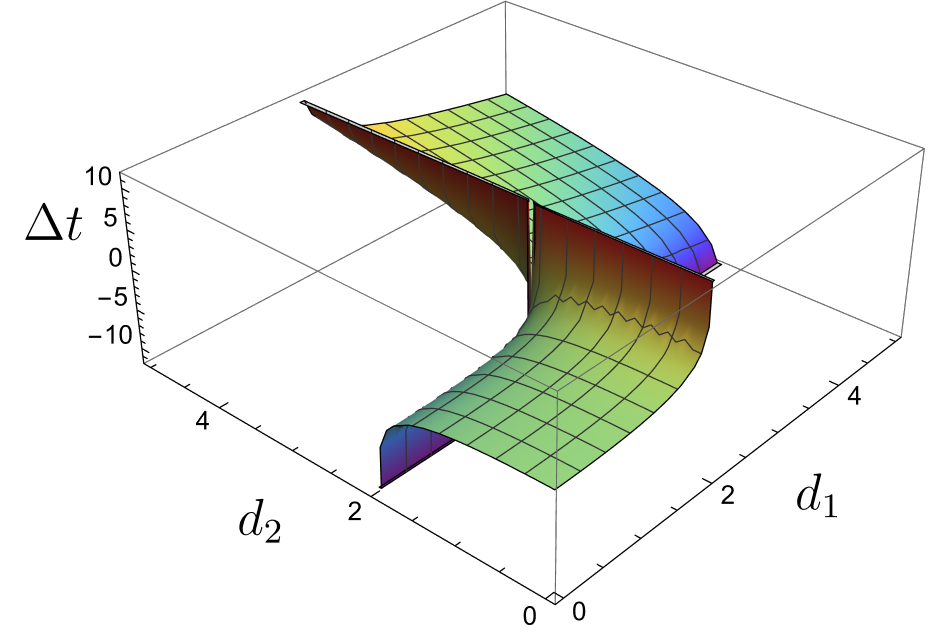}
     \includegraphics[scale=0.5]{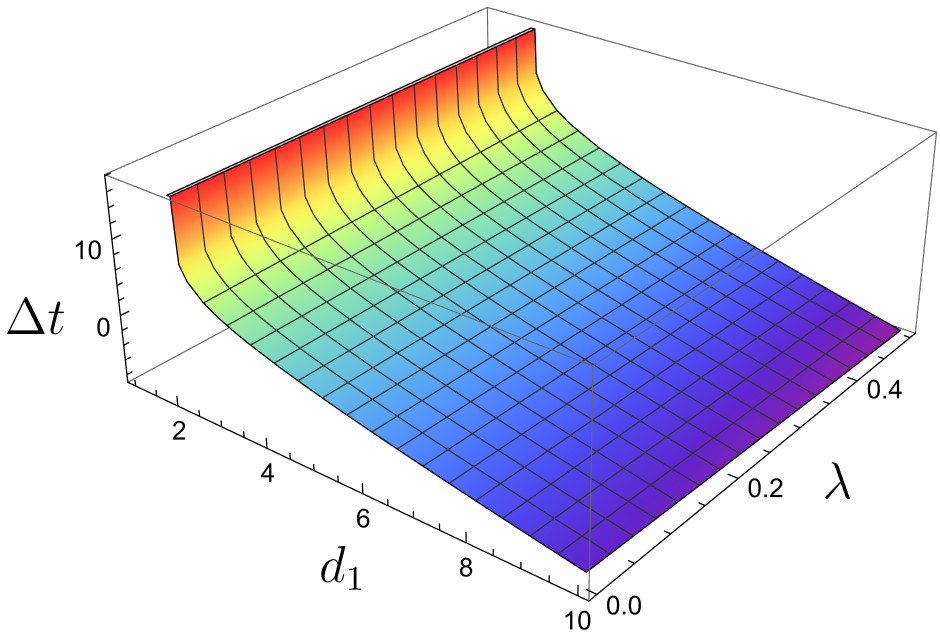}
    \caption{The dependence of the time delay $\Delta t$ on the parameters $d_{1}$, $d_{2}$, and $\lambda$ is illustrated. {The left panel shows the sensitivity to the radial interval, whereas the right panel displays the effect of varying the Lorentz-violating parameter for fixed $d_{2}$.}}
    \label{tddfff}
\end{figure}

Figure \ref{tddfff} presents the behavior of the time delay $\Delta t$ for massless particles under distinct parameter choices. In the left plot, $\Delta t$ is evaluated as a function of $d_{1}$ and $d_{2}$ while the parameter $\lambda$ is held constant at $0.1$. The right plot, on the other hand, displays $\Delta t$ against $d_{1}$ and $\lambda$, with the second distance fixed at $d_{2}=5$.

{It is also useful to quantify the sensitivity of the result to small variations of the input parameters. Defining}
\ie
{
F(d_{1},d_{2};M)
=
d_{2}-d_{1}
+2M\ln\left(\frac{d_{2}-2M}{d_{1}-2M}\right),
}
\fe
{one has $\Delta t=\sqrt{1+\lambda}\,F$. Hence,}
\ie
{
\frac{\partial \Delta t}{\partial \lambda}
=
\frac{F}{2\sqrt{1+\lambda}},
\qquad
\frac{\partial \Delta t}{\partial d_{2}}
=
\sqrt{1+\lambda}\left(1+\frac{2M}{d_{2}-2M}\right),
\qquad
\frac{\partial \Delta t}{\partial d_{1}}
=
-\sqrt{1+\lambda}\left(1+\frac{2M}{d_{1}-2M}\right).
}
\fe
{Therefore, when $d_{1},d_{2}\gg2M$, the delay is dominated by $d_{2}-d_{1}$, while the logarithmic mass correction is suppressed by the small ratio $M/d_{i}$. The dependence on $\lambda$ is also weak for $\lambda\ll1$, since $\sqrt{1+\lambda}\simeq1+\lambda/2$.}

In addition, starting from the time-delay function
\ie
\Delta t(d_{1},d_{2};M,\lambda)
=
\sqrt{\lambda+1}\left[
d_{2}-d_{1}
+2M\ln\left(\frac{d_{2}-2M}{d_{1}-2M}\right)
\right],
\fe
let us evaluate its magnitude in an astrophysical setting. The central black hole of the Milky Way, Sagittarius A$^{*}$, is taken as the reference, with an estimated mass of
\ie
M_{\mathrm{SgrA}^{*}}
\approx 4\times10^{6}M_{\odot}
\approx 1.2\times10^{10}\,\mathrm{m}.
\fe
Two radial positions are considered:
\ie
d_{1}
\approx 9.461\times10^{15}\,\mathrm{m},
\fe
which corresponds to approximately $1\,\mathrm{light\mbox{-}year}$, and
\ie
d_{2}
\approx 9.470\times10^{15}\,\mathrm{m},
\fe
which corresponds to approximately $1.001\,\mathrm{light\mbox{-}years}$. The Lorentz-violating parameter is fixed to
\ie
\lambda = 2.2\times10^{-9},
\fe
consistent with the bounds reported in Ref.~\cite{14}. Restoring SI units, with $c=3\times10^{8}\,\mathrm{m/s}$, the delay is
\ie
\Delta t_{\mathrm{seconds}}
=
\frac{\Delta t_{\mathrm{meters}}}{c}
\sim
8\,\mathrm{h}\,20\,\mathrm{min}.
\fe
{This value is mainly set by the chosen radial separation, since $d_{2}-d_{1}\simeq9.0\times10^{12}\,\mathrm{m}$ already gives}
\ie
{
\frac{d_{2}-d_{1}}{c}
\simeq
3.0\times10^{4}\,\mathrm{s}
\simeq
8.33\,\mathrm{h}.
}
\fe
{By contrast, the logarithmic correction contributes only}
\ie
{
\frac{2M}{c}
\ln\left(\frac{d_{2}-2M}{d_{1}-2M}\right)
\sim
7.6\times10^{-2}\,\mathrm{s},
}
\fe
{for the numerical values used above. The correction due to $\lambda$ is even smaller, since $\sqrt{1+\lambda}-1\simeq1.1\times10^{-9}$, corresponding to a change of order $10^{-5}\,\mathrm{s}$ in the quoted estimate. In other words, the numerical value $8\,\mathrm{h}\,20\,\mathrm{min}$ should be understood as an illustrative propagation-time difference determined primarily by the selected radii. It is worth mentioning that a genuine energy--dependent delay would require keeping the finite--$m$ or energy--dependent terms before taking the massless limit.}


\subsection{Spectral density: locally modified Planck law}
\label{subsec:spectral_density}

As we have shown in the previous subsections, the modified dispersion relation
provides $p(E,r)$ and $\mathrm{d}p/\mathrm{d}E$ directly from the Hamiltonian (shown in Eq. (\ref{hamiltonian})) rearrangement, {which now follows from $E^2=A(r)\,\left[m^2+B(r)\,p^2\right]$. Hence,}
\begin{equation}
{p^2(E,r)=\frac{E^2-m^2\,A(r)}{A(r)\,B(r)} .}
\end{equation}
{For the bumblebee geometry considered here, $A(r)=f(r)$ and $B(r)=f(r)/(1+\lambda)$, with $f(r)=1-2M/r$. Therefore, outside the event horizon, $r>2M$, the positive branch of the momentum must be selected.}
In the massless sector ($m\to 0$), one finds
\begin{equation}
p(E,r)=\frac{\sqrt{1+\lambda}}{f(r)}\,E,
\qquad 
\frac{\mathrm{d}p}{\mathrm{d}E}=\frac{\sqrt{1+\lambda}}{f(r)} .
\label{eq:p_massless}
\end{equation}
The density of one--particle states per unit volume (in momentum space) is therefore
\begin{equation}
g(E;r)=\frac{1}{2\pi^2}\,p^2(E,r)\,\frac{\mathrm{d}p}{\mathrm{d}E}
=\frac{1}{2\pi^2}\,\frac{(1+\lambda)^{3/2}}{f(r)^3}\,E^2 .
\label{eq:dos_massless}
\end{equation}
Using Bose--Einstein statistics, the local spectral number density and the local spectral
energy density follow as
\begin{align}
n_E(r) \equiv \frac{\mathrm{d}n}{\mathrm{d}E}
&=\frac{g(E;r)}{e^{\beta E}-1}
=\frac{1}{2\pi^2}\,\frac{(1+\lambda)^{3/2}}{f(r)^3}\,
\frac{E^2}{e^{\beta E}-1},
\label{eq:spectral_number}\\[4pt]
u_E(r) \equiv \frac{\mathrm{d}U}{\mathrm{d}E}
&=\frac{E\,g(E;r)}{e^{\beta E}-1}
=\frac{1}{2\pi^2}\,\frac{(1+\lambda)^{3/2}}{f(r)^3}\,
\frac{E^3}{e^{\beta E}-1}.
\label{eq:spectral_energy}
\end{align}
Equations \eqref{eq:spectral_number}--\eqref{eq:spectral_energy} show that the Planckian
shape in $E$ is preserved, while the bumblebee sector produces a multiplicative deformation
through $(1+\lambda)^{3/2}/f(r)^3$. In particular, the near--horizon behavior $f(r)\to 0^+$
leads to a strong enhancement of the local spectral densities, whereas the asymptotic limit
$f(r)\to 1$ yields finite plateaus proportional to $(1+\lambda)^{3/2}$.

\subsection{Transport in the kinetic picture: flux, diffusion and thermal conductivity}
\label{subsec:transport}

A convenient kinetic description of transport follows by combining: (i) the group velocity
from the Modified Dispersion Relation (MDR) and (ii) a phenomenological mean free path $\ell_{\rm mfp}(r)$.
For massless modes, {the MDR obtained in Eq.~(\ref{metric}) implies} the group velocity as shown in Eq. (\ref{masslessvelocity}),
\begin{equation}
v_g(r)=\frac{f(r)}{\sqrt{1+\lambda}},
\nonumber
\end{equation}
which approaches $v_g\to 1/\sqrt{1+\lambda}$ asymptotically ($r\to\infty$) and vanishes at
the horizon.

\paragraph{Energy flux.}
For an isotropic gas of massless bosons, the kinetic estimate for the radial energy flux reads
\begin{equation}
J_U(r)\simeq -\kappa_{\rm th}(r)\,\partial_r T(r),
\end{equation}
where $\kappa_{\rm th}$ is the thermal conductivity. In terms of the local heat capacity at
constant volume, $C_V(r)$, one may use the standard kinetic approximation
\begin{equation}
\kappa_{\rm th}(r)\simeq \frac{1}{3}\,C_V(r)\,v_g(r)\,\ell_{\rm mfp}(r).
\label{eq:kappa_kinetic}
\end{equation}

\paragraph{Diffusion coefficient.}
The thermal diffusion coefficient (or, more generally, a diffusion coefficient for energy carriers)
is estimated by
\begin{equation}
D(r)\simeq \frac{1}{3}\,v_g(r)\,\ell_{\rm mfp}(r)
=\frac{f(r)}{3\sqrt{1+\lambda}}\,\ell_{\rm mfp}(r).
\label{eq:diffusion}
\end{equation}

\paragraph{Scaling of $\kappa_{\rm th}$ from the MDR thermodynamics.}
In the massless sector, the ensemble analysis yields thermodynamic quantities with the same
geometric block $(1+\lambda)^{3/2}/f(r)^3$ that appears in the spectral density. In particular,
since $U(r)\propto [(1+\lambda)^{3/2}/f(r)^3]\,T^4$ in the present background, one has
\begin{equation}
C_V(r)=\left(\frac{\partial U}{\partial T}\right)_V
\propto \frac{(1+\lambda)^{3/2}}{f(r)^3}\,T^3.
\label{eq:cv_scaling}
\end{equation}
Combining \eqref{masslessvelocity}, \eqref{eq:kappa_kinetic}, and \eqref{eq:cv_scaling},
the thermal conductivity scales as
\begin{equation}
\kappa_{\rm th}(r)\propto
\left(\frac{(1+\lambda)^{3/2}}{f(r)^3}\,T^3\right)
\left(\frac{f(r)}{\sqrt{1+\lambda}}\right)\,\ell_{\rm mfp}(r)
=
\frac{1+\lambda}{f(r)^2}\,T^3\,\ell_{\rm mfp}(r).
\label{eq:kappa_scaling}
\end{equation}
{Equation \eqref{eq:kappa_scaling} shows the competing roles of the two factors:} the bumblebee parameter enhances transport through $(1+\lambda)$, while the near--horizon regime is dominated by the strong redshift factor $f(r)^{-2}$.

\subsection{GUP coupled to the MDR}
\label{subsec:gup_mdr}

The MDR fixes $p(E,r)$, so generalized--uncertainty-principle (GUP) corrections can be incorporated in a direct way. We adopt the isotropic GUP
\begin{equation}
\Delta x\,\Delta p \ge \frac{1}{2}\left(1+\beta_{\rm GUP}\,(\Delta p)^2\right),
\label{eq:gup_basic}
\end{equation}
where $\beta_{\rm GUP}$ is the GUP deformation parameter (not to be confused with the inverse temperature $\beta\equiv 1/T$ used in Bose--Einstein factors). A standard implementation amounts to deforming the momentum--space measure as
\begin{equation}
\mathrm{d}^3p \ \to\ \frac{\mathrm{d}^3p}{\left(1+\beta_{\rm GUP}\,p^2\right)^3}.
\label{eq:measure_deform}
\end{equation}
Accordingly, the density of states in \eqref{eq:dos_massless} becomes
\begin{equation}
g_{\rm GUP}(E;r)=\frac{g(E;r)}{\left(1+\beta_{\rm GUP}\,p(E,r)^2\right)^3}.
\label{eq:dos_gup_def}
\end{equation}
Using \eqref{eq:p_massless}, namely $p(E,r)^2=(1+\lambda)\,E^2/f(r)^2$, one obtains
\begin{equation}
g_{\rm GUP}(E;r)=
\frac{1}{2\pi^2}\,\frac{(1+\lambda)^{3/2}}{f(r)^3}\,E^2
\left[1+\beta_{\rm GUP}\,\frac{(1+\lambda)\,E^2}{f(r)^2}\right]^{-3}.
\label{eq:dos_gup_explicit}
\end{equation}
The GUP--deformed spectral densities are therefore
\begin{equation}
n_E^{\rm GUP}(r)=\frac{g_{\rm GUP}(E;r)}{e^{\beta E}-1},
\qquad
u_E^{\rm GUP}(r)=\frac{E\,g_{\rm GUP}(E;r)}{e^{\beta E}-1}.
\label{eq:spectra_gup}
\end{equation}
For $\beta_{\rm GUP}\,(1+\lambda)\,E^2/f(r)^2\ll 1$, one finds the controlled expansion
\begin{equation}
g_{\rm GUP}(E;r)\approx g(E;r)\left[
1-3\,\beta_{\rm GUP}\,\frac{(1+\lambda)\,E^2}{f(r)^2}
+\mathcal{O}(\beta_{\rm GUP}^2)
\right].
\label{eq:dos_gup_expand}
\end{equation}
This route propagates consistently into thermodynamic integrals and transport coefficients
through the replacement $g\to g_{\rm GUP}$. A compact estimate follows by identifying the
characteristic localization length with a horizon--scale proper length. Since the radial sector
is stretched by $\sqrt{1+\lambda}$, we take
\begin{equation}
\Delta x \sim 2M\,\sqrt{1+\lambda}.
\label{eq:dx_choice}
\end{equation}
Saturating \eqref{eq:gup_basic} and solving perturbatively for $\Delta p$ in the regime
$\beta_{\rm GUP}/\Delta x^2\ll 1$ yields
\begin{equation}
\Delta p \approx \frac{1}{2\,\Delta x}+\frac{\beta_{\rm GUP}}{8\,\Delta x^3}
+\mathcal{O}(\beta_{\rm GUP}^2).
\label{eq:dp_gup_expand}
\end{equation}
Associating $E\sim \Delta p$ and using $T\sim E/(2\pi)$, one finds
\begin{equation}
T_{\rm eff}^{\rm GUP}\approx
\frac{1}{8\pi\,M\,\sqrt{1+\lambda}}
\left[
1+\frac{\beta_{\rm GUP}}{16\,M^2\,(1+\lambda)}+\mathcal{O}(\beta_{\rm GUP}^2)
\right].
\label{eq:Teff_gup}
\end{equation}
The leading term reproduces the bumblebee suppression $T_{\rm eff}\propto 1/\sqrt{1+\lambda}$, while the GUP contribution introduces a higher--order correction controlled by $\beta_{\rm GUP}$.

\subsection{GUP coupled to the MDR}
\label{subsec:gup_mdr}

The MDR fixes $p(E,r)$, so generalized--uncertainty-principle (GUP) corrections can be incorporated in a direct way. We adopt the isotropic GUP
\begin{equation}
\Delta x\,\Delta p \ge \frac{1}{2}\left(1+\beta_{\rm GUP}(\Delta p)^2\right),
\label{eq:gup_basic}
\end{equation}
where $\beta_{\rm GUP}$ is the GUP deformation parameter (not to be confused with the inverse temperature $\beta\equiv 1/T$ used in Bose--Einstein factors). A standard implementation amounts to deforming the momentum--space measure as
\begin{equation}
\mathrm{d}^3p \ \to\ \frac{\mathrm{d}^3p}{\left(1+\beta_{\rm GUP}\, p^2\right)^3}.
\label{eq:measure_deform}
\end{equation}
Accordingly, the density of states in \eqref{eq:dos_massless} becomes
\begin{equation}
g_{\rm GUP}(E;r)=\frac{g(E;r)}{\left(1+\beta_{\rm GUP}\, p(E,r)^2\right)^3}.
\label{eq:dos_gup_def}
\end{equation}
Using \eqref{eq:p_massless}, namely $p(E,r)^2=(1+\lambda)E^2/f(r)^2$, one obtains
\begin{equation}
g_{\rm GUP}(E;r)=
\frac{1}{2\pi^2}\,\frac{(1+\lambda)^{3/2}}{f(r)^3}\,E^2
\left[1+\beta_{\rm GUP}\,\frac{(1+\lambda)E^2}{f(r)^2}\right]^{-3}.
\label{eq:dos_gup_explicit}
\end{equation}
The GUP--deformed spectral densities are therefore
\begin{equation}
n_E^{\rm GUP}(r)=\frac{g_{\rm GUP}(E;r)}{e^{\beta E}-1},
\qquad
u_E^{\rm GUP}(r)=\frac{E\,g_{\rm GUP}(E;r)}{e^{\beta E}-1}.
\label{eq:spectra_gup}
\end{equation}
For $\beta_{\rm GUP}(1+\lambda)E^2/f(r)^2\ll 1$, one finds the controlled expansion
\begin{equation}
g_{\rm GUP}(E;r)\approx g(E;r)\left[
1-3\beta_{\rm GUP}\,\frac{(1+\lambda)E^2}{f(r)^2}
+\mathcal{O}(\beta_{\rm GUP}^2)
\right].
\label{eq:dos_gup_expand}
\end{equation}
This route propagates consistently into thermodynamic integrals and transport coefficients through the replacement $g\to g_{\rm GUP}$. A compact estimate follows by identifying the characteristic localization length with a horizon--scale proper length. Since the radial sector is stretched by $\sqrt{1+\lambda}$, we take
\begin{equation}
\Delta x \sim 2M\sqrt{1+\lambda}.
\label{eq:dx_choice}
\end{equation}
Saturating \eqref{eq:gup_basic} and solving perturbatively for $\Delta p$ in the regime $\beta_{\rm GUP}/\Delta x^2\ll 1$ yields
\begin{equation}
\Delta p \approx \frac{1}{2\Delta x}+\frac{\beta_{\rm GUP}}{8\Delta x^3}
+\mathcal{O}(\beta_{\rm GUP}^2).
\label{eq:dp_gup_expand}
\end{equation}
Associating $E\sim \Delta p$ and using $T\sim E/(2\pi)$, one finds
\begin{equation}
T_{\rm eff}^{\rm GUP}\approx
\frac{1}{8\pi M\sqrt{1+\lambda}}
\left[
1+\frac{\beta_{\rm GUP}}{16M^2(1+\lambda)}+\mathcal{O}(\beta_{\rm GUP}^2)
\right].
\label{eq:Teff_gup}
\end{equation}
The leading term reproduces the bumblebee suppression $T_{\rm eff}\propto 1/\sqrt{1+\lambda}$, while the GUP contribution introduces a higher--order correction controlled by $\beta_{\rm GUP}$.


\section{Interparticle potential}
\label{sec3}

The interaction energy between particles is determined by first identifying the propagation properties of the underlying field. The starting point is the dispersion relation implied by Eq.~(\ref{hamiltonian}), which fixes the analytic behavior of the propagator. Once this structure is specified, the interparticle potential emerges from the pole contributions of the Green’s function, since these singularities control the spatial dependence of the interaction. This procedure yields the explicit form of $V(r)$ without assuming it beforehand and applies equally to excitations with zero mass and to modes with nonvanishing mass. The construction is completed by expressing the propagator in a representation that makes the pole structure manifest
\ie
G(p) = \frac{1}{m^{2} \mathrm{A}(r) {+} \mathrm{A}(r)\mathrm{B}(r) p^{2}} = \frac{1}{\alpha(r)^{2} + \beta(r)^{2}}.
\fe
Introducing the radial functions $\alpha(r)^2 = m^2 \mathrm{A}(r)$ and $\beta(r)^2 = \mathrm{A}(r)\mathrm{B}(r)\,p^2$, the interaction potential is constructed from the Green function by changing its representation. The propagator is first written in momentum space and then mapped to configuration space through a Fourier transform. This operation transfers the momentum dependence into an explicit radial dependence, from which the interparticle potential $V(r)$ follows directly. Carrying out this procedure leads to the expression reported below, in agreement with standard integral representations and previously established results \cite{blackledge2005digital,gradshteyn2014table,filho2025modified,filho2025particlemotion}:
\ie
\begin{split}
\label{interparticlepotential}
V(r) &= \int \frac{\mathrm{d}^{3}p}{(2\pi)^{3}} e^{i{\bf{p}} \cdot {\bf{r}}} G(p)\\
& = \frac{1}{(2\pi)^3} \int_{0}^{\infty} \mathrm{d}p \, p^2 \int_{0}^{\pi} \int^{2\pi}_{0} \mathrm{d} \varphi \, \mathrm{d}\theta \, \sin(\theta) \, e^{i p r \cos(\theta)} \, G(p) \\
& = \frac{1}{2 r \pi^2} \int_{0}^{\infty} \mathrm{d} p \, p \sin(p\, r) \, G(p) \\
&= \frac{1}{2 r \pi^2} \int_{0}^{\infty} \mathrm{d} p \, p \sin(p \,r) \, \left[   \frac{1}{\alpha(r)^{2} + \beta(r)^{2} p^{2}} \right] \\
& = \frac{e^{-\frac{\alpha(r) r}{\beta(r)} }}{4\pi r \beta(r)^{2}} = \frac{(\lambda +1) r\, e^{-m \sqrt{\frac{(\lambda +1) r^3}{{-}2 M{+}r}}}}{4 \pi  (r-2 M)^2}, \quad \forall \quad (r-2 M){>}0.
\end{split}
\fe
The above result shows that, for the massive case ($m>0$), the interparticle potential remains confined {outside} the black hole—at least within the assumptions adopted in this work. It is also worth noting that, if one were to consider $r - 2M {<} 0$ instead, the exponential term would acquire an imaginary component, leading to potentially nonphysical results. Nevertheless, in the massless case ($m=0$), the corresponding interparticle potential $V_{0}(r)$ does not exhibit this behavior, as we shall see below. In other words, a well--defined positive value persists outside the black hole{, in agreement with the analysis of $V(r)$}.

In the considered black hole scenario, the spacetime curvature induces an effective interaction that combines features of Coulomb and Yukawa potentials when massive particles are involved. The overall intensity of this mixed interaction is regulated by the Lorentz--violating parameter $\lambda$, which acts as a coupling measure. Notably, the resulting potential $V(r)$ approaches zero at both extremes of the radial coordinate, fulfilling the asymptotic conditions $V(r \to 0) = 0$ and $V(r \to \infty) = 0$.

\begin{figure}
    \centering
     \includegraphics[scale=0.6]{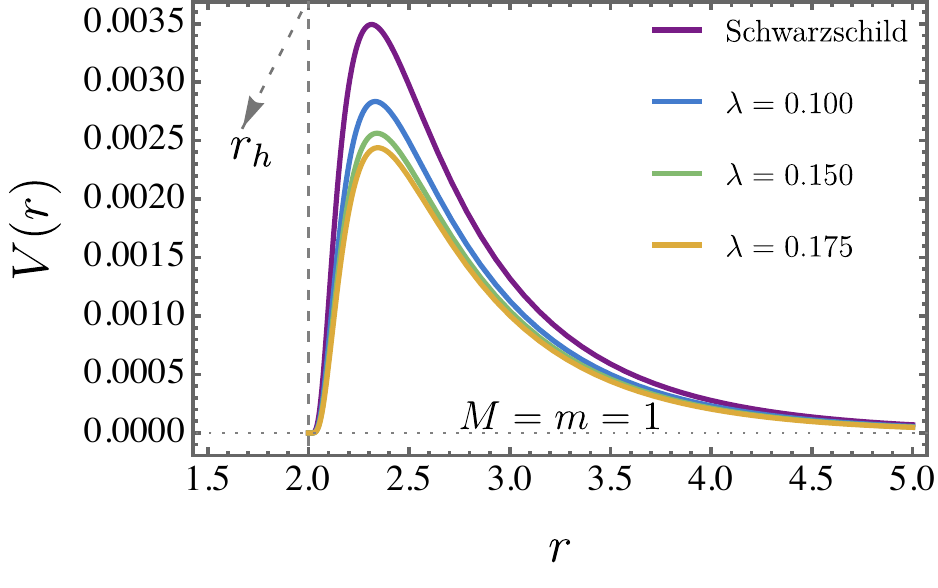}
    \caption{The plot displays how the interparticle potential $V(r)$ behaves for massive modes across different values of the Lorentz--violating parameter $\lambda$, showing its dependence on the radial coordinate $r$. To highlight deviations induced by $\lambda$, the conventional Schwarzschild configuration is also represented for direct comparison.}
    \label{ineftegrnpgnarticvblepvbotentialmassive}
\end{figure}

Figure~\ref{ineftegrnpgnarticvblepvbotentialmassive} depicts the radial profile of the potential $V(r)$ for various choices of $\lambda$, emphasizing how the interparticle force varies throughout the black hole spacetime. Within the optical-–mechanical framework, $V(r)$ represents the effective energy configuration encountered by a test particle propagating in a bumblebee black hole geometry. One {important} aspect of the behavior is that the potential remains entirely confined {outside} the event horizon, which effectively acts as a barrier preventing massive particles from escaping. Consistently, the boundary conditions $V(0)=0$ and $V(r_h)=0$ are satisfied.

\begin{figure}
    \centering
     \includegraphics[scale=0.55]{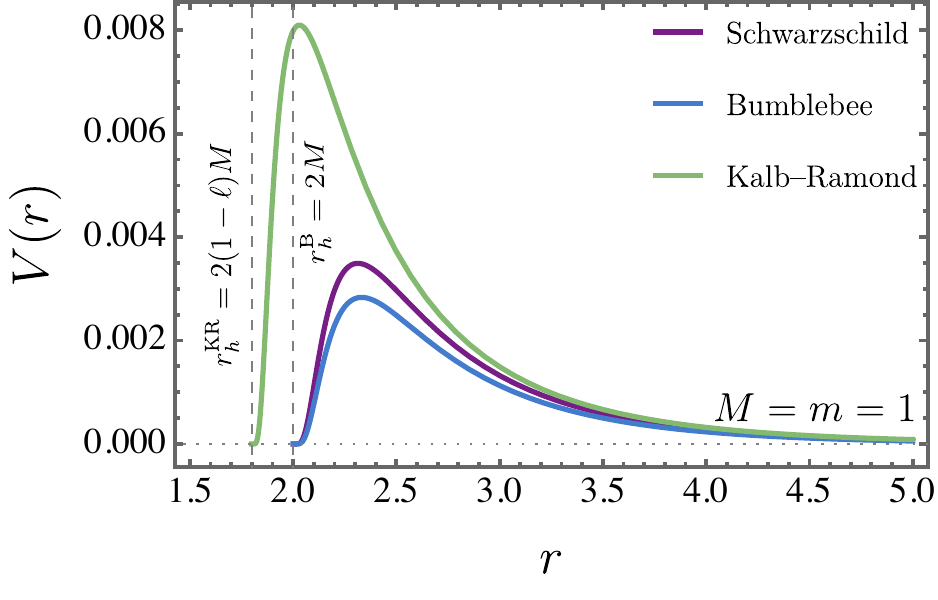}
    \caption{Comparison of the potential $V(r)$ for Schwarzschild, Bumblebee, and Kalb--Ramond black holes, with the Lorentz--violating parameter set to $\lambda = 0.1$ for the latter two cases.}
    \label{intercomp}
\end{figure}

As the parameter $\lambda$ increases, the potential $V(r)$ {decreases} until it reaches its maximum values at $r \approx 2.35$ for $\lambda = 0.100$, $r \approx 2.35$ for $\lambda = 0.150$, and $r \approx 2.35$ for $\lambda = 0.175$. Figure~\ref{intercomp} provides a comparative view of $V(r)$ for the present model alongside the corresponding profiles for the Kalb–Ramond and Schwarzschild geometries. Among them, the Kalb–Ramond black hole presents the highest potential intensity, exceeding those found in the bumblebee and Schwarzschild configurations. It is important to note that all Lorentz--violating parameters were fixed at $\lambda = \ell= 0.1$ for consistency. Here, $\ell$ represents the Lorentz--violating parameter coming from Kalb--Ramond gravity.

An immediate question that naturally arises concerns the behavior of the potential when the particle mass is absent. To address this point, we consider the limit $m_0 \rightarrow 0$, leading to
\ie
V_{0}(r) = \frac{(\lambda + 1) r}{4\pi (r - 2M)^2}.
\label{potentialm0}
\fe

When the rest mass vanishes, the potential transitions to a purely massless form, $V_{0}(r)$, which manifests a modified interaction resembling a Coulomb--like dependence. In this regime, the underlying black hole geometry allows for nontrivial interactions even among massless particles, such as possible photon--photon coupling effects.

To better understand the behavior of $V_0(r)$, its asymptotic properties can be examined. As the radial coordinate approaches the event horizon ($r \to r_h$), the potential diverges, signaling a singular behavior near that region. On the other hand, in the limit $r \to 0$, $V_0(r)$ approaches zero, indicating the absence of singularities inside the horizon and suggesting a suppressed gravitational influence near the core. At large distances ($r \to \infty$), the potential again tends to zero, as expected from the decay of gravitational effects at infinity.

\begin{figure}
    \centering
     \includegraphics[scale=0.74]{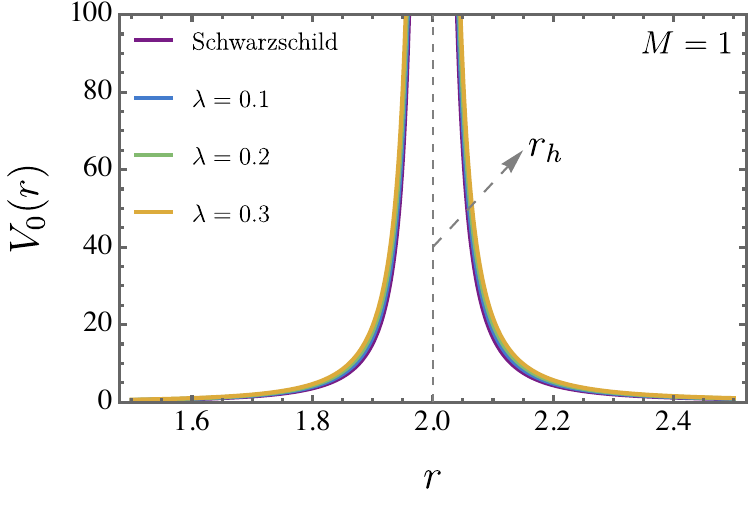}
      \includegraphics[scale=0.72]{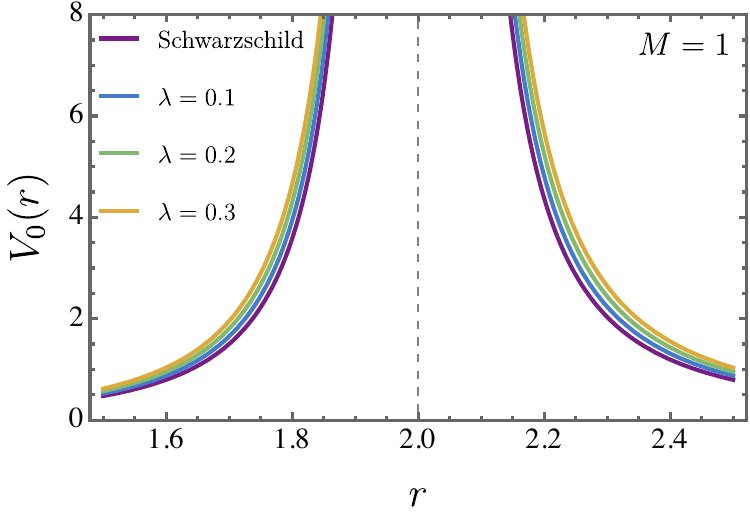}
    \caption{The profile of the massless--mode potential $V_0(r)$ is displayed as a function of the radial coordinate $r$ for different values of the Lorentz--violating parameter $\lambda$. The Schwarzschild case is shown alongside as a baseline for comparison.}
    \label{ihntegrpfarsticlepaotaentaiaalmaaasslesscase}
\end{figure}

These features are illustrated in Fig.~\ref{ihntegrpfarsticlepaotaentaiaalmaaasslesscase}, where the potential profile is shown for different values of the Lorentz--violating parameter $\lambda$. As previously discussed, the plots show that variations in $\lambda$ leave the event horizon position unchanged—since $\lambda$ does not contribute to the horizon radius—while progressively increasing the overall {intensity} of the potential $V_0(r)$ with larger $\lambda$ values.

\section{Electron scattering cross section}
\label{sec4}

The procedure adopted for the forthcoming analysis follows the general framework developed in Ref.~\cite{Touati:2024tqy}. To evaluate the Lorentz--violating scattering amplitude, denoted by $f(\lambda,q)$, we make use of the potential previously obtained in Eq.~(\ref{interparticlepotential}) within the Born approximation scheme. Accordingly, the scattering amplitude takes the form
\begin{equation}
\label{ampliscat}
f(q) = -\frac{2m}{\hbar^{2} q} \int_{0}^{\infty} \! V(r)\, r \sin(q\,r)\, \mathrm{d}r,
\end{equation}
where $q = 2\kappa \sin\!\left(\tfrac{\theta}{2}\right)$, with $\kappa$ denoting the wave number of the incoming particle, which, in this case, is an electron.
Nevertheless, by direct inspection, substituting Eq.~(\ref{interparticlepotential}) into Eq.~(\ref{ampliscat}) reveals that an analytical expression cannot be obtained in closed form. To address this difficulty, we expand the interparticle potential up to first order in $M$, which corresponds to adopting the weak--field approximation \cite{gibbons2008applications}. Under this assumption, it takes the form
\ie
\label{approVVV}
V(r) \approx \,\, \frac{(\lambda +1) e^{- \sqrt{\lambda +1} m r}}{4 \pi  r} + \frac{(\lambda +1) M e^{- \sqrt{\lambda +1} m r} \left(4- \sqrt{\lambda +1} m r\right)}{4 \pi  r^2} + \mathcal{O}(M^{2}). 
\fe

{In addition}, by performing the integration by substituting Eq.~(\ref{approVVV}) into Eq.~(\ref{ampliscat}), we obtain
\ie
\label{fappro}
\begin{split}
f(\lambda,q) = &  {\frac{(\lambda +1) m \left(\frac{q \left(\sqrt{\lambda +1} m M-1\right)}{(\lambda +1) m^2+q^2}-4 M \cot ^{-1}\left(\frac{\sqrt{\lambda +1} m}{q}\right)\right)}{2 \pi  \hbar^2 q}} \\
& \approx \, \, {\frac{m \left(\frac{q (m M-1)}{m^2+q^2}-4 M \tan ^{-1}\left(\frac{q}{m}\right)\right)}{2 \pi  \hbar^2 q}}\\
& {+\frac{\lambda  \left(5 m^4 M q-8 m M \left(m^2+q^2\right)^2 \tan ^{-1}\left(\frac{q}{m}\right)+m q^3 (7 m M-2)\right)}{4 \pi  \hbar^2 q \left(m^2+q^2\right)^2}}
\end{split}
\fe

{
Although the bumblebee black hole metric contains the mass parameter $M$, the Lorentz--violating contribution governed by $\lambda$ also appears in the mass--independent sector. After the redefinition of the metric signature, the weak--field expression in Eq.~(\ref{fappro}) shows that the scattering amplitude contains an $\mathcal{O}(M^0)$ contribution, a pure mass correction, and a mixed term proportional to $\lambda M$. In other words, the expansion keeps the gravitational sector while organizing the amplitude into distinct contributions associated with the bumblebee deformation, the black hole mass, and their combined effect.

The $\mathcal{O}(M^0)$ sector shows that Lorentz symmetry breaking already modifies the momentum--transfer dependence of the amplitude even in the absence of the explicit black hole mass contribution. The terms proportional to $M$ describe the gravitational correction, whereas the $\lambda M$ contribution measures how the bumblebee deformation changes the mass--dependent part of the scattering process. }

\begin{figure}
    \centering
     \includegraphics[scale=0.55]{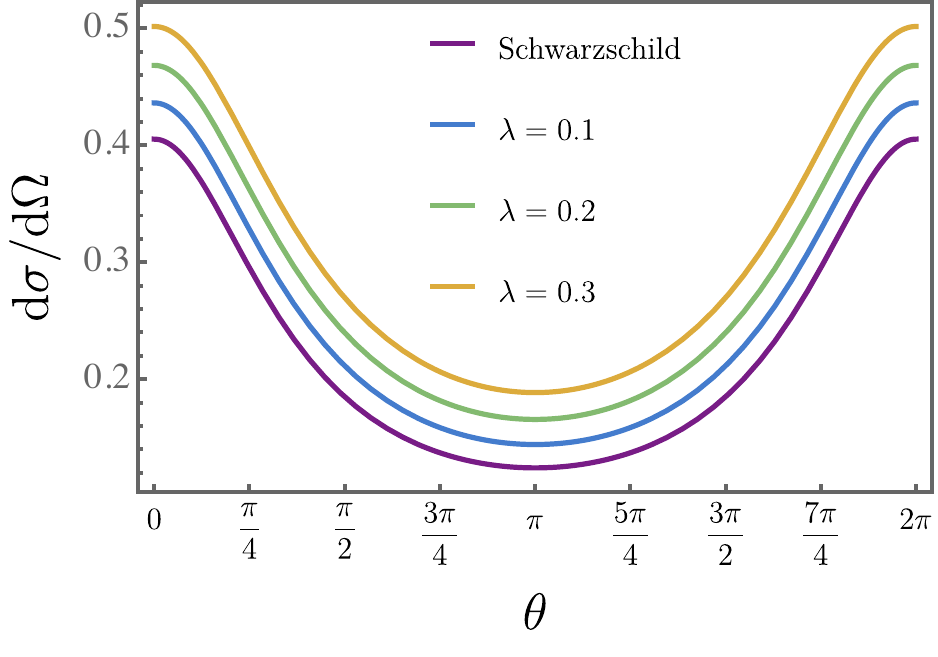}
    \caption{Dependence of the differential cross section $\mathrm{d}\sigma/\mathrm{d}\Omega$ on the scattering angle $\theta$ for different values of the Lorentz–violating parameter $\lambda$.  } 
    \label{dsigma}
\end{figure}

Based on these previous argumentations, let us proceed with the calculations.
\ie
\label{crosssection}
\frac{\mathrm{d}\sigma}{\mathrm{d}\Omega} = |f(\lambda,q)|^{2},
\fe
where $\mathrm{d}\Omega$ represents the differential element of solid angle. To proceed further, let us substitute Eq. (\ref{fappro}) in Eq. (\ref{crosssection}) so that
\ie
\begin{split}
\frac{\mathrm{d}\sigma}{\mathrm{d}\Omega} & =  \left| { \frac{m \left(\frac{q (m M-1)}{m^2+q^2}-4 M \tan ^{-1}\left(\frac{q}{m}\right)\right)}{2 \pi  \hbar^2 q}} \right.\\
& \left. { +\frac{\lambda  \left(5 m^4 M q-8 m M \left(m^2+q^2\right)^2 \tan ^{-1}\left(\frac{q}{m}\right)+m q^3 (7 m M-2)\right)}{4 \pi  \hbar^2 q \left(m^2+q^2\right)^2} } \right|^2.
\end{split}
\fe
It is worth emphasizing that, from Eq.~(\ref{fappro})—or equivalently from Eq.~(\ref{crosssection})— {no} divergence arises {in this case}. Figure~\ref{dsigma} depicts the variation of $\mathrm{d}\sigma/\mathrm{d}\Omega$ with respect to $\theta$ for different values of $\lambda$, {while keeping $m=h=\kappa = M=1$.

}

\begin{figure}
    \centering
     \includegraphics[scale=0.7]{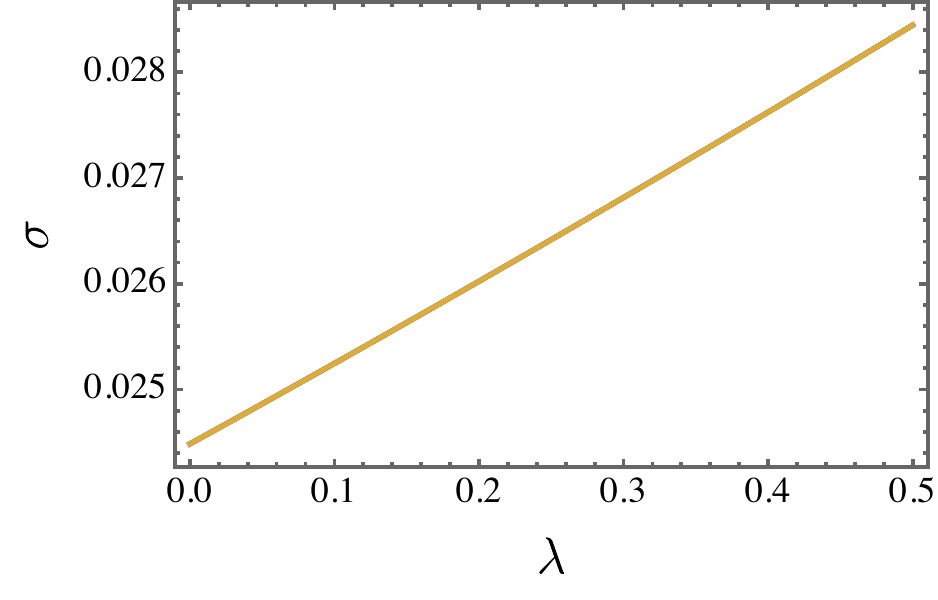}
    \caption{Total cross section $\sigma$ plotted as a function of the Lorentz–violating parameter $\lambda$.}
    \label{sigmaaa}
\end{figure}

In addition, notice that Eq. (\ref{crosssection}) can be rewritten as 
\ie
\sigma=\int \mathrm{d}\Omega\,|f(\theta)|^{2}
= \int_{0}^{\pi} 2\pi \sin\theta\,|f(\theta)|^{2}\,\mathrm{d}\theta,
\fe
with $q=2\kappa\sin(\tfrac{\theta}{2})\Rightarrow \mathrm{d}q=\kappa\cos(\tfrac{\theta}{2})\,\mathrm{d}\theta$ and
$ \mathrm{d}\Omega=2\pi\sin\theta\,\mathrm{d}\theta
=4\pi\sin\!\tfrac{\theta}{2}\cos\!\tfrac{\theta}{2}\,\mathrm{d}\theta
=\frac{2\pi}{\kappa^{2}}\,q\,\mathrm{d}q.$ In this manner,  the total cross section within the framework of the first--order Born approximation can be determined through the integral formulation given by
\ie
\sigma = \frac{2\pi}{\kappa^{2}} \int^{2\kappa}_{0} |f(\lambda,q)|^{2} q\, \mathrm{d}q,
\fe
and substituting Eq. (\ref{fappro}) in above expression {and considering the limiting case where $M\to 0$}, we obtain
\ie
\begin{split}
\sigma & = \frac{2\pi}{\kappa^{2}} \int^{2\kappa}_{0} |f(\lambda,q)|^{2} q\, \mathrm{d}q \\
& = \frac{16 \kappa ^4 (\lambda  (\lambda +3)+3)+3 m^4 {+} 12 \kappa ^2 (\lambda +2) m^2}{3 \pi  \hbar^4 \left(m^2 {+} 4 \kappa ^2\right)^3},\\
& \forall\, \{m, \kappa, \lambda, \hbar, q\}:\, 
m > 0,\, \kappa > 0,\, \lambda \in \mathbb{R},\, \hbar > 0,\, 
0 \le q \le 2\kappa,\, 2\kappa < m.
\end{split}
\fe
Next, we analyze the dependence of the total cross section $\sigma$ on the Lorentz--violating parameter $\lambda$, as depicted in Fig.~\ref{sigmaaa}. The figure shows that $\sigma$ {increases} progressively as $\lambda$ increases.

From the result obtained above, several remarks can be made. To begin with, in the absence of Lorentz violation, that is, in the limit $\lambda \to 0$, the cross section reduces to
\ie
\lim_{\lambda \to 0} \sigma = \frac{1}{\pi  \hbar^4 m^2-4 \pi  \hbar^4 \kappa ^2}.
\fe
Secondly, in the limit corresponding to a massless configuration, $\lim_{m \to 0} \sigma$, the expression becomes
\ie
\lim_{m \to 0} \sigma = \frac{\lambda  (\lambda +3)+3}{12 \pi  \hbar^4 \kappa ^2}.
\fe
Thirdly, when both the massless limit is taken and the Lorentz–violating effects are absent, the cross section reduces to
\ie
\lim_{m \to 0, \, \, \lambda \to 0} \sigma = -\frac{1}{4 \pi  \hbar^4 \kappa ^2}.
\fe

For investigating the behavior of $\sigma$ as a function of energy, let us redefine the variable conveniently, as shown below:
\begin{equation}
x \equiv \frac{2\kappa}{m} \in [0,1)
\;\Longleftrightarrow\;
E = \frac{\hbar^{2}\kappa^{2}}{2m}
= \frac{\hbar^{2} m}{8}\, x^{2}.
\end{equation}

Accordingly, the cross section takes the form
\begin{equation}
\sigma(x) =
\frac{1}{3\,\hbar^{4}\pi\,m^{2}}\,
\frac{\,3 {+} 3x^{2}(2 + \lambda)
+ x^{4}\big(3 + \lambda (3 + \lambda)\big)\,}
{{(1 + x^{2})^{3}}},
\qquad 0 \le x < 1.
\end{equation}

\begin{figure}
    \centering
     \includegraphics[scale=0.7]{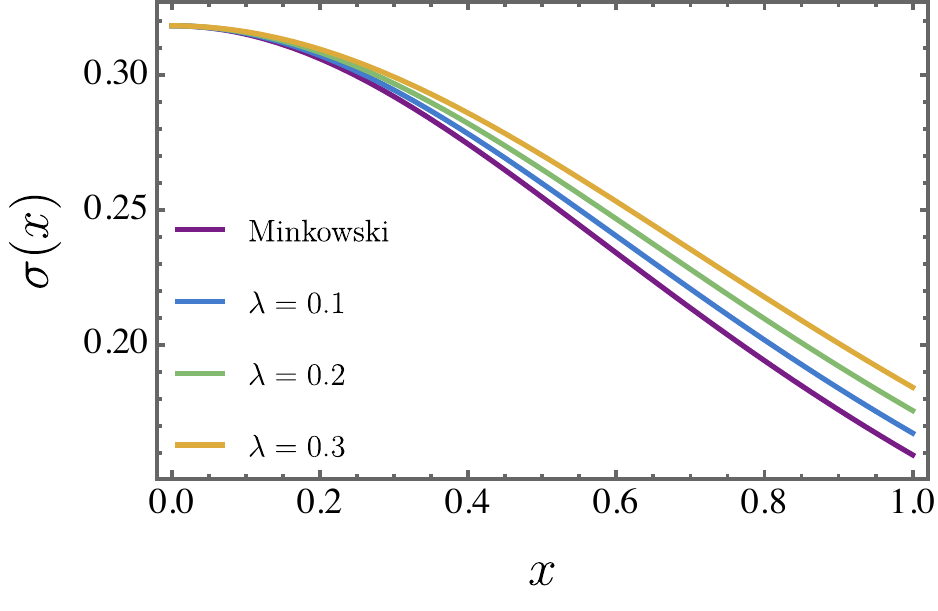}
    \caption{The cross section $\sigma(x)$ is plotted as a function of $x$ for various values of $\lambda$. The Minkowski case is also included for reference.}
    \label{sigxx}
\end{figure}

Here, the parameter $x$ accounts for a dimensionless measure of the particle kinetic energy, restricted to $x < 1$, which corresponds to the endpoint condition $2\kappa=m$. {We notice that $\sigma(x)$ approaches the finite value when}
\begin{equation}
{
\lim_{x\to 1^{-}}\sigma(x)
=
\frac{\lambda^{2}+6\lambda+12}{24\,\pi\,\hbar^{4}m^{2}} .
}
\end{equation}
{ 

The endpoint $x\to 1^{-}$ corresponds to the limiting value  $E=\hbar^{2}m/8$, associated with the condition $2\kappa<m$.  Since $\sigma(x)$ remains finite in this limit, this endpoint does not  represent a resonance or a singular point of the cross section. For the phenomenological range $0\leq \lambda <1$, $\sigma(x)$ decreases monotonically with $x$, while remaining positive and finite throughout the whole interval $0\leq x<1$. In the Minkowski limit $\lambda=0$, one obtains}
\begin{equation}
{
\sigma(x)\big|_{\lambda=0}
=
\frac{1}{\pi\,\hbar^{4}m^{2}(1+x^{2})},
}
\end{equation}
{which also decreases from $\sigma(0)=1/(\pi\hbar^{4}m^{2})$ to $\sigma(1^{-})=1/(2\pi\hbar^{4}m^{2})$. Fig.~\ref{sigxx} displays the behavior of $\sigma(x)$ as a function of $x$ for various values of the parameter $\lambda$. For comparison, the case $\lambda=0$, corresponding to the Minkowski spacetime limit, is also shown.}

\section{Thermal properties in the ensemble approach }
\label{sec5}

The thermodynamic properties associated with particle motion near a bumblebee black hole are examined by adopting a statistical viewpoint. The analysis is framed in terms of ensemble variables, where the momentum plays a essential role in characterizing the accessible microstates. For this purpose, the Hamiltonian relation obtained earlier is not taken in its original form; it is algebraically reorganized so that the kinetic contribution emerges as the leading quantity. Only after this rearrangement does the squared momentum appear explicitly, which is the form required for constructing the ensemble description. Consequently, Eq. (\ref{hamiltonian}) is transformed to isolate $p^{2}$, leading to
\ie
p^{2} = \frac{\big(m^{2} \,\mathrm{A}(r) - E^{2}\big)}{\mathrm{A}(r)\mathrm{B}(r)}.
\fe

A variety of gravitational and field–theoretic frameworks already place thermodynamic effects driven by modified dispersion relations at the center of their analyses, ranging from rainbow gravity \cite{Filho:2023ydb} and quantum gases \cite{araujo2022does} to Lorentz--violating electrodynamics in the Myers--Pospelov scheme \cite{Anacleto:2018wlj}, Einstein--æther theory \cite{araujo2023thermodynamics}, and the thermodynamics of Ellis--type wormholes \cite{araujo2023thermodynamical}. Related developments also appear in black hole scenarios governed by bumblebee gravity \cite{filho2025modified} and by Kalb--Ramond fields \cite{AraujoFilho:2025fwd}.

Within this context, modified dispersion relations serve as an effective tool for accessing the thermodynamic behavior of high--energy astrophysical environments \cite{amelino2013quantum,Wagner:2023fmb,araujo2022thermal}. Rather than extending standard relativistic prescriptions, the deformation of the energy--momentum sector opens new routes for interpreting observational data. Observable consequences emerge in the timing structure of gamma--ray bursts, where energy--dependent delays may arise \cite{jacob2008lorentz}, in the spectral features of ultra--high--energy cosmic rays \cite{anchordoqui2003ultrahigh}, and in atypical emission patterns reported for active galactic nuclei \cite{anchordoqui2003ultrahigh}.

Rather than reproducing the familiar photon--like dispersion law, the Hamiltonian relationship in Eq. (\ref{hamiltonian}) defines the energy--momentum structure in curved spacetime. From this, a set of physical implications naturally emerges and will be addressed in the subsequent analysis. Solving the relation does not lead to a unique branch: two mathematical possibilities arise. However, only the branch that remains real and positive throughout the relevant domain is compatible with physical requirements. For this reason, the discussion that follows is restricted to that suitable solution
\ie
p  = \sqrt{\frac{ {-} m^{2} \,\mathrm{A}(r) {+} E^{2}}{\mathrm{A}(r)\mathrm{B}(r)}}.
\fe

To ``control'' the momentum measure appearing in the integrals, the differential with respect to $p$ is rewritten in a form suited to the chosen parametrization, leading to
\ie
\mathrm{d} p =  \frac{E \, \mathrm{d}E}{ \sqrt{\mathrm{A}(r) \mathrm{B}(r)} \sqrt{ {-} m^{2}\mathrm{A}(r) {+} E^2}}.
\label{vol}
\fe
The analysis then turns to the construction of the phase–space volume. Instead of keeping the momentum variable explicit, the momentum sector is integrated out, which yields the total count of available microstates. This quantity, denoted by $\Omega$, takes the form
\ie
\Omega =  \int\!\!\!\int_{0}^{\infty} p^{2}\,\mathrm{d}p \,\mathrm{d}^{3}q.\label{ms2}
\fe
Adopting this perspective reshapes the relation in Eq. (\ref{ms2}), which now reads
\begin{widetext}
\ie
\begin{split}
\Omega(r,\lambda) & =  \int\!\!\!\int_{0}^{\infty} \frac{E \sqrt{{-}\mathrm{A}(r) m^2 {+} E^2}}{\mathrm{A}(r)^{3/2} \mathrm{B}(r)^{3/2}} \,\mathrm{d}E \,\mathrm{d}^{3}q.
\end{split}
\fe
\end{widetext}

The discussion adopts natural units from the outset, so that all thermodynamic variables are expressed in a dimensionless and transparent form. Rather than beginning with specific observables, the formulation is anchored in the statistical foundation of the system. In this spirit, the analysis opens by defining the classical partition function in its most general form, in accordance with the standard ensemble approach presented in Ref.~\cite{greiner2012thermodynamics}
\ie
Z = \frac{1}{N!h^{3N}} \int\!\!\!\int \mathrm{d}q^{3N}\mathrm{d}p^{3N} e^{-\beta H(q,p)}  \equiv \int \mathrm{d}E \,\Omega(r,\lambda) e^{-\beta E}. \label{partti1}
\fe

The statistical model is built for a collection of indistinguishable particles, where internal spin states are initially left aside. Instead of listing parameters upfront, the description is organized around the phase--space dynamics: the system is characterized by a Hamiltonian $H(p,q)$ defined over generalized coordinates $q$ and their conjugate momenta $p$. The ensemble contains a fixed number $N$ of constituents, and the thermal weight is controlled by the inverse temperature $\beta = 1/(\kappa_{B}T)$, with $\kappa_{B}$ denoting the Boltzmann constant. The quantum of action is measured by Planck’s constant $h$. Because the reference expression in Eq. (\ref{partti1}) neglects any internal degeneracy, an additional factor must later be introduced to account for the spin contribution, following the prescription discussed in \cite{isihara2013statistical,wannier1987statistical,salinas1999introduccao,vogt2017statistical,mandl1991statistical}
\ie
\mathrm{ln}[Z(r,\lambda)] = - \int\!\!\!\int \mathrm{d}E \mathrm{d}^{3}q \,\Omega(r,\lambda) \mathrm{ln} [ 1 \pm e^{-\beta E}].
\fe
Notice that the choice of sign in the logarithmic factor, $\ln\!\left[1 \pm e^{-\beta E}\right]$, takes into account the quantum statistics of the particles: the lower sign implements Bose--Einstein behavior, whereas the upper one enforces Fermi--Dirac counting. Once this distinction is fixed, the statistical sum specializes accordingly. In particular, selecting the bosonic branch leads to the partition function written as
\begin{widetext}
\ie\label{lnz1}
\begin{split}
\mathrm{ln}[Z(r,\lambda)] & = {-} \int\!\!\!\int  \frac{E \sqrt{{-}\mathrm{A}(r) m^2 {+}E^2} \ln \left(1-e^{-\beta  E}\right)}{\mathrm{A}(r)^{3/2} \mathrm{B}(r)^{3/2}} \mathrm{d}E\,\mathrm{d}^{3}q.
\end{split}
\fe
\end{widetext}

The subsequent analysis develops the thermodynamic description implied by the relation obtained above. It is worth mentioning that, for the sake of simplicity, all thermodynamic quantities are evaluated per unit proper spatial volume, with the volume element defined as $\mathrm{d}^{3}q$. Accordingly, the functions derived throughout this section should be interpreted as local thermodynamic densities, evaluated at fixed radial coordinate $r$. This local treatment is appropriate in gravitational settings, where thermodynamic observables are naturally defined with respect to local static observers, while global quantities would require an additional integration over the proper spatial volume.

Also, the treatment acknowledges from the outset that massive modes ($m\neq0$) prevent the emergence of closed analytic expressions, independent of whether bosonic or fermionic statistics are imposed. For this reason, the discussion proceeds by establishing the fundamental thermodynamic observables in a formal manner, which then serve as the basis for the results presented in the sections that follow
\ie
\begin{split}
  & P (r,\lambda)= \frac{1}{\beta} \mathrm{ln}\left[Z(r,\lambda)\right], \\
 & U(r,\lambda)=-\frac{\partial}{\partial\beta} \mathrm{ln}\left[Z(r,\lambda)\right], \\
 & S(r,\lambda)=k_B\beta^2\frac{\partial}{\partial\beta}F(r,\lambda), \\
 & C(r,\lambda)=-k_B\beta^2\frac{\partial}{\partial\beta}U(r,\lambda).
\label{properties}
\end{split}
\fe

To make analytic progress possible, the discussion is confined to the limit of vanishing mass, $m \to 0$. In this regime, the thermodynamic functions admit closed expressions, in contrast with the massive case, which resists analytic treatment. Although both Bose--Einstein and Fermi--Dirac gases become tractable when $m=0$, their partition functions differ only by an overall numerical offset, equal to $-7/360$. For this reason, and in line with the strategy adopted in Refs.~\cite{filho2025modified,AraujoFilho:2025fwd}, the analysis is restricted to bosonic modes. The set of observables addressed in what follows consists of the pressure $P(r,\lambda)$, the internal energy $U(r,\lambda)$, the entropy $S(r,\lambda)$, and the heat capacity at constant volume $C_V(r,\lambda)$, with the Helmholtz free energy introduced through the relation $F(r,\lambda)=-P(r,\lambda)$.

A brief clarification is in order regarding Ref.~\cite{filho2025modified}. After presenting the partition function, that work states that analytical results are available, without explicitly separating the massive and massless sectors. Such wording could suggest that the expression given there for the general case, namely Eq.~(51), also leads to closed forms. In fact, no analytic solution exists when $m\neq0$, and numerical methods become unavoidable. The statement is correct only in the massless sector, where all thermodynamic quantities can indeed be derived in analytic form. The present analysis follows the same logic: explicit formulas arise solely in the {\bf massless} limit, while the massive regime remains beyond analytic reach.


\subsection{Pressure }

With the statistical framework fixed to bosonic excitations and the partition function already specified, the analysis now turns to the extraction of the thermodynamic observables. In this manner, the discussion is organized around three representative domains: the vicinity of the event horizon, the region of the photon sphere, and the asymptotic far field. This approach starts with the evaluation of the pressure, which takes the form
\ie
\begin{split}
\label{pressure}
P(r,\lambda) & = \frac{1}{\beta} \mathrm{ln}\left[Z(r,\lambda)\right] = \frac{1}{\beta}  \int  \frac{E \sqrt{\mathrm{A}(r) m^2-E^2} \ln \left(1-e^{-\beta  E}\right)}{\mathrm{A}(r)^{3/2} \mathrm{B}(r)^{3/2}} \mathrm{d}E \\
& = \frac{\pi ^4 (\lambda +1)^{3/2} r^3}{45 \beta ^4 (r-2 M)^3}.
\end{split}
\fe

\begin{figure}
    \centering
     \includegraphics[scale=0.51]{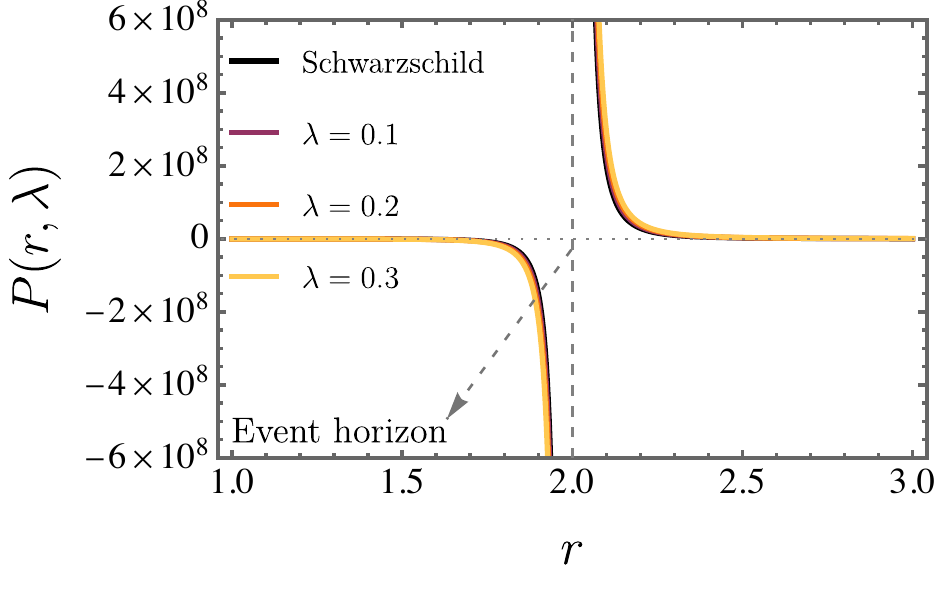}
     \includegraphics[scale=0.51]{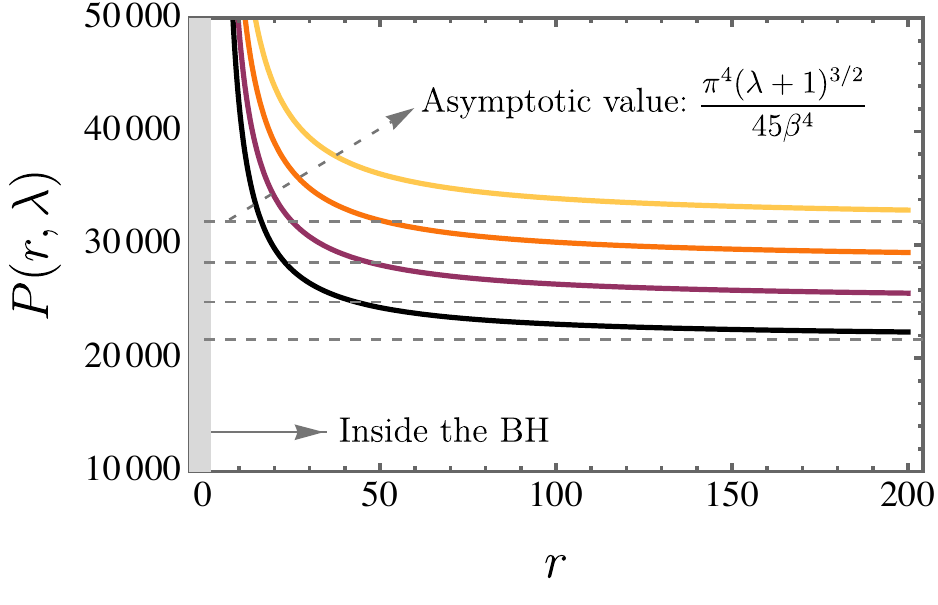}
    \caption{Radial dependence of the pressure $P(r,\lambda)$. }
    \label{pressurerr}
\end{figure}

The structure encoded in Eq.~(\ref{pressure}) gives rise to a rich set of features that become apparent once its dependence on the radial coordinate $r$ and on the parameter $\lambda$ is explored, as displayed in Fig.~\ref{pressurerr}. The discussion first identifies the qualitative behavior: the pressure can switch sign as the parameters vary, which signals the possible presence of phase--transition--like behavior. This behavior is, however, bounded by the event horizon, which acts as a natural limit for the physically accessible domain. As the horizon is approached, the pressure diverges, and immediately outside it $P(r,\lambda)$ decreased until reaching its corresponding asymptotic values (as shown in the right side of Fig.~\ref{pressurerr}). For reference, the Schwarzschild result is shown as well.

In the representation shown in the left panel, the curves seem to decay to zero at large $r$, but this impression is misleading and originates solely from the extremely wide vertical range adopted, $[-10^{9},10^{9}]$. The right panel removes this ambiguity and makes the asymptotic trend explicit: far from the black hole, the pressure does not vanish but instead converges to the finite value $\pi^{4}(\lambda+1)^{3/2}/(45\,\beta^{4})$. This plot also makes clear that physically admissible positive pressures occur only in the exterior region, $r>r_{h}$.

The analysis now shifts from radial profiles to the thermal response of the system. In this part, the pressure is explored as a function of the temperature $T$, evaluated at three representative locations: a point extremely close to the horizon, fixed at $r=1.001\,\times\, r_h$, the region associated with the photon sphere at $r=r_{ph}$, and the far--field limit at large distances. These choices are meant to sample the near--horizon, intermediate, and asymptotic regimes of the geometry. The photon sphere is adopted here merely as a convenient benchmark for the near--horizon domain, but no special role is attached to it, since any nearby radius would lead to the same qualitative behavior.


\subsubsection{Extreme near–horizon regime }

The focus now moves from spatial variations to thermal dependence. After fixing the radial coordinate and examining how the pressure responds across the geometry, the analysis proceeds to study how it evolves with the temperature $T$ in the immediate neighborhood of the event horizon. For this near--horizon regime, the pressure is therefore written in the form
\ie
\lim\limits_{r \to 1.001\times r_{h}} P(r, \lambda) = 2.17115\times 10^9 (\lambda +1)^{3/2} T^4.
\fe

\begin{figure}
    \centering
     \includegraphics[scale=0.51]{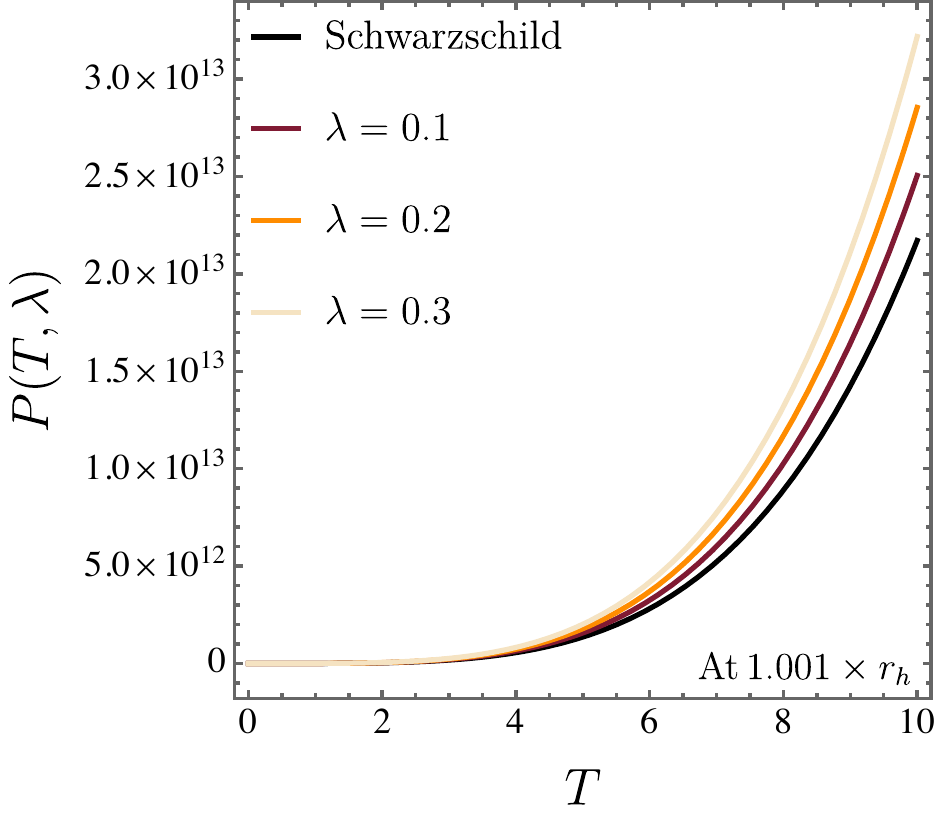}
    \caption{Pressure $P(T,\lambda)$ as a function of the temperature $T$ in the near--horizon region, fixed at $r=1.001\,\times\, r_h$.}
    \label{pressurevery}
\end{figure}

The structure of this relation becomes clear once it is confronted with the plot profiles displayed in Fig.~\ref{pressurevery}. The curves show that the pressure grows in magnitude as the Lorentz--violating parameter $\lambda$ increases. Alongside these results, the Schwarzschild case is plotted as a baseline. In other words, a direct comparison reveals that the bumblebee geometry consistently produces larger values of $P(T,\lambda)$ in the near--horizon region.

\subsubsection{Near–horizon regime }

\begin{figure}
    \centering
     \includegraphics[scale=0.51]{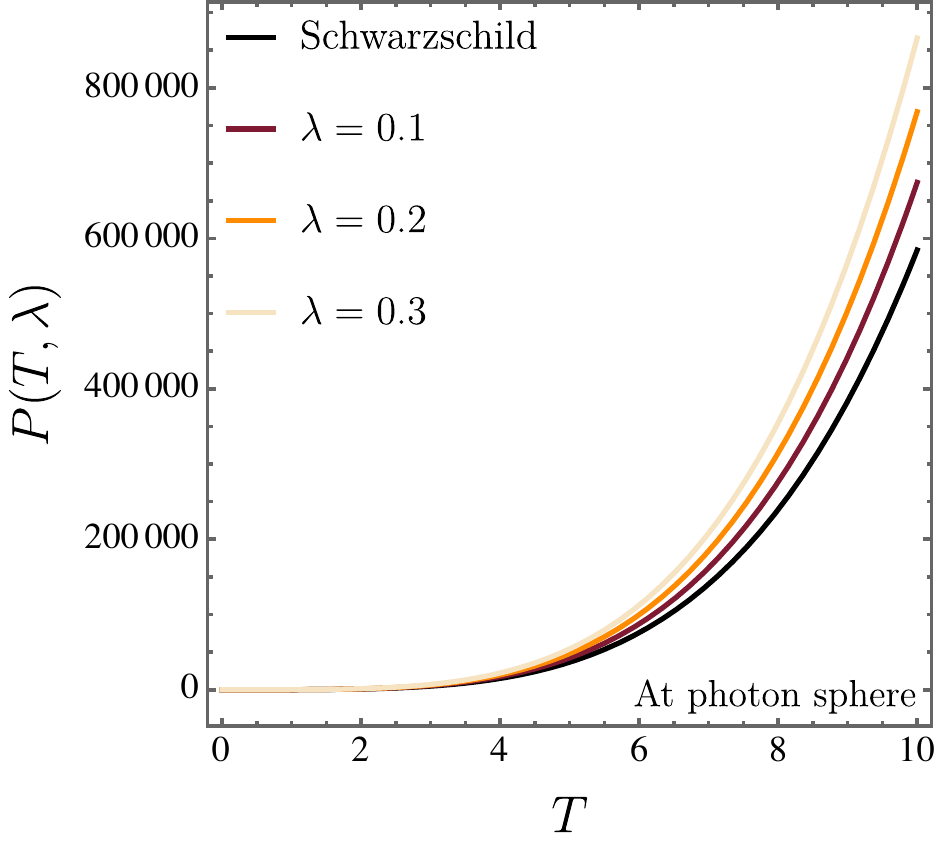}
    \caption{Pressure $P(T,\lambda)$ versus temperature at the photon sphere location.}
    \label{pressurephoton}
\end{figure}

Evaluating the pressure at the photon sphere $r_{ph}=3M$ \cite{Heidari:2025oop} leads to
\ie
\lim\limits_{r \to r_{ph}} P(r, \lambda) = \frac{3}{5} \pi ^4 (\lambda +1)^{3/2} T^4.
\fe
The thermal response at the photon sphere becomes transparent once the profiles displayed in Fig.~\ref{pressurephoton} are examined. The curves reveal that strengthening the Lorentz--violating parameter $\ell$ essentially increases the magnitude of the pressure $P(T,\lambda)$ in this region. Notice that this trend mirrors the behavior observed near the event horizon, indicating that Lorentz--violating effects act to higher the pressure throughout the inner zones of the geometry.

\subsubsection{Asymptotic regime }

The far--field behavior of the pressure becomes relevant once the spatial coordinate moves to large $r$. As already suggested by the radial profiles, $P(r,\lambda)$ does not decay to zero but instead settles to a constant limit. To extract this asymptotic value in closed form, the analysis therefore turns to the limit
\ie
\lim\limits_{r \to \infty} P(r, \lambda) = \frac{\pi ^4 (\lambda +1)^{3/2}}{45} T^{4}.
\fe

\begin{figure}
    \centering
     \includegraphics[scale=0.51]{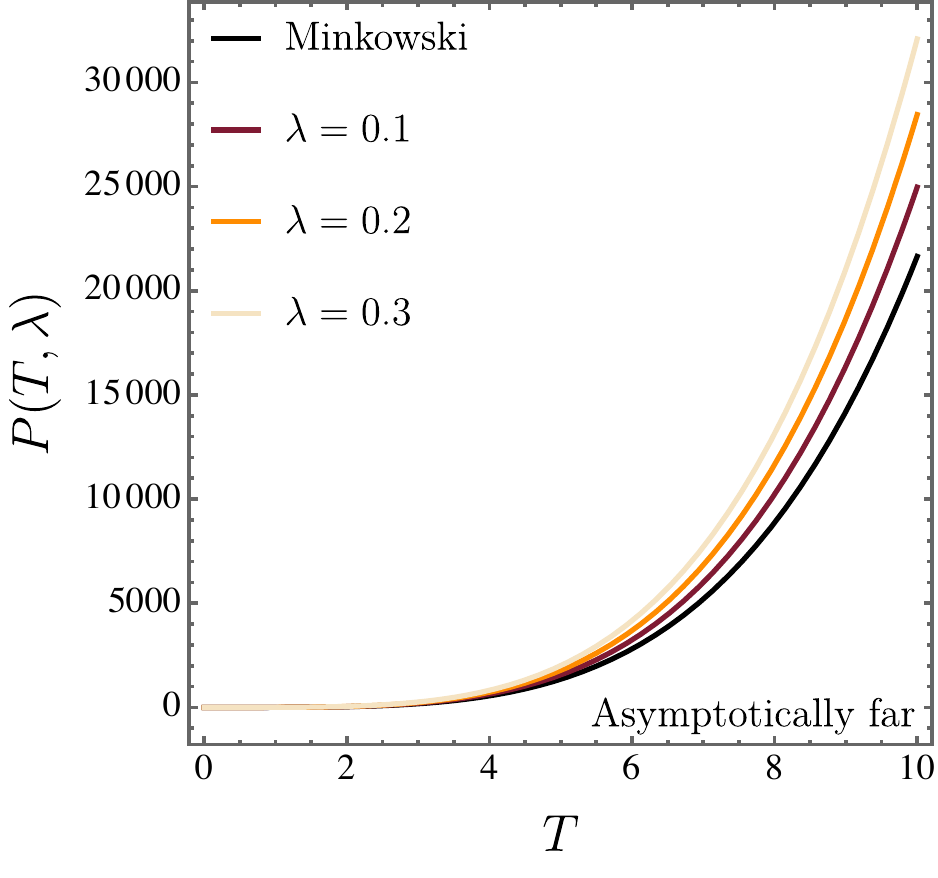}
    \caption{Temperature dependence of the pressure $P(T,\lambda)$ in the asymptotically far region.}
    \label{pressureassimp}
\end{figure}

The asymptotic behavior becomes clear once the profiles shown in Fig.~\ref{pressureassimp} are inspected. In this far--field regime, the pressure responds monotonically to the Lorentz--violating parameter: larger values of $\lambda$ lead to higher asymptotic values of $P(T,\lambda)$. This trend follows the same pattern already identified near the horizon and at the photon sphere. In contrast to those inner regions, however, the benchmark adopted here is not the Schwarzschild geometry but flat Minkowski spacetime, which provides the natural reference at large distances.

A comparison across all three regimes considered in this work—very close to the horizon, in the photon–sphere neighborhood, and in the asymptotic domain—reveals a behavior opposite to that reported for the Kalb--Ramond black hole in Ref.~\cite{AraujoFilho:2025fwd}. In the present bumblebee scenario, increasing the Lorentz--violating parameter $\lambda$ systematically enhances the pressure, whereas in the Kalb--Ramond case the pressure decreases as the corresponding parameter $\ell$ grows. This contrast highlights the distinct thermodynamic response induced by different realizations of Lorentz symmetry breaking.

\subsection{Mean Energy }

To probe the thermodynamic behavior across the spacetime, the analysis is organized by separating the geometry into three characteristic zones, as we did before: the asymptotic sector, the neighborhood of the photon orbit, and the immediate vicinity of the horizon. In this framework, the mean energy is introduced as
\ie
\begin{split}
\label{meanenergy}
U(r, \lambda) & = - \frac{\partial}{\partial \beta} \mathrm{ln}\left[Z(r, \lambda)\right] = \frac{\pi ^4 (\lambda +1)^{3/2} r^3}{15  (r-2 M)^3} T^{4}.
\end{split}
\fe

\begin{figure}
    \centering
     \includegraphics[scale=0.51]{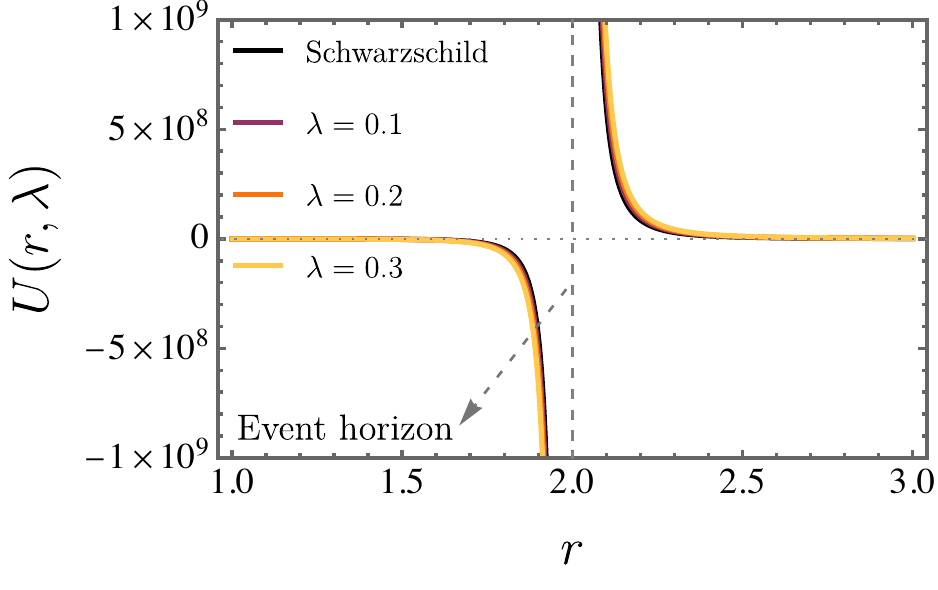}
     \includegraphics[scale=0.51]{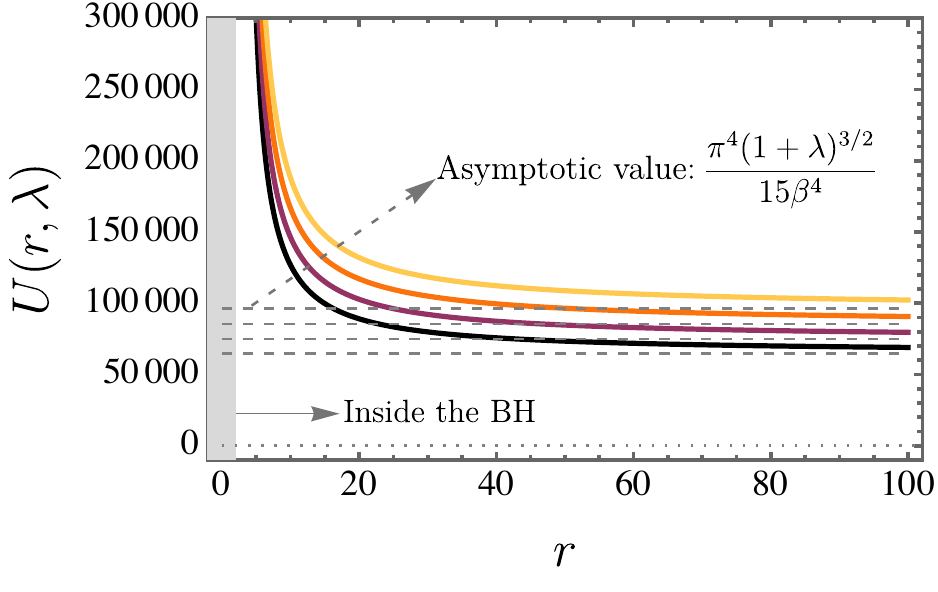}
    \caption{The radial dependence of the mean energy $U(r,\lambda)$ is displayed. }
    \label{Urr}
\end{figure}

Equation (\ref{meanenergy}) encodes a rich dependence of the average energy on the spacetime geometry and on the Lorentz--violating coupling. The discussion focuses on how this quantity responds to changes in the radius and in $\lambda$, as illustrated in Fig. \ref{Urr}. The plot reveals that the sign of $U(r,\lambda)$ is not fixed: depending on the chosen parameters, the mean energy takes either positive or negative values, indicating the coexistence of qualitatively different thermodynamic sectors.

The horizon plays a central role in this behavior. As the radial coordinate approaches $r_h$, the mean energy decreases without bound, and the curve develops a divergence that separates the physically accessible exterior from the interior region. Only for $r>r_h$ does $U(r,\lambda)$ remain positive, which identifies the exterior spacetime as the domain of thermodynamic relevance. Alongside these curves, the Schwarzschild case is included as a reference to highlight the effect of Lorentz violation.

Although the left panel gives the impression that $U(r,\lambda)$ fades away at large distances, this impression arises solely from the very broad vertical scale adopted in the figure, spanning $[-10^{9},10^{9}]$. A closer inspection, provided in the right panel, shows instead that the mean energy does not vanish asymptotically. In the far region, it approaches a constant plateau, $U(r,\lambda)\to \pi^{4}(1+\lambda)^{3/2}/(15\,\beta^{4})$, which confirms the persistence of a finite energy content even at spatial infinity. In this configuration, larger values of $\lambda$ lead to higher energy.

The analysis now examines the thermal response of the system by tracking how the mean energy depends on the temperature $T$ in representative sectors of the geometry. Rather than following the earlier presentation, three reference locations are selected to sample distinct regimes: a point placed slightly above the horizon at $r = 1.001\, r_h$, the photon orbit at $r = r_{ph}$, and a distant region where asymptotic behavior dominates. The photon sphere serves here only as a convenient marker of the near--horizon domain; replacing it with any other radius in the same neighborhood would lead to the same qualitative trends and would not modify the conclusions.


\subsubsection{Extreme near–horizon regime }

Having established the radial behavior at fixed temperature, the discussion now turns to the thermal dependence of the mean energy in the near--horizon domain. Rather than keeping the radius as the control variable, the temperature is taken as the driving parameter while the location is held just outside the event horizon. Within this setting, the mean energy is written as
\ie
\lim\limits_{r \to 1.001\times r_{h}} U(r, \lambda) = 6.51344\times 10^9 (1+ \lambda)^{3/2} T^4.
\fe

\begin{figure}
    \centering
     \includegraphics[scale=0.51]{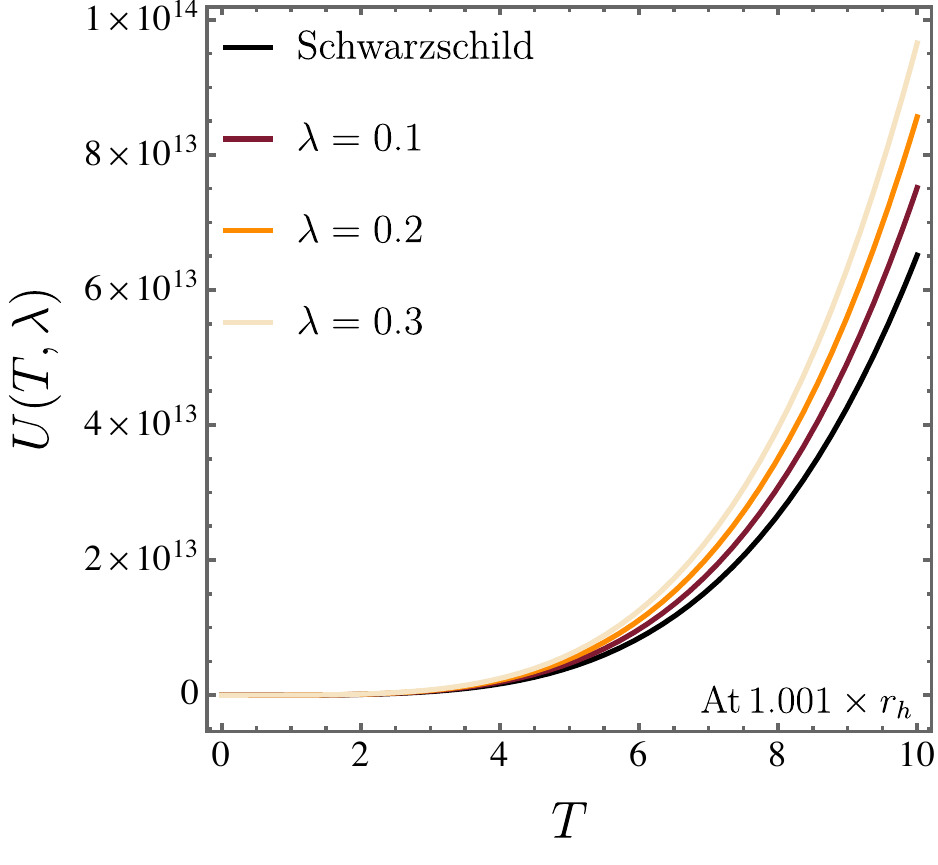}
    \caption{Mean energy $U(T,\lambda)$ versus temperature $T$ evaluated just outside the horizon at $r = 1.001\,\times\, r_h$.}
    \label{Meanvery}
\end{figure}

The behavior encoded in the previous formula is examined through its graphical profile in Fig. \ref{Meanvery}. The attention centers on the role played by the Lorentz--violating coupling $\lambda$ in shaping the thermal response. The curves show that increasing $\lambda$ systematically increases the magnitude of the mean energy $U(T,\lambda)$. For comparison, the corresponding Schwarzschild result is included to highlight how Lorentz violation alters the standard scenario.

\subsubsection{Near–horizon regime }

\begin{figure}
    \centering
     \includegraphics[scale=0.51]{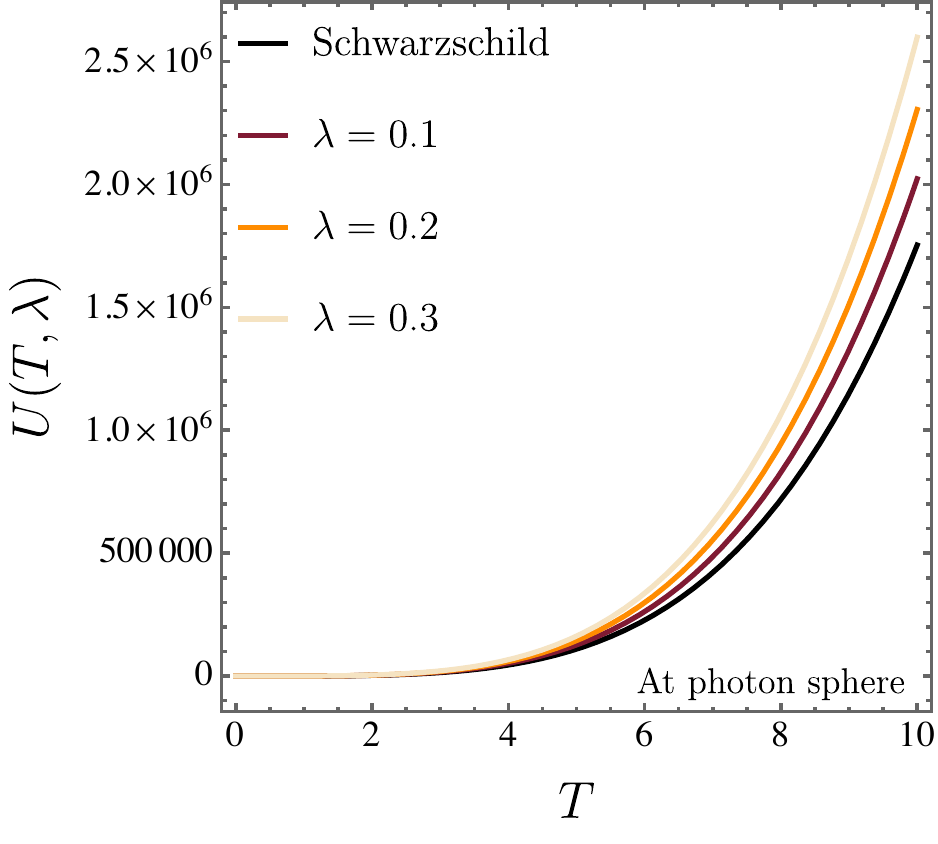}
    \caption{Mean energy $U(T,\lambda)$ versus temperature $T$ evaluated at the photon sphere.}
    \label{Meanphoton}
\end{figure}

The analysis now turns to the photon orbit, where the mean energy is evaluated at $r_{ph}=3M$. Fixing the radius at this location, the expression takes the form
\ie
\lim\limits_{r \to r_{ph}} U(r, \lambda) = \frac{9}{5} \pi ^4 (\lambda +1)^{3/2} T^4.
\fe
The implications of this expression are illustrated in Fig. \ref{Meanphoton}, where the thermal profile at the photon orbit is presented. The curves show that larger values of $\ell$ systematically amplify the magnitude of the mean energy $U(T,\lambda)$ at the photon sphere.

\subsubsection{Asymptotic regime }


The large--radius behavior of the mean energy again reveals a saturation effect, echoing the pattern found in the radial analysis of $U(r,\lambda)$. The asymptotic form is extracted directly by taking the limit of the expression, which leads to
\ie
\lim\limits_{r \to \infty} U(r, \lambda) = \frac{\pi ^4 (1+\lambda)^{3/2}}{15 \beta ^4}.
\fe

\begin{figure}
    \centering
     \includegraphics[scale=0.51]{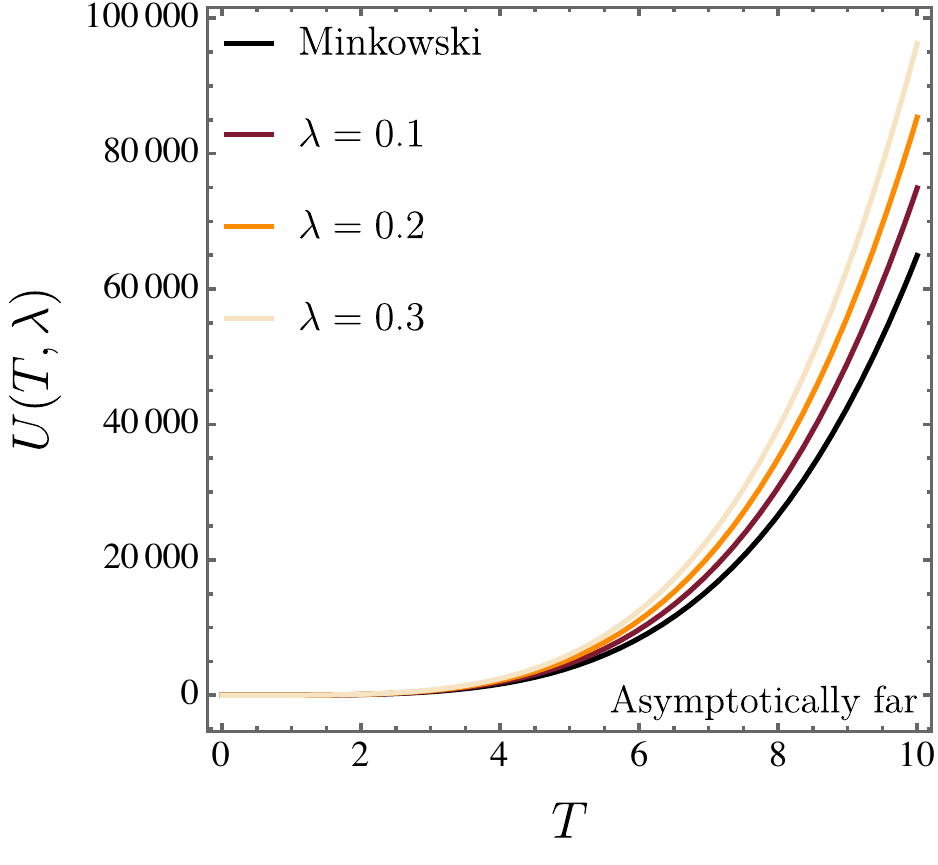}
    \caption{Mean energy $U(T,\lambda)$ versus temperature $T$ in the asymptotic region.}
    \label{Meanassimp}
\end{figure}

The behavior implied by the asymptotic formula is examined through the curves displayed in Fig. \ref{Meanassimp}. The far--field region reinforces the tendency already seen near the horizon and at the photon orbit: increasing the Lorentz--violating parameter $\lambda$ progressively increases the magnitude of the mean energy $U(T,\lambda)$. In this regime, however, the benchmark is no longer the Schwarzschild geometry. Instead, the comparison is carried out with flat Minkowski spacetime, which provides the natural reference background at large distances.

\subsection{Entropy }

The discussion now organizes the spacetime into representative sectors to examine the entropic behavior of the system. With this partition in place, the analysis begins by introducing the expression for the entropy, written as
\ie
\begin{split}
\label{entropy}
S(r, \lambda) &   = \frac{4 \pi ^4 (\lambda +1)^{3/2} r^3}{45 (r-2 M)^3} T^{3}.
\end{split}
\fe

\begin{figure}
    \centering
     \includegraphics[scale=0.51]{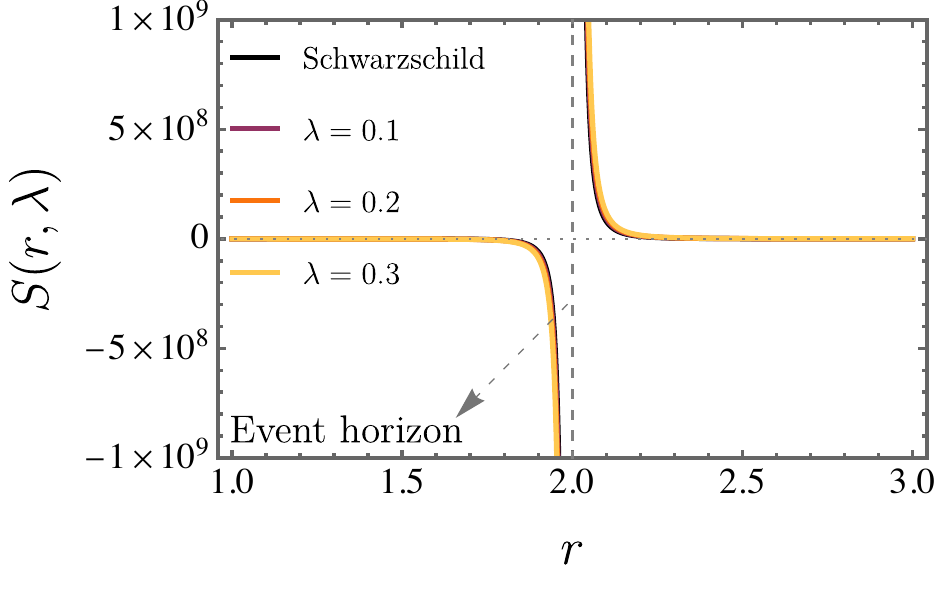}
     \includegraphics[scale=0.51]{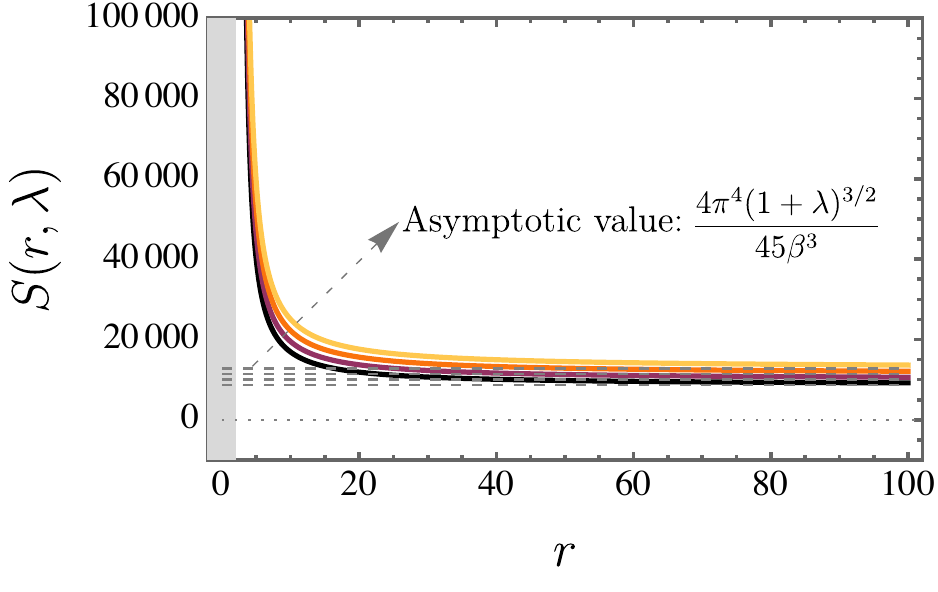}
    \caption{Radial profile of the entropy $S(r,\lambda)$. }
    \label{entropyrr}
\end{figure}

The structure encoded in Eq. (\ref{entropy}) is examined through its radial and parametric behavior, as illustrated in Fig. \ref{entropyrr}. As we did before, the discussion centers on how the entropy responds to changes in the radius and in the Lorentz--violating coupling $\lambda$. The profiles reveal that $S(r,\lambda)$ does not preserve a fixed sign: depending on the chosen parameters, it becomes either positive or negative, which signals the presence of distinct thermodynamic sectors and hints at possible phase--like transitions.

A central role is played by the event horizon. As the radius approaches $r_h$, the entropy develops a divergence, and immediately outside this surface it rises sharply, reproducing the behavior already encountered for the other thermodynamic quantities (pressure and mean energy). The horizon therefore marks the boundary separating the physically meaningful exterior from the interior region. For orientation, the corresponding Schwarzschild curve is shown together with the Lorentz--violating case. Although the left panel seems to suggest that the entropy fades away at large $r$, this visual effect results from the very broad vertical scale adopted in the plot, extending from $-10^{9}$ to $10^{9}$.
The asymptotic trend becomes clear in the right panel. Instead of vanishing, the entropy settles into a finite plateau at spatial infinity, $S \to 4\pi^{4}(1+\lambda)^{3/2}/(45\,\beta^{3})$. This representation also makes evident that only the exterior region, $r>r_h$, supports physically admissible (positive) entropy values.

The analysis now turns to the thermal behavior of the entropy by tracking its dependence on the temperature $T$ in representative sectors of the spacetime. Here, we also consider: a point placed just outside the horizon at $r = 1.001\,\times\, r_h$, the photon orbit at $r = r_{ph}$, and a distant region characterizing the asymptotic limit. The photon sphere is adopted here only as a convenient marker of the near--horizon domain.


\subsubsection{Extreme near–horizon regime }

Having established the radial behavior of the entropy at fixed temperature, the discussion now turns to its thermal response in the near-horizon region. The temperature is taken as the driving parameter while the location is held just outside the event horizon. Within this setting, the entropy is written as
\ie
\lim\limits_{r \to 1.001\times r_{h}} S(r, \lambda) = 8.68459\times 10^9 (\lambda +1)^{3/2} \,T^3.
\fe

\begin{figure}
    \centering
     \includegraphics[scale=0.51]{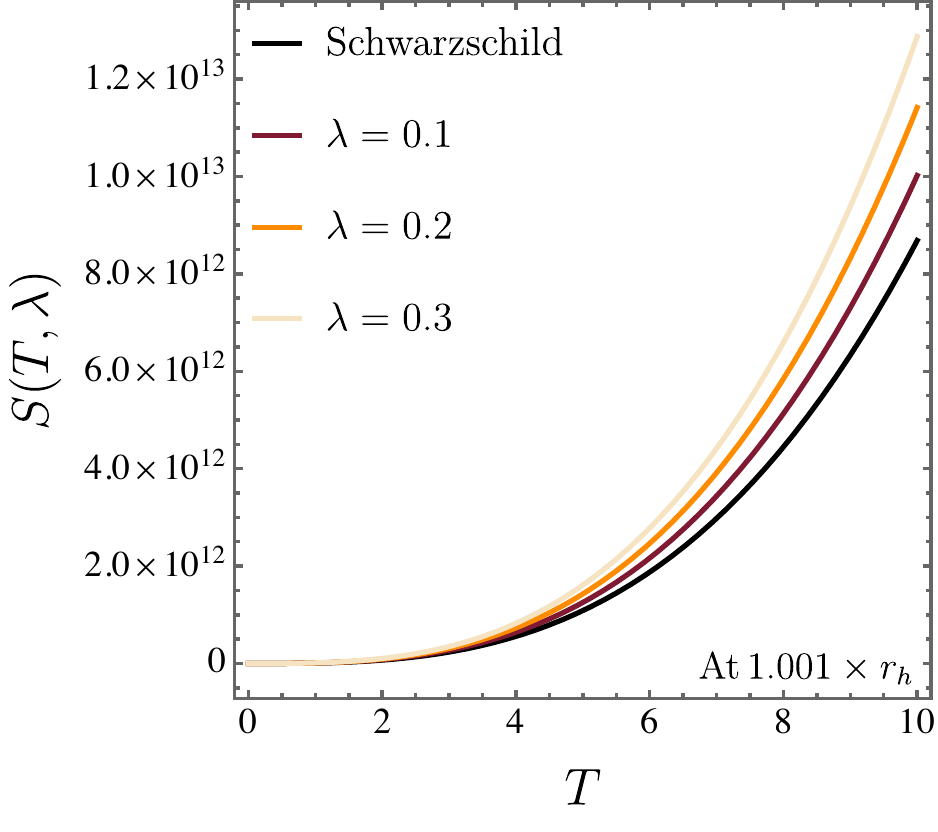}
    \caption{Entropy $S(T,\lambda)$ versus temperature $T$ evaluated just outside the horizon at $r = 1.001\,\times\, r_h$.}
    \label{entropyvery}
\end{figure}

The behavior encoded in the preceding formula is examined through the curves displayed in Fig. \ref{entropyvery}. The figure indicates that larger values of $\lambda$ systematically increases the entropy $S(T,\lambda)$. To highlight this departure from the standard case, the corresponding Schwarzschild curve is shown for comparison.

\subsubsection{Near–horizon regime }

\begin{figure}
    \centering
     \includegraphics[scale=0.51]{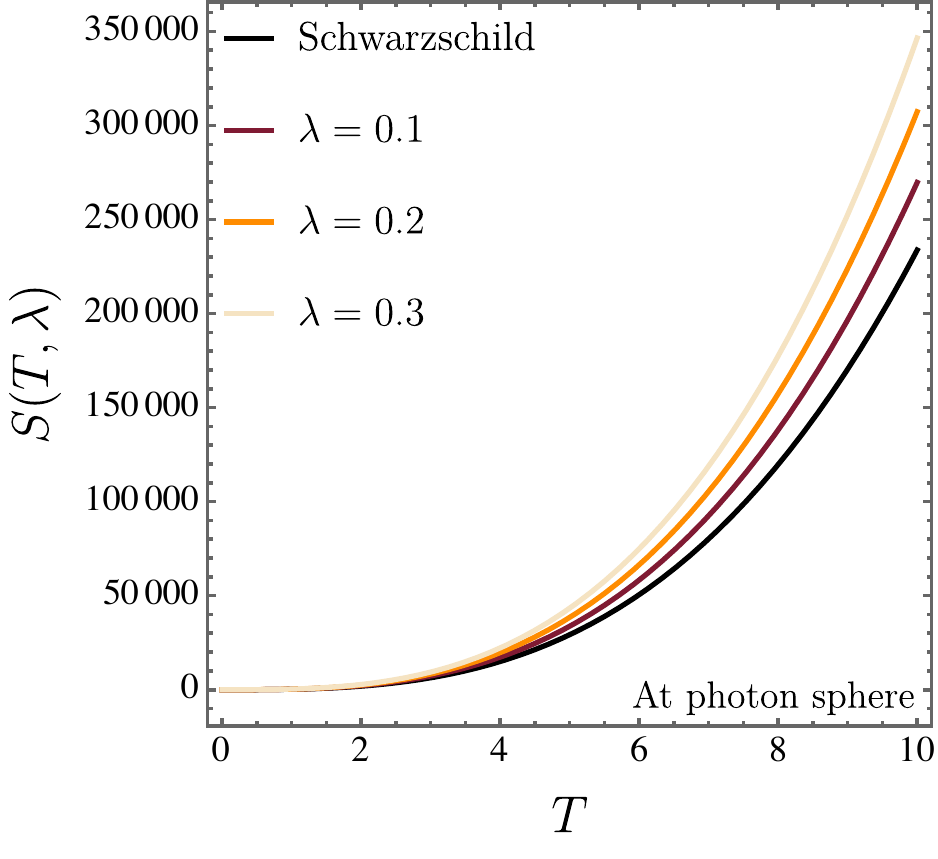}
    \caption{Entropy $S(T,\lambda)$ versus temperature $T$ evaluated at the photon sphere.}
    \label{entropyphoton}
\end{figure}

The discussion now turns to the photon orbit, where the entropy is evaluated at $r_{ph}=3M$. Fixing the radius at this location, the entropy takes the form
\ie
\lim\limits_{r \to r_{ph}} S(r, \lambda) = \frac{12}{5} \pi ^4 (\lambda +1)^{3/2} \,T^3.
\fe
The behavior at the photon orbit is displayed in Fig. \ref{entropyphoton}, which serves to visualize the thermal response in this region. The curves confirm the pattern already seen close to the horizon: increasing the Lorentz--violating parameter $\lambda$ systematically enhances the entropy $S(T,\lambda)$ at the photon sphere.

\subsubsection{Asymptotic regime }


The large--radius behavior of the entropy again reveals a saturation pattern, echoing the trend identified in the radial analysis of $S(r,\lambda)$. The far--field value is extracted directly by taking the asymptotic form of the expression, which leads to
\ie
\lim\limits_{r \to \infty} S(r, \lambda) = \frac{4 \pi ^4 (1 + \lambda)^{3/2}}{45} T^{3}.
\fe

\begin{figure}
    \centering
     \includegraphics[scale=0.51]{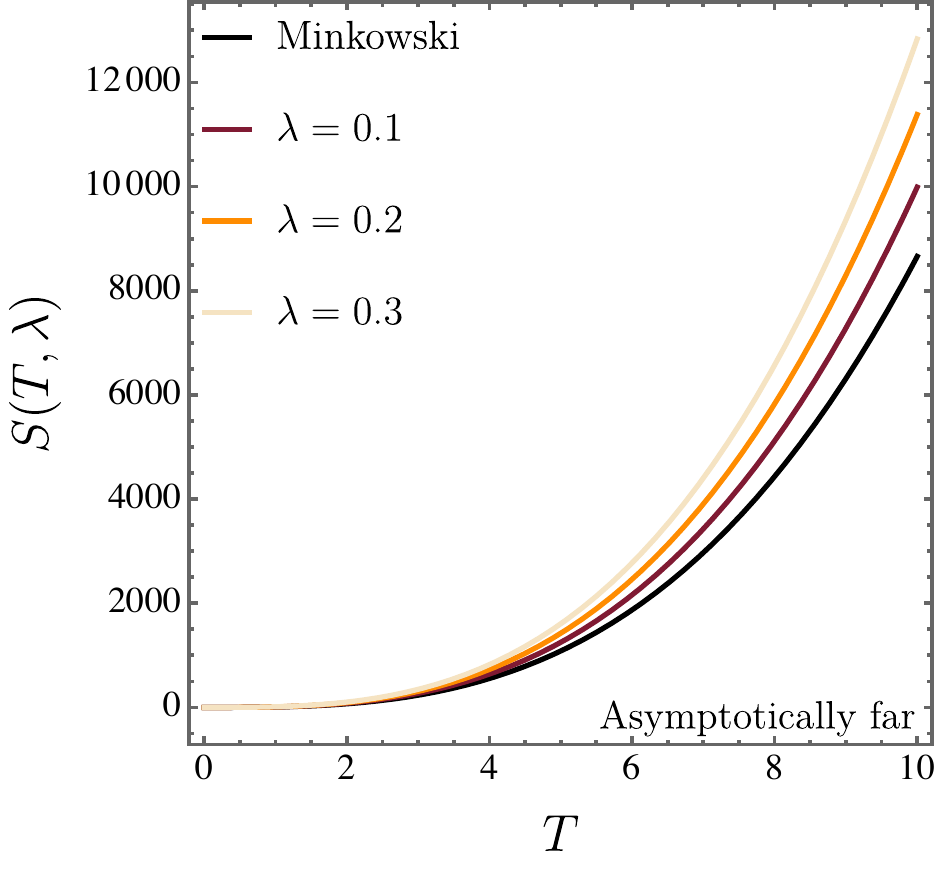}
    \caption{Entropy $S(T,\lambda)$ versus temperature $T$ in the asymptotic region.}
    \label{entropyassimp}
\end{figure}

The asymptotic trend is examined through the curves displayed in Fig. \ref{entropyassimp}.  Increasing the Lorentz--violating coupling $\lambda$ steadily increases the entropy. In this regime, however, the natural benchmark is no longer the Schwarzschild solution. The comparison is now performed against flat Minkowski spacetime, which governs the physics at large distances.

\subsection{Heat capacity }

The spacetime is now organized into representative domains to investigate the next thermodynamic quantity. With this partition in place, the analysis opens by introducing the expression for the entropy, written as
\ie
\begin{split}
\label{heat}
C_{V}(r, \lambda) & =  \frac{4 \pi ^4 (\lambda +1)^{3/2} r^3}{15 \beta ^3 (r-2 M)^3}.
\end{split}
\fe

\begin{figure}
    \centering
     \includegraphics[scale=0.51]{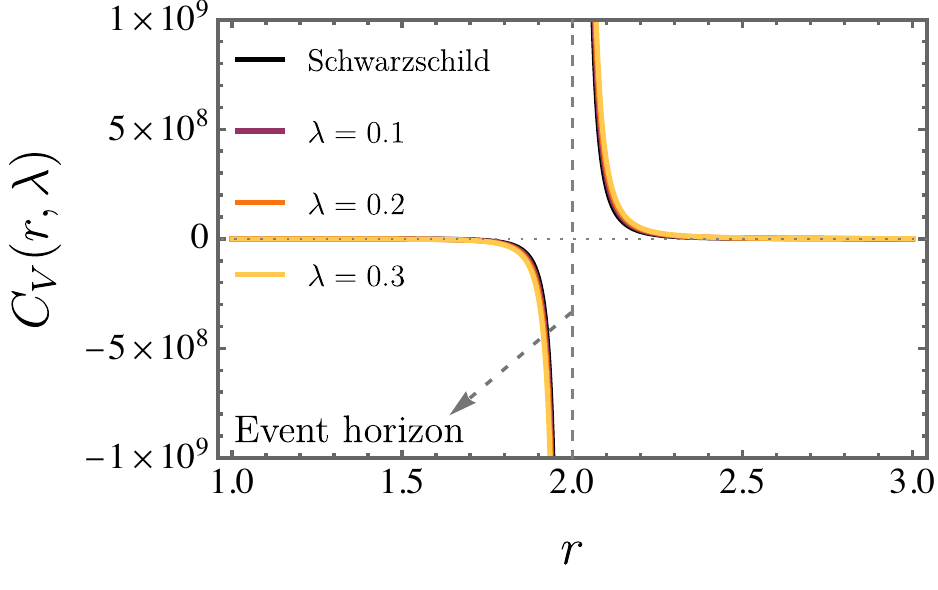}
     \includegraphics[scale=0.51]{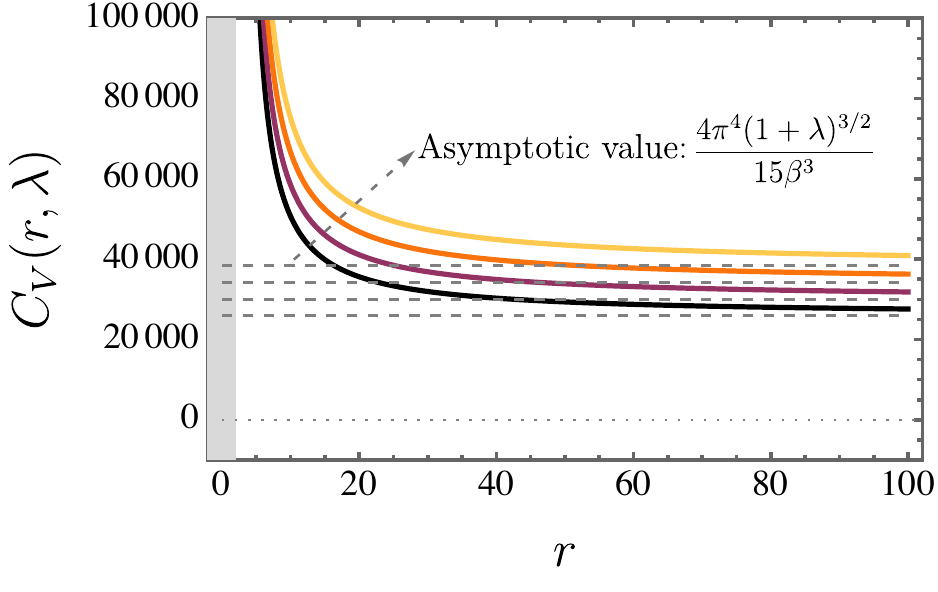}
    \caption{Heat capacity $C_V(r,\lambda)$ versus the radial coordinate $r$. }
    \label{heatrr}
\end{figure}

The properties encoded in Eq. (\ref{heat}) are examined through their radial and parametric behavior, as illustrated in Fig. \ref{heatrr}.  The profiles reveal that $C_V(r,\lambda)$ does not keep a fixed sign: depending on the chosen values of $r$ and $\lambda$, it becomes either positive or negative, signaling the coexistence of distinct thermodynamic sectors and the possible onset of phase--like transitions.

The event horizon plays a decisive role in shaping this structure. As $r$ approaches $r_h$, the heat capacity develops a divergence, and immediately outside this surface it decreases without bound, marking the horizon as the boundary that constrains the physically admissible behavior. Alongside these curves, the Schwarzschild case is displayed to provide a reference for the impact of Lorentz violation. Although the left panel seems to suggest that $C_V(r,\lambda)$ fades away at large radii, this impression arises from the very broad vertical scale adopted in the plot, extending from $-10^{9}$ to $10^{9}$.

The true asymptotic trend becomes evident in the right panel. Instead of vanishing at spatial infinity, the heat capacity settles into a finite plateau, $C_V \to 4\pi^{4}(1-\ell)^{3}/(15,\beta^{3})$. This representation also makes clear that physically acceptable, positive values of $C_V(r,\lambda)$ occur only in the exterior region, namely for $r>r_h$.

The analysis now turns to the thermal behavior of the heat capacity by tracking its dependence on the temperature $T$ in representative sectors of the spacetime, as we did before.


\subsubsection{Extreme near–horizon regime }

Having established the radial profile of the heat capacity at constant temperature, the discussion now turns to its thermal response in the near--horizon region. In this setting, the heat capacity is written as
\ie
\lim\limits_{r \to 1.001\times r_{h}} C_{V}(r, \lambda) =\frac{2.60538\times 10^{10} (\lambda +1)^{3/2}}{\beta ^3}.
\fe

\begin{figure}
    \centering
     \includegraphics[scale=0.51]{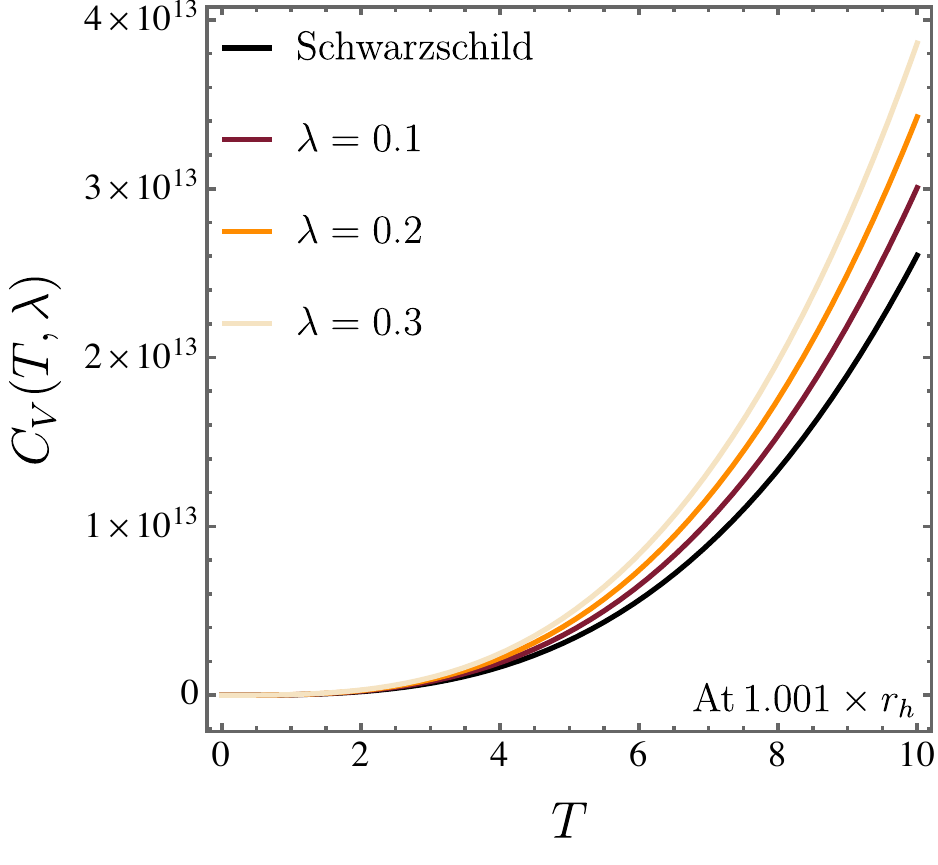}
    \caption{Heat capacity $C_V(T,\lambda)$ versus temperature $T$ evaluated just outside the horizon at $r = 1.001\,\times\, r_h$.}
    \label{heatvery}
\end{figure}

The behavior encoded in the preceding formula is examined through the curves shown in Fig. \ref{heatvery}. Attention is directed to the role of the Lorentz--violating coupling in shaping the thermal profile. The figure indicates that increasing $\lambda$ progressively increases the magnitude of the heat capacity $C_V(T,\lambda)$. To highlight this departure from the standard scenario, the Schwarzschild curve is included as a reference. In contrast with the general--relativistic case, the bumblebee black hole exhibits an overall enhanced heat-capacity profile.

\subsubsection{Near–horizon regime }

\begin{figure}
    \centering
     \includegraphics[scale=0.51]{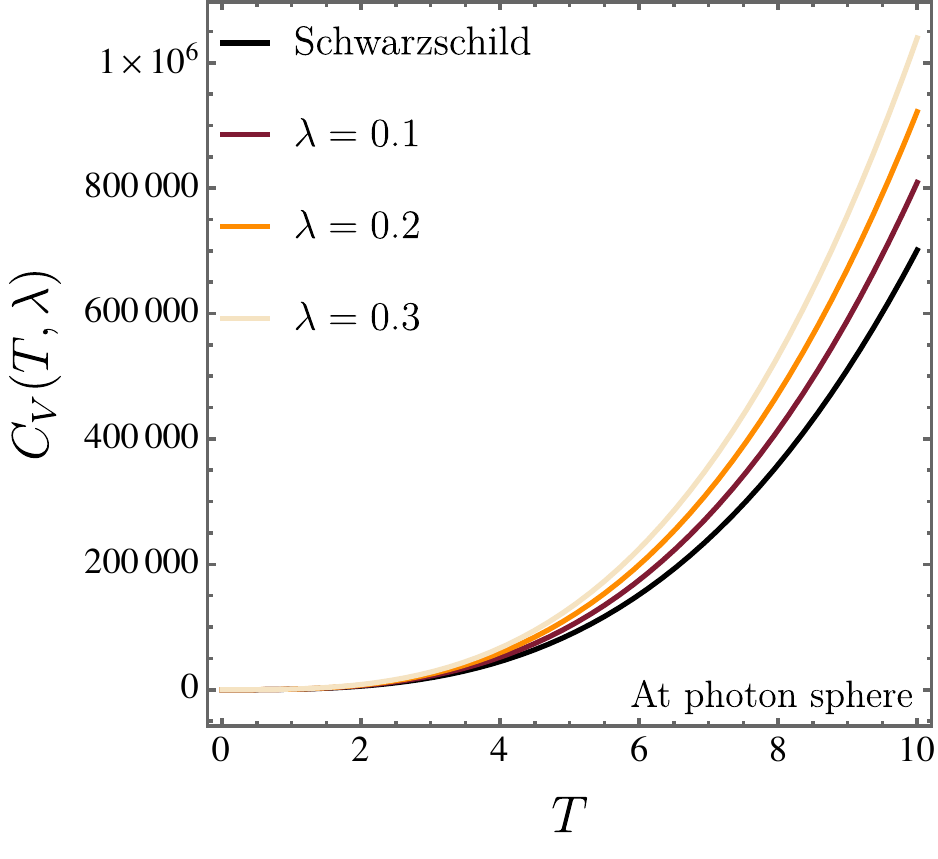}
    \caption{Heat capacity $C_V(T,\lambda)$ versus temperature $T$ evaluated at the photon sphere.}
    \label{heatphoton}
\end{figure}

The discussion now examines the heat capacity at the photon orbit, where the radius is fixed at $r_{ph}=3M$. In this case, the expression takes the form
\ie
\lim\limits_{r \to r_{ph}} C_{V}(r, \lambda) = \frac{36}{5} \pi ^4 (1+\lambda)^{3/2} T^3.
\fe
The thermal profile at the photon orbit is illustrated in Fig. \ref{heatphoton}, which clarifies the implications of the above expression. The curves show that strengthening the Lorentz--violating coupling $\lambda$ systematically increases the magnitude of the heat capacity $C_V(T,\lambda)$ in this region, following the same tendency identified close to the event horizon.

\subsubsection{Asymptotic regime }


The behavior of the heat capacity at large distances again reveals a saturation effect, in line with the trend found in its radial profile $C_V(r,\lambda)$. To make this feature explicit, the asymptotic value is extracted directly by taking the far--field limit of the expression, which gives
\ie
\lim\limits_{r \to \infty} C_{V}(r, \lambda) = \frac{4\pi ^4 (1+\lambda)^{3/2}}{15 \beta^3}.
\fe

\begin{figure}
    \centering
     \includegraphics[scale=0.51]{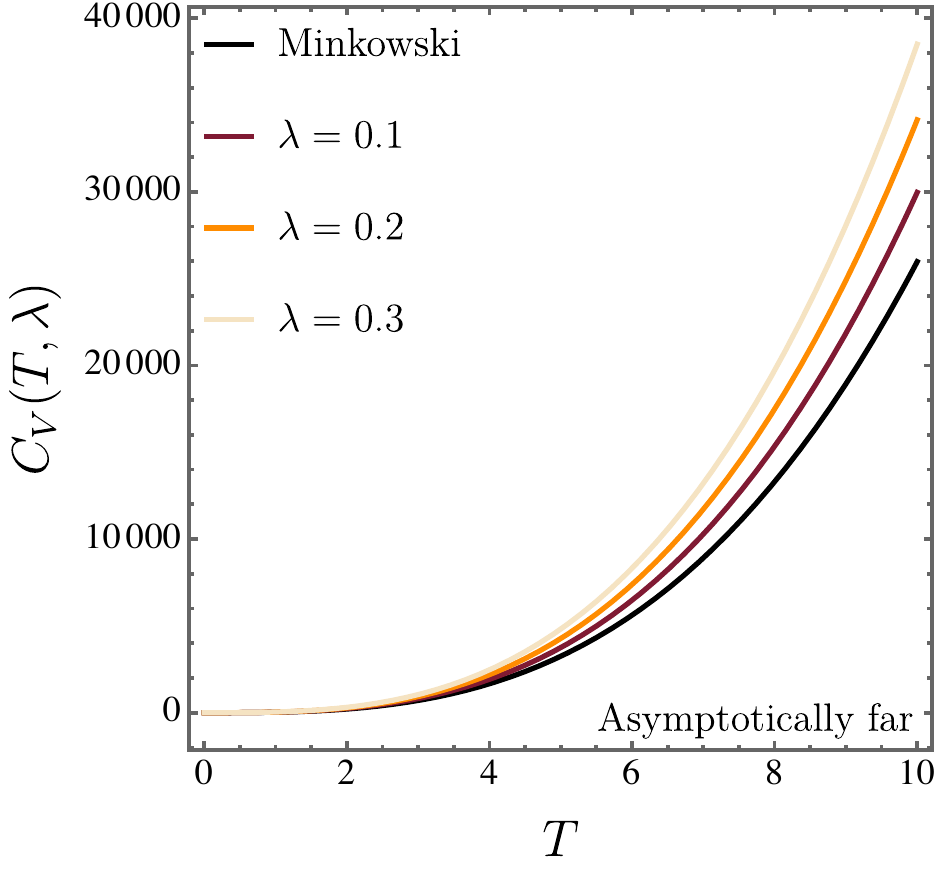}
    \caption{Heat capacity $C_V(T,\lambda)$ versus temperature $T$ in the asymptotic region.}
    \label{heatassimp}
\end{figure}

The asymptotic behavior implied by the above formula is illustrated in Fig. \ref{heatassimp}. The curves confirm that the Lorentz--violating parameter $\lambda$ continues to increases the heat capacity in the far--field region, extending the pattern already identified close to the horizon and at the photon orbit. In this regime, however, the natural reference is no longer the Schwarzschild geometry; the comparison is instead carried out with flat Minkowski spacetime, which governs the asymptotic limit.

{It is useful to collect the far--field limits of the quantities obtained above in a single place. Although the spectral densities and the thermodynamic observables were derived in separate subsections, their asymptotic behavior follows the same pattern. In the massless sector, the modified dispersion relation introduces the common geometric factor $(1+\lambda)^{3/2}/f(r)^3$, with $f(r)=1-2M/r$. In this manner, in the limit $r\to\infty$, one has $f(r)\to 1$, and all local quantities approach finite plateaus controlled by the same Lorentz--violating scaling.


\section{The vacuum state and the Casimir energy in curved spacetime }
\label{sec6}

Before addressing the global Casimir effect in the vicinity of a static black hole, it is essential to first examine how a massive scalar field evolves within the bumblebee black hole. The subsequent analysis explores the field equation and its corresponding regularized vacuum energy at zero temperature, followed by the evaluation of the thermal Casimir quantities—namely, the Helmholtz free energy, mean energy, entropy, and heat capacity.


\subsection{The field solution }

To analyze the dynamics of a scalar field in the spacetime under study, it becomes necessary to determine the solutions of the covariant Klein--Gordon equation. This equation governs the propagation of a massive, uncharged scalar particle that interacts minimally with the gravitational background. In a curved geometry, its general form can be expressed as
\ie
\Bigg[\frac{1}{\sqrt{-g}}\partial_{\mu}(g^{\mu\nu}\sqrt{-g}\partial_{\nu}) {-} m^{2} \Bigg]\Phi = 0.\label{Th4e3kleqin5Go55rdon}
\fe 
 
Conversely, the spacetime geometry associated with a static, neutral black hole within the framework of bumblebee gravity is described by the corresponding bumblebee metric, given by
\ie
\begin{split}
\label{model2}
\mathrm{d}s^{2} = & - \left( 1 - \frac{2M}{r}   \right) \mathrm{d}t^{2} + \frac{(1 + \lambda)}{1 - \frac{2M}{r} } \, \mathrm{d}r^{2} + r^{2}\mathrm{d}\theta^{2}  + r^{2} \sin^{2}\theta\mathrm{d}\varphi^{2}.
\end{split}
\fe

It should be emphasized that, while examining the effects of backreaction within this framework would certainly be of interest, such an investigation lies beyond the scope of the present work. The analysis here is restricted to treating the scalar field as a minor perturbation on the background geometry. Under this assumption, Eq.~(\ref{Th4e3kleqin5Go55rdon}) takes the form
\begin{equation}
\Phi(t, r, \theta, \varphi) = \sum_{\ell=0}^{\infty} \sum_{m=-\ell}^{\ell} Y_{\ell\,m}(\theta, \varphi) \, \frac{\Psi(t, r)}{r}.
\end{equation}
Considering 
\ie
\mathrm{d}s^{2} = -f(r)\,\mathrm{d}t^{2}+\frac{(1+\lambda)}{f(r)}\,\mathrm{d}r^{2}+r^{2}\mathrm{d}\Omega^{2},\qquad
f(r)=1-\frac{2M}{r},
\fe
and after carrying out the necessary algebraic rearrangements, the Klein--Gordon equation reads
\ie
\frac{\partial^{2}\Psi}{\partial t^{2}}
-\frac{f(r)}{1+\lambda}\!\left[
f(r)\,\frac{\partial^{2}\Psi}{\partial r^{2}}
+f'(r)\!\left(\frac{\partial\Psi}{\partial r}-\frac{\Psi}{r}\right)
\right]
+f(r)\!\left[m^{2}+\frac{\ell(\ell+1)}{r^{2}}\right]\Psi=0.
\label{eq:KG_time_bumblebee_correct}
\fe

By introducing a time–harmonic decomposition expressed as
\begin{equation}
\Psi(t,r)=e^{-i\omega t}\,\psi(r),
\end{equation}
and substituting the metric function $f(r)=1-2M/r$ into Eq.~\eqref{eq:KG_time_bumblebee_correct}, we have
\begin{equation}
\frac{\mathrm{d}}{\mathrm{d}r}\!\left(\frac{f(r)}{1+\lambda}\,\frac{\mathrm{d}\psi}{\mathrm{d}r}\right)
-\frac{f'(r)}{1+\lambda}\,\frac{\psi}{r}
+\left(\frac{\omega^{2}}{f(r)}-m^{2}-\frac{\ell(\ell+1)}{r^{2}}\right)\psi(r)=0.
\label{eq:radial_KG_bumblebee_final_reduced}
\end{equation}

In the limit $\lambda \to 0$, Eq.~\eqref{eq:radial_KG_bumblebee_final_reduced} exactly reproduces the Schwarzschild equation
\ie
\frac{\mathrm{d}}{\mathrm{d}r}\!\big(f(r)\,\psi'(r)\big)
- f'(r)\,\frac{\psi(r)}{r}
+\left(\frac{\omega^{2}}{f(r)}-m^{2}-\frac{\ell(\ell+1)}{r^{2}}\right)\psi(r)=0
\fe
as obtained in Ref.~\cite{Muniz:2015jba}.


To express Eq.~\eqref{eq:radial_KG_bumblebee_final_reduced} in the canonical confluent Heun form~\cite{fiziev2009novel,bezerra2014klein}, we consider
\begin{equation}
\frac{\mathrm{d}^{2}U}{\mathrm{d}x^{2}}
+\left(\alpha+\frac{\beta+1}{x}+\frac{\gamma+1}{x-1}\right)\frac{\mathrm{d}U}{\mathrm{d}x}
+\left(\frac{\mu}{x}+\frac{\nu}{x-1}\right)U=0,
\label{eq:Heun_confluente_forma_canonica}
\end{equation}
whose solutions are $U(x)=\mathrm{HeunC}(\alpha,\beta,\gamma,\delta,\eta;x)$.
The auxiliary parameters $\mu$ and $\nu$ are related to $\alpha$, $\beta$, $\gamma$, $\delta$, and $\eta$ through \cite{fiziev2009novel,bezerra2014klein}
\begin{equation}
\mu=\tfrac{1}{2}(\alpha-\beta-\gamma+\alpha\beta-\beta\gamma)-\eta,
\label{eq:mu_Heun_conlfuente_2}
\end{equation}
\begin{equation}
\nu=\tfrac{1}{2}(\alpha+\beta+\gamma+\alpha\gamma+\beta\gamma)+\delta+\eta.
\label{eq:nu_Heun_conlfuente_2}
\end{equation}

Using the {dimensionless transformation}
\begin{equation}
{
x=1-\frac{r}{2M}},
\end{equation}
{which maps the event horizon to $x=0$ and the curvature singularity to $x=1$}, together with the gauge choice
\begin{equation}
\psi(r)=e^{\frac{\alpha x}{2}}\,x^{\frac{\beta}{2}}\,{(1-x)}\,U(x),
\end{equation}
one obtains
\begin{subequations}
\begin{align}
\alpha &= {+\,4M\sqrt{1+\lambda}\,\sqrt{m^{2}-\omega^{2}}},\\
\beta  &= i\,4M\sqrt{1+\lambda}\,\omega,\\
\gamma &= 0,\\
\delta &= 4M^{2}(1+\lambda)\,(m^{2}-2\omega^{2}),\\
\eta   &= {
-\,(1+\lambda)\ell(\ell+1)-4M^{2}(1+\lambda)\,(m^{2}-2\omega^{2})}.
\end{align}
\label{eq:parameters_bumblebee_corrected}
\end{subequations}
The corresponding radial function can therefore be expressed in the general form
\begin{equation}
\psi(r)=e^{\frac{\alpha x}{2}}\,x^{\frac{\beta}{2}}\,{(1-x)}
\Big[
C_{1}\,\mathrm{HeunC}(\alpha,\beta,\gamma,\delta,\eta;x)
+C_{2}\,x^{-\beta}\,\mathrm{HeunC}(\alpha,-\beta,\gamma,\delta,\eta;x)
\Big],
\label{eq:Heun_solution_bumblebee_reduced}
\end{equation}
with {$x=1-r/(2M)$} and {$2M<r<\infty$, corresponding to $x<0$}. In the limit $\lambda\to 0$, the parameters in Eq.~\eqref{eq:parameters_bumblebee_corrected} reduce to the Schwarzschild set, with the angular contribution correctly retained through the term $-\ell(\ell+1)$, in agreement with Ref.~\cite{Muniz:2015jba}.

To determine the global Casimir energy, one must first obtain the discrete energy spectrum corresponding to the stationary configurations of the massive scalar field in the black hole background. This requires enforcing appropriate boundary conditions on the field at spatial infinity. Since the confluent Heun functions exhibit irregular behavior in that region, a polynomial form for $U(x)$ must be imposed. Following the procedure outlined in Ref.~\cite{fiziev2009novel}, this truncation is implemented through the $\delta_N$ condition and the determinant condition
\begin{eqnarray}
\frac{\delta}{\alpha}+\frac{\beta+\gamma}{2}+1&=&-n, \label{condddi}\\
\Delta_{N+1}&=&0,
\end{eqnarray}
where $n$ denotes a positive integer. The $\delta_N$ condition fixes the leading spectral relation, while $\Delta_{N+1}=0$ selects the allowed polynomial sector. In the weak--binding approximation used below, the former condition already gives the gravitational Bohr--type spectrum. From Eq.~\eqref{condddi}, we have
\ie
\begin{split}
\frac{\delta}{\alpha}
&= \frac{4M^{2}(1+\lambda)(m^{2}-2\omega^{2})}
{4M\sqrt{1+\lambda}\,\sqrt{m^{2}-\omega^{2}}} \\
&= M\sqrt{1+\lambda}\,
\frac{m^{2}-2\omega^{2}}{\sqrt{m^{2}-\omega^{2}}}
= -\,M\sqrt{1+\lambda}\,
\frac{2\omega^{2}-m^{2}}{\sqrt{m^{2}-\omega^{2}}},\\
\frac{\beta+\gamma}{2}
&= i\,2M\sqrt{1+\lambda}\,\omega .
\end{split}
\fe

By inserting these expressions into Eq.~\eqref{condddi}, one obtains
\begin{equation}
{
n+1+i\,2M\sqrt{1+\lambda}\,\omega
- M\sqrt{1+\lambda}\,
\frac{2\omega^{2}-m^{2}}{\sqrt{m^{2}-\omega^{2}}}
=0.}
\label{eq:delta_condition_bumblebee}
\end{equation}
For the real bound--state spectrum, the phase contribution proportional to $i\omega$ is discarded. In the weak--binding regime,
\begin{equation}
{
0<m-\omega\ll m,\qquad \omega\simeq m,}
\end{equation}
so that $2\omega^{2}-m^{2}\simeq m^{2}$. Under this approximation, Eq.~\eqref{eq:delta_condition_bumblebee} gives
\ie
{
M\sqrt{1+\lambda}\,
\frac{2\omega^{2}-m^{2}}{\sqrt{m^{2}-\omega^{2}}}
\simeq n+1.}
\fe
Absorbing the shift $+1$ into the definition of the principal quantum number, one obtains
\ie
{
M\sqrt{1+\lambda}\,m^{2}
\simeq n\,\sqrt{m^{2}-\omega^{2}}.}
\fe
Solving for $\omega$ then gives
\begin{equation}
E_{n} = \omega_{n}
= m\,\sqrt{1-\frac{(1+\lambda)\,m^{2}M^{2}}{n^{2}}},\qquad n=1,2,3,\dots \, .
\label{eq:omega_bumblebee}
\end{equation}
Restoring the physical constants, the resulting discrete energy spectrum takes the form
\begin{equation}
E_{n}
= m c^{2}\,
\sqrt{1-\frac{(1+\lambda)\,G^{2}m^{2}M^{2}}
{\hbar^{2}c^{2}n^{2}}}, \qquad n=1,2,3,\dots \, .
\label{eq:energy_levels_bumblebee}
\end{equation}

It should be emphasized that the same expression was independently obtained in Ref.~\cite{barranco2014schwarzschild} through an alternative procedure in the limit $\lambda \to 0$. The bound--state energy, defined as $E_{n,b} = E_n - mc^2$, is
\begin{equation}
E_{n,b}
=mc^{2}\left[
\sqrt{1-\frac{(1+\lambda)G^{2}m^{2}M^{2}}
{\hbar^{2}c^{2}n^{2}}}-1
\right].
\end{equation}
In the weak--binding regime, it reduces to
\begin{equation}
E_{n,b}\simeq
-\frac{(1+\lambda)G^{2}M^{2}m^{3}}{2\hbar^{2}n^{2}}.
\end{equation}
The reality condition for the spectrum is
\begin{equation}
{
n^{2}\geq
(1+\lambda)\frac{G^{2}m^{2}M^{2}}{\hbar^{2}c^{2}}.}
\end{equation}
Equivalently, in terms of the Compton wavelength $\lambda_C=\hbar/(mc)$ and the horizon radius $r_h=2GM/c^2$, the condition for the $n$th level is
\begin{equation}
{
n\,\lambda_C\geq \sqrt{1+\lambda}\,\frac{GM}{c^2}
=\sqrt{1+\lambda}\,\frac{r_h}{2}.}
\end{equation}
In particular, if the lowest level $n=1$ is required to be present, one must impose
\begin{equation}
{
\lambda_C\geq \sqrt{1+\lambda}\,\frac{r_h}{2}.}
\end{equation}
This criterion keeps the energy spectrum real within the bound--state approximation and prevents the discrete levels from crossing into the black--hole interior sector of the model. Also, we can clearly see that, in the nonrelativistic limit $\mathcal{O}(1/c^2) \to 0$ and after removing the particle’s rest energy, the resulting expression coincides with the Bohr--like energy spectrum of a gravitational analog of the hydrogen atom (up to the Lorentz--violating term $\lambda$), namely
\begin{eqnarray}
E_{n}\approx \,- \frac{(1+\lambda)M^2\,G^2 m^3}{2\hbar^2n^2}.
\end{eqnarray}
A comparable expression was also derived in Refs.~\cite{laptev2006electromagnetic,lasenby2005bound} for the Schwarzschild geometry through distinct methodologies. This agreement reinforces the reliability and coherence of the present formulation employed in this paper.


\subsection{Regularized zero--temperature vacuum energy }

From Eq.~(\ref{eq:omega_bumblebee}), the expression that defines the zero–temperature quantum vacuum energy associated with a massive scalar field can be written as
\begin{eqnarray} E^{(0)}=\frac{1}{2}\sum_{n=1}^{\infty}n^2\omega_n = \frac{m}{2}\sum_{n= {1}}^{\infty}n^2\sqrt{1-\frac{(1+\lambda)M^2 m^2}{n^2}}.
\label{lasdl}
\end{eqnarray}
Here, the calculation is carried out once more in natural units. The multiplicative factor $n^{2}$ arises from the degeneracy of the corresponding energy levels. The summation in Eq.~(\ref{lasdl}) turns out to be divergent; therefore, a regularization scheme must be employed to render it finite. To accomplish this, the Riemann zeta--function method is adopted. As a first step, the square root will be expanded using its binomial series, leading to
\begin{equation}
E^{(0)}=\frac{m}{2}\sum_{n=1}^{\infty}n^2\left[1+\sum_{k=1}^{\infty}\binom{1/2}{k}(-1)^k   (1+\lambda)^{k}  \left(\frac{mM}{n}\right)^{2k}\right].
\end{equation}
After applying the zeta--function regularization, the expression becomes
\begin{equation}
E_{\rm reg}^{(0)}=\frac{m\sqrt{\pi}}{4}\sum_{k=1}^{\infty}\frac{(-1)^k (1+\lambda)^k (mM)^{2k}\zeta(2k-2)}{\Gamma(k+1)\Gamma(3/2-k)},
\label{Ereg}
\end{equation}
in which the relation $\zeta(-2)=0$ has been applied to eliminate the first summation term on the right–hand side. The resulting expression remains convergent provided that {$(1+\lambda)(mM)^{2}<1$}. The $k=1$ component represents the nonrelativistic gravitational Bohr levels, which dominate the regularized vacuum energy in regions far from the event horizon. It is worth emphasizing that possible scattering states in those outer domains are not included in this analysis.

Figure~\ref{energyregularized} displays the behavior of the regularized vacuum energy, emphasizing the influence of Lorentz symmetry breaking. In this evaluation, the summation was extended up to $k=60$ to ensure high numerical precision. It is observed that within specific intervals of the mass parameter, the regularized vacuum energy $E^{(0)}_{\text{reg}}$ attains positive values.
Note also that although several values have been attributed in Fig. \ref{energyregularized}, the LIV factor converges quickly to $\lambda=0$ result for values to $10^{-1}>\lambda$, which shows us the sensibility of the LIV parameter in this geometry.

\begin{figure}[h!]
    \centering
     \includegraphics[scale=0.61]{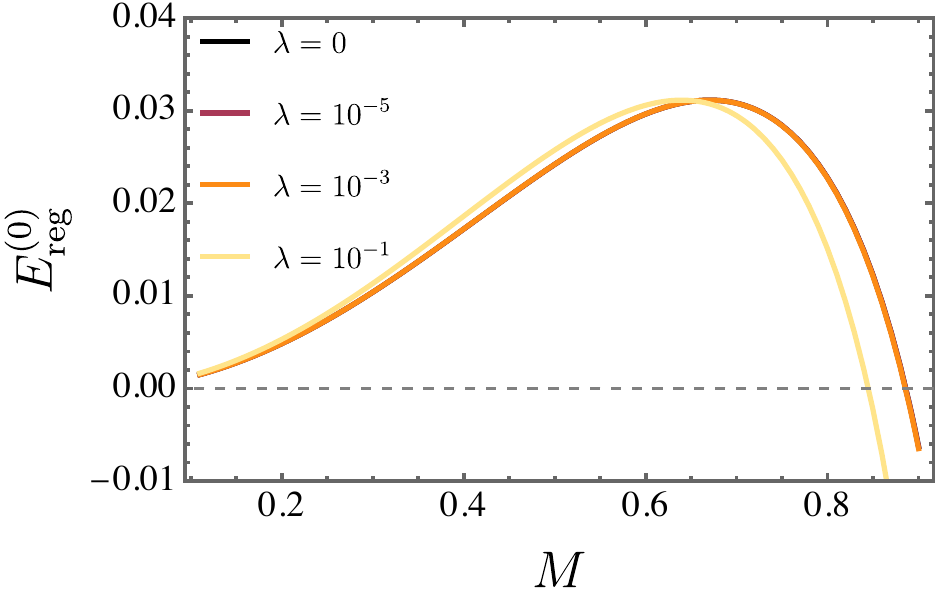}
    \caption{The Casimir--regularized energy $E^{(0)}_{\rm reg}$, Eq. \eqref{Ereg}, as a function of $M$ for various configurations of the Lorentz–violating parameter $\lambda$.}
    \label{energyregularized}
\end{figure}

It should be emphasized that the Casimir energy generally depends on a characteristic geometric scale of the system. In this context, it varies with the horizon radius, reflecting the fact that the black hole mass enters the expression through the Schwarzschild radius, $r_{h}=2M$. Moreover, such energy gives rise to a surface tension, interpreted as the reversible work required to form a unit area of the surface, expressed by $\tau = \partial E / \partial S$, with $S$ denoting the surface area. Consequently, the corresponding tension at the event horizon, whose area is $S_{h}=16\pi M^{2}$, can be written as
\begin{equation}
\tau_h=\frac{\partial E_{reg}^{(0)}}{\partial S_h}=\frac{m}{64\sqrt{\pi}M^2}\sum_{k=1}^{\infty}\frac{(-1)^{k} (1+\lambda)^k k(mM)^{2k}\zeta(2k-2)}{\Gamma(k+1)\Gamma(3/2-k)}.
\label{Talh}
\end{equation}
It can be observed that the leading term of the expansion is independent of the black hole mass, implying that in the limit $M \rightarrow 0$, the surface tension approaches $\tau_{h}= {(1+\lambda) } m^{3}/128\pi$. This contribution may be interpreted as originating from the singularity itself. Although the Casimir energy alone does not explicitly reveal the underlying nontrivial topology induced by the black hole’s singularity, the persistence of a finite horizon tension in the vanishing--mass limit points to the presence of such a singular structure and, consequently, to the nontrivial geometry of the spacetime. In contrast, configurations characterized by regular topologies—such as the surface of an ordinary sphere—do not exhibit any residual Casimir tension as their radius tends to infinity, that is, as the spacetime becomes flat \cite{mostepanenko1988casimir,Muniz:2015jba}.

In Fig.~\ref{tau}, we show the behavior of $\tau_{h}$ as a function of $\lambda$. The chosen range of $\lambda$ values is consistent with the most recent bounds reported in the literature~\cite{14}. The observed decrease of $\tau_h(\lambda)$ indicates that Lorentz–symmetry violation weakens the interaction between the vacuum energy and the black hole horizon. As $\lambda$ increases, the quantum vacuum contributes less—and may even contribute negatively—to the horizon’s thermodynamic response, highlighting therefore a suppression of vacuum effects induced by the bumblebee field. On the right panel, a small irregular feature—visible as a localized bump—is noticeable around the point $(0.485,,0.00139)$.

\begin{figure}[h!]
    \centering
     \includegraphics[scale=0.61]{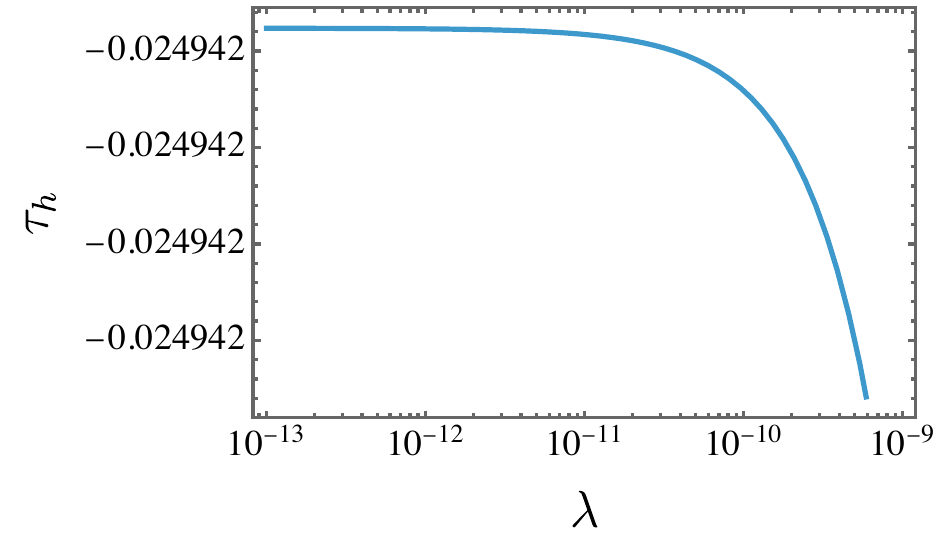}
    \caption{The horizon tension $\tau_{h}$, Eq. \eqref{Talh}, is plotted as a function of the Lorentz–violating parameter $\lambda$, considering summation terms up to $k = 30$.}
    \label{tau}
\end{figure}

\begin{figure}[h!]
    \centering
     \includegraphics[scale=0.51]{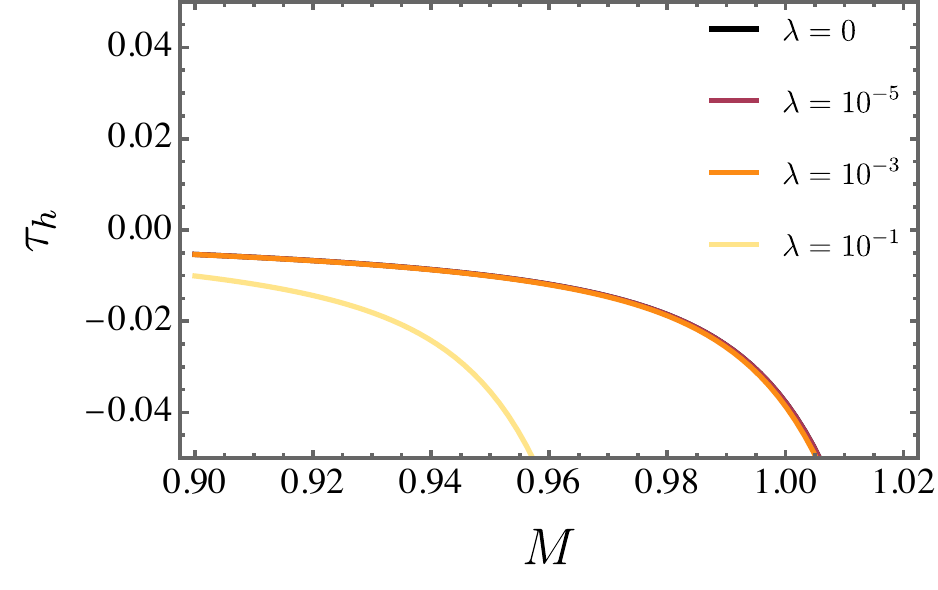}
     \includegraphics[scale=0.51]{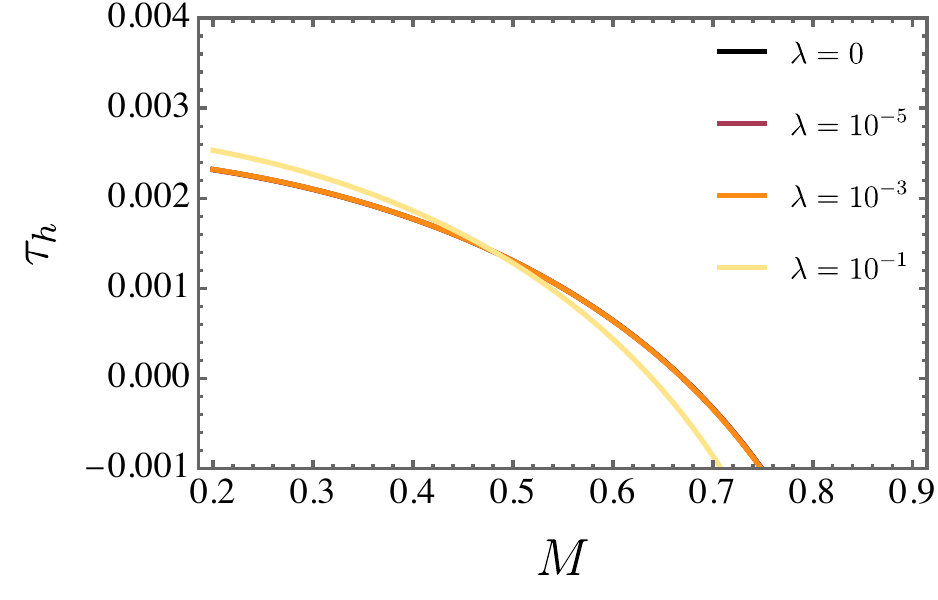}
    \caption{The horizon tension $\tau_{h}$, Eq. \eqref{Talh}, is displayed as a function of $M$ for several values of the Lorentz–violating parameter $\lambda$, with the summation extended up to $k = 60$.}
    \label{tau2}
\end{figure}

In addition, observe that, as occurred with the behavior of Eq. \eqref{Ereg} presented in Fig. \ref{energyregularized}, the horizon tension shown by Fig. \ref{tau2} also has a quick convergence to a non-LIV result for small $\lambda$ independent of the black hole mass M. Hence, in an eventual detection, the LIV characteristic may be present as a minor correction factor since we expected a little LIV influence on the system.


\subsection{Thermal effects }

Next, the thermal contributions to the Casimir energy are examined through the evaluation of the Helmholtz free energy, expressed as \cite{mostepanenko1988casimir,Muniz:2015jba}
\begin{equation} \label{FreeEnergy}
F^{(0)}=\beta^{-1}\sum_{n=1}^{\infty}n^2 \ln{\left[1-\exp{\left(-\beta  \omega_n\right)}\right]}=\beta^{-1}\sum_{n=1}^{\infty}n^2 \ln{\left[1-\exp{\left(-\beta m \sqrt{1-\frac{(1+\lambda) m^2M^2}{n^2}}\right)}\right]}.
\end{equation}
Here, $\beta=1/k_{B}T$. The main objective is to examine how the thermal corrections behave in the limit where the black hole mass approaches zero, allowing us to infer the manifestation of the singularity. To this end, we focus on the regime $mM \ll 1$, under which the Helmholtz free energy reduces to
\begin{equation}
F^{(0)}\approx -\beta^{-1}\sum_{n=1}^{\infty} n^2\sum_{k=1}^{\infty} \frac{e^{-\beta k m}}{k}\left(1+\frac{(1+\lambda) \beta m^3M^2k}{2n^2}\right).
\end{equation}
In this step, the logarithmic term has been expanded in series form. After applying the Riemann zeta--function regularization and using once more the condition $\zeta(-2)=0$, the resulting expression is obtained as
\begin{equation}
F^{(0)}(T,\lambda)\approx- \sum_{k=1}^{\infty}e^{-\beta k m}\frac{(1+\lambda) m^3M^2}{2}\zeta(0)=\frac{(1+\lambda) m^3M^2}{4
 \left[e^{ \frac{m}{\kappa T}}-1\right]}.
 \label{freeE}
\end{equation}

\begin{figure}[h!]
    \centering
     \includegraphics[scale=0.61]{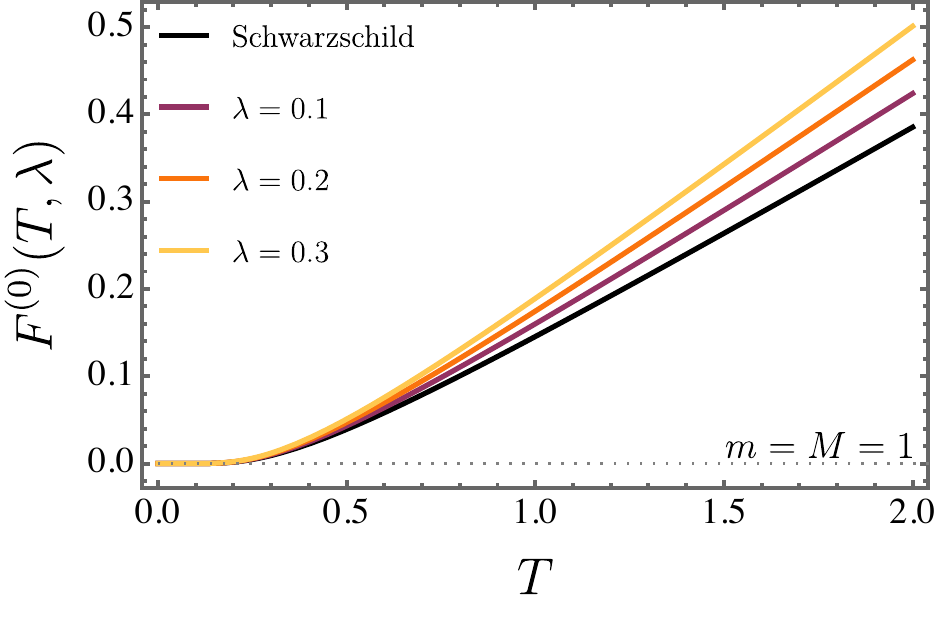}
    \caption{The Casimir--Helmholtz free energy $F^{(0)}(T,\lambda)$ is shown as a function of $T$ for different configurations of the Lorentz--violating parameter $\lambda$.}
    \label{fzero}
\end{figure}

In Fig.~\ref{fzero}, the Casimir--Helmholtz free energy is shown as a function of temperature $T$ for different values of the Lorentz–violating parameter $\lambda$. In contrast to the zero temperature Casimir energy, in a finite temperature system, as $\lambda$ increases, the corresponding $F^{(0)}(T,\lambda)$ also grows. Similarly, an increase in temperature leads to a higher value of $F^{(0)}(T,\lambda)$. It is worth noting that the occurrence of positive values for the Helmholtz free energy is an unusual feature, as it implies negative values for subsequent thermal quantities. Nonetheless, this behavior poses no issue, since only the Casimir contributions associated with the vacuum of stationary quantum field states are being considered.

The thermal contribution to the Casimir energy, corresponding to the internal energy of the system, can be expressed as
\begin{equation}
U^{(0)}(T,\lambda)= F^{(0)} + \beta \frac{\partial F^{(0)}}{\partial \beta} \approx \, \, \frac{(\lambda +1) m^3 M^2}{4 \left[e^{m/(\kappa T)}-1\right]}-\frac{(\lambda +1) m^4 M^2 e^{m/(\kappa T)}}{4 T \left[e^{m/(\kappa T)}-1\right]^2}.
\end{equation}
It can be readily confirmed that, in the limit $M \rightarrow 0$, the thermal correction terms vanish. Moreover, when the temperature approaches infinity, $T \to \infty$, the expression reduces to
\ie
\lim_{T \to \infty } U^{(0)}(T,\lambda) = -\frac{1}{8} (\lambda +1) m^3 M^2.
\fe
In Fig.~\ref{uzero}, the Casimir mean energy $U^{(0)}(T,\lambda)$ is presented as a function of the temperature $T$ for different values of the Lorentz--violating parameter $\lambda$. The asymptotic limits $\lim_{T \to \infty} U^{(0)}(T,\lambda)$ are also indicated. As $\lambda$ increases, the corresponding mean energy rises accordingly. As anticipated from the analysis of the Casimir–Helmholtz free energy, the energy here assumes negative values. This behavior is not problematic, since the discussion is restricted to the vacuum energy configuration associated with the stationary modes of the quantum field.

\begin{figure}[h!]
    \centering
     \includegraphics[scale=0.61]{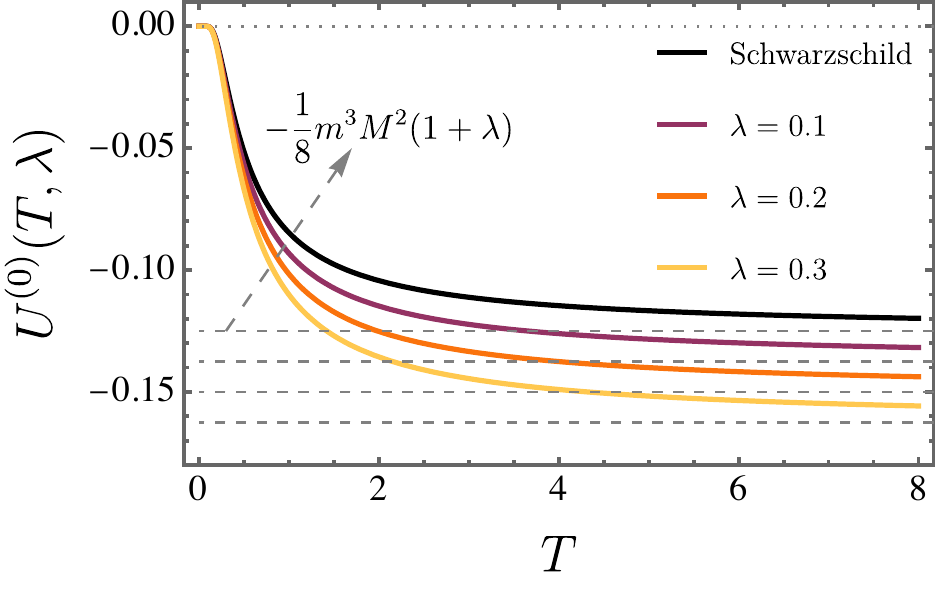}
    \caption{The Casimir mean energy $F^{(0)}(T,\lambda)$ is plotted as a function of $T$ for several configurations of the Lorentz–violating parameter $\lambda$. The asymptotic values are highlighted with dashed lines.}
    \label{uzero}
\end{figure}

Another relevant thermodynamic quantity to be evaluated is the Casimir entropy, which can be written as
\begin{equation}
S^{(0)}(T,\lambda) = - \frac{\partial F^{(0)}(T,\lambda)}{\partial T} = \, \, -\frac{(\lambda +1) m^4 M^2 e^{m/T}}{4 T^2 \left(e^{m/T}-1\right)^2}.
\end{equation}
In the limit of high temperatures, the Casimir entropy approaches a constant value given by
\begin{equation}
S^{(0)}_{T\to \infty}= -\frac{(1+\lambda)  m^2M^2}{4}.
\end{equation}
It is noteworthy that the Casimir entropy scales with the horizon area, mirroring the dependence found in the Hawking entropy. Figure~\ref{szero} illustrates the variation of the Casimir entropy, including its asymptotic limits. As previously discussed, the appearance of negative entropy values reflects the fact that the present analysis accounts solely for the vacuum contribution associated with the stationary modes of the quantum field, without incorporating the complete thermodynamic evolution of the system.

\begin{figure}[h!]
    \centering
     \includegraphics[scale=0.61]{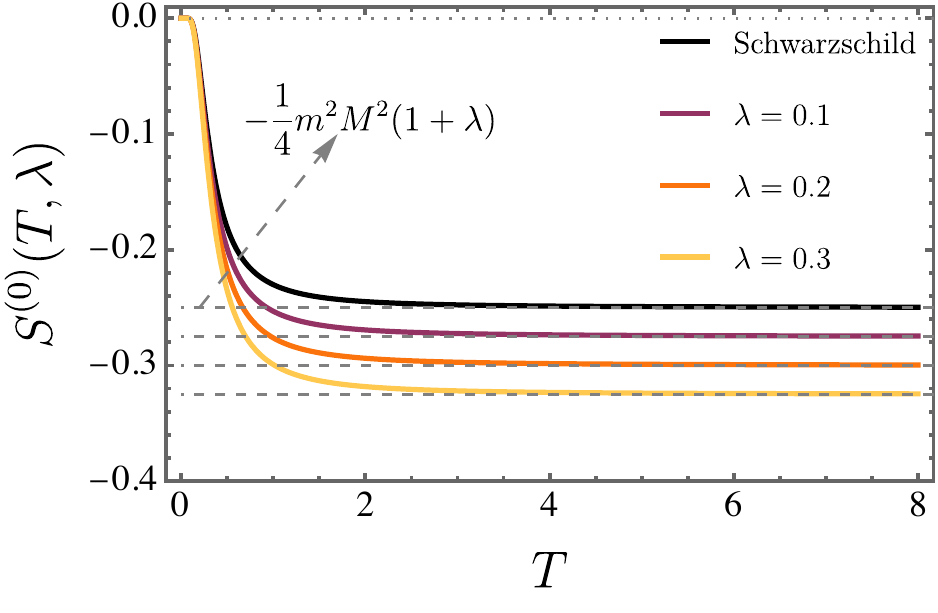}
    \caption{The Casimir entropy $S^{(0)}(T,\lambda)$ is plotted as a function of $T$ for several configurations of the Lorentz–violating parameter $\lambda$.}
    \label{szero}
\end{figure}

Conversely, the heat capacity at constant volume is defined as
\begin{equation}
C_{V} = T\left(\frac{\partial{S^{(0)}}}{\partial T}\right)_V  = -\frac{(\lambda +1) m^4 M^2 \left(m \coth \left(\frac{m}{2 T}\right)-2 T\right) \text{csch}^2\left(\frac{m}{2 T}\right)}{16 T^3}.
\end{equation}
According to Fig.~\ref{hzero}, the heat capacity at constant volume remains negative throughout the range considered, indicating that the quantum vacuum in this configuration might be thermodynamically unstable. The presence of turning points in the curve may, however, signal possible phase transition regions.

\begin{figure}[h!]
    \centering
     \includegraphics[scale=0.61]{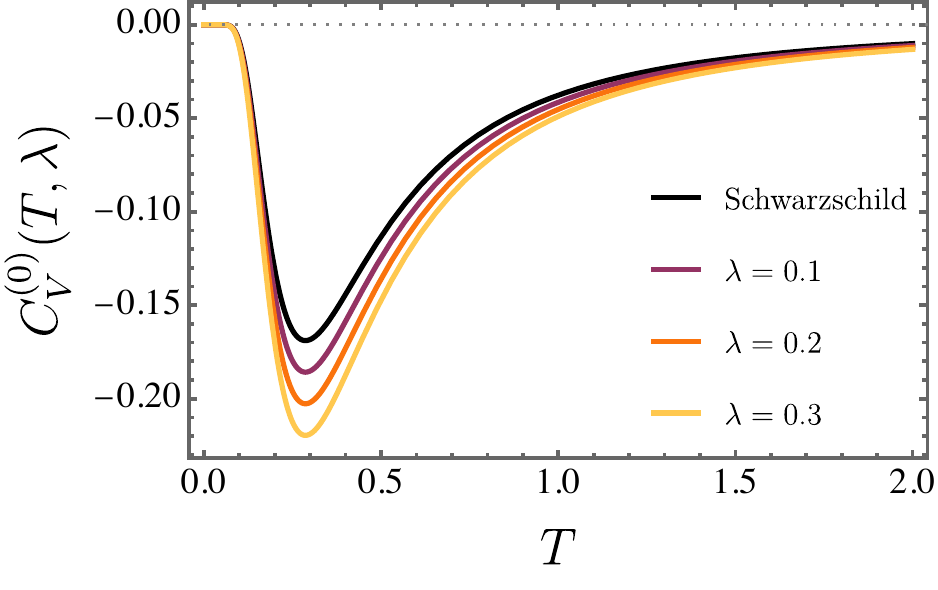}
    \caption{The Casimir heat capacity $C_{V}^{(0)}(T,\lambda)$ is plotted as a function of $T$ for several configurations of the Lorentz–violating parameter $\lambda$.}
    \label{hzero}
\end{figure}

In an overall perspective, as shown by the results obtained in this section, temperature affected the Casimir energy and highlighted possible LIV signatures in the thermodynamic quantities, since these quantities increased proportionally with the Lorentz--violating parameter $\lambda$.


\section{Conclusion }
\label{sec7}

In this work, we discussed the implications of Lorentz invariance violation within the framework of bumblebee gravity. Starting from the modified dispersion relation given by Eq. \eqref{hamiltonian}, we identified the impact of the LIV parameter on optical properties, including the refractive index, the group velocity of massive and massless modes, and time--delay effects that became pronounced near the black hole horizon. These kinematic signatures showed how the LIV parameter reshaped particle propagation even in regimes where the geometry closely approached the Schwarzschild case. Each situation was examined for several values of $\lambda$ and compared with the Schwarzschild limit by taking $\lambda \to 0$, as displayed in Figs. \ref{nindex}, \ref{normalgv}, and \ref{masslessgv}.

The static interaction potential was then investigated through its Green--function formulation. For a massive case, we obtained a mixed Coulomb--Yukawa profile confined within the horizon, as given in Eq. \eqref{interparticlepotential}, whereas for massless excitations the potential became long--ranged and retained sensitivity to Lorentz violation, Eq. \eqref{potentialm0}. The radial behavior of the potential was shown in Fig. \ref{ineftegrnpgnarticvblepvbotentialmassive}, where the Schwarzschild case $(\lambda=0)$ was compared with LIV--influenced configurations, and in Fig. \ref{intercomp}, where a comparison with other black hole model, Kalb–Ramond case, was presented.

Next, the scattering amplitude was analyzed, and it exhibited explicit LIV--dependent corrections, which allowed us to derive analytical expressions for both the differential and total cross sections, as illustrated in Fig. \ref{sigxx}. These results demonstrated that even in the weak--field regime, where spacetime curvature became negligible, spontaneous Lorentz violation left observable traces in quantum scattering processes. The physically relevant limits, including the massless case and the non--LIV scenario, were also examined.

The thermodynamic sector was then carried out within the ensemble framework. In this context, pressure, mean energy, entropy, and heat capacity were investigated. In all cases, the event horizon $r_h$ acted as a barrier that signaled phase transitions. Analytical expressions were obtained for all thermodynamic quantities in the massless bosonic case. The analysis was performed by varying the radial position $r$ and, in addition, by fixing the location very close to the horizon, at the photon sphere, and in the asymptotic region, while studying the dependence on the temperature $T$. In general terms, across all thermodynamic functions, the Lorentz--violating parameter $\lambda$ was found to increase their magnitudes.

Finally, we examined the behavior of the vacuum state and the Casimir energy for a scalar field in the bumblebee background. By solving the Klein--Gordon equation in terms of confluent Heun functions, Eq. \eqref{eq:Heun_solution_bumblebee_reduced}, we constructed the mode spectrum and evaluated both the zero--temperature vacuum energy given by Eq. \eqref{Ereg} and its finite--temperature extension in Eq. \eqref{freeE}, which exhibited explicit dependence on the Lorentz--violating parameter, as shown in Figs. \ref{energyregularized} and \ref{fzero}. In the thermal case, the LIV parameter contributed more effectively to the energy. The horizon tension was also obtained for the zero--temperature system in Eq. \eqref{Talh}, with its behavior displayed in Figs. \ref{tau} and \ref{tau2}. The internal energy, Casimir entropy, and heat capacity were calculated as well, and, as in the case of the finite-temperature Casimir energy, the LIV parameter increased these quantities proportionally to $\lambda$, as illustrated in Figs. \ref{uzero}, \ref{szero}, and \ref{hzero}.

As a further perspective, carrying out similar analyses for other Lorentz--violating black holes, such as the metric--affine extension of the bumblebee solution \cite{filho2023vacuum} and the new bumblebee black holes recently proposed in the literature \cite{Liu:2025oho,Zhu:2025fiy}, appeared to be a fruitful direction to pursue.


\section*{Acknowledgments}
\hspace{0.5cm}

A. A. Araújo Filho is supported by Conselho Nacional de Desenvolvimento Cient\'{\i}fico e Tecnol\'{o}gico (CNPq) and Fundação de Apoio à Pesquisa do Estado da Paraíba (FAPESQ), project numbers 150223/2025-0 and 1951/2025. K.E.L.F would like to thank the Paraíba State Research Support Foundation FAPESQ  for financial support. F.A.B and E.P acknowledge support from CNPq (Grant nos. 306398/2021-4,
309092/2022-1, 304290/2020-3). A.R.Q work is supported by FAPESQ-PB. A.R.Q also acknowledges support by CNPq under process number 310533/2022-8. A. \"O. would like to acknowledge networking support of the COST Action CA21106 - COSMIC WISPers in the Dark Universe: Theory, astrophysics and experiments (CosmicWISPers), the COST Action CA22113 - Fundamental challenges in theoretical physics (THEORY-CHALLENGES), the COST Action CA21136 - Addressing observational tensions in cosmology with systematics and fundamental physics (CosmoVerse), the COST Action CA23130 - Bridging high and low energies in search of quantum gravity (BridgeQG), and the COST Action CA23115 - Relativistic Quantum Information (RQI) funded by COST (European Cooperation in Science and Technology). A. \"O. would also like to acknowledge the funding support of SCOAP3, Switzerland and TUBITAK, Turkiye.V.B. Bezerra is partially supported by the Conselho
Nacional de Desenvolvimento Científico e Tecnológico (CNPq), grant number 307211/2020-
7. Also, H. H. is grateful to Excellence project FoS UHK 2203/2025-2026 for the financial support. {The authors also thank the anonymous referee for the valuable suggestions provided.}

\section{Data Availability Statement}

Data Availability Statement: No Data associated in the manuscript


\bibliographystyle{ieeetr}
\bibliography{main}

\begin{thebibliography}{10}

\bibitem{1}
V.~Kostelecky and S.~Samuel, ``Spontaneous breaking of lorentz symmetry in
  string theory,'' {\em Phys. Rev. D}, vol.~39, p.~683, 1989.

\bibitem{4}
S.~Carroll, J.~Harvey, V.~Kostelecky, C.~Lane, and T.~Okamoto, ``Noncommutative
  field theory and lorentz violation,'' {\em Phys. Rev. Lett.}, vol.~87,
  p.~141601, 2001.

\bibitem{2}
J.~Alfaro, H.~Morales-Tecotl, and L.~Urrutia, ``Loop quantum gravity and light
  propagation,'' {\em Phys. Rev. D}, vol.~65, p.~103509, 2002.

\bibitem{6}
S.~Dubovsky, P.~Tinyakov, and I.~Tkachev, ``Massive graviton as a testable cold
  dark matter candidate,'' {\em Phys. Rev. Lett.}, vol.~94, p.~181102, 2005.

\bibitem{3}
P.~Horava, ``Quantum gravity at a lifshitz point,'' {\em Phys. Rev.}, vol.~79,
  p.~084008, 2009.

\bibitem{7}
G.~Bengochea and R.~Ferraro, ``Dark torsion as the cosmic speed-up,'' {\em
  Phys. Rev. D}, vol.~79, p.~124019, 2009.

\bibitem{8}
A.~Cohen and S.~Glashow, ``Very special relativity,'' {\em Phys. Rev. Lett.},
  vol.~97, p.~021601, 2006.

\bibitem{5}
T.~Jacobson and D.~Mattingly, ``Gravity with a dynamical preferred frame,''
  {\em Phys. Rev. D}, vol.~64, p.~024028, 2001.

\bibitem{bluhm2006overview}
R.~Bluhm, ``Overview of the standard model extension: implications and
  phenomenology of lorentz violation,'' in {\em Special Relativity: Will it
  Survive the Next 101 Years?}, pp.~191--226, Springer, 2006.

\bibitem{araujo2025impact}
A.~A. Ara{\'u}jo~Filho, N.~Heidari, J.~A. A.~S. Reis, and H.~Hassanabadi, ``The
  impact of an antisymmetric tensor on charged black holes: evaporation
  process, geodesics, deflection angle, scattering effects and quasinormal
  modes,'' {\em Classical and Quantum Gravity}, vol.~42, no.~6, p.~065026,
  2025.

\bibitem{Liu:2025bpp}
X.~Liu, W.~Liu, Z.~Liu, and J.~Wang, ``{Harvesting correlations from BTZ black
  hole coupled to a Lorentz-violating vector field},'' {\em JHEP}, vol.~08,
  p.~094, 2025.

\bibitem{liu2024shadow}
W.~Liu, D.~Wu, and J.~Wang, ``Shadow of slowly rotating kalb-ramond black
  holes,'' {\em arXiv preprint arXiv:2407.07416}, 2024.

\bibitem{filho2023vacuum}
A.~A. Ara{\'u}jo~Filho, J.~R. Nascimento, A.~Y. Petrov, and P.~J.
  Porf{\'\i}rio, ``Vacuum solution within a metric-affine bumblebee gravity,''
  {\em Physical Review D}, vol.~108, no.~8, p.~085010, 2023.

\bibitem{12}
Q.~Bailey and V.~Kostelecky, ``Signals for lorentz violation in post-newtonian
  gravity,'' {\em Phys. Rev. D}, vol.~74, p.~045001, 2006.

\bibitem{AraujoFilho:2024ykw}
A.~A. Ara\'ujo~Filho, J.~R. Nascimento, A.~Y. Petrov, and P.~J. Porf\'\i{}rio,
  ``{An exact stationary axisymmetric vacuum solution within a metric-affine
  bumblebee gravity},'' {\em JCAP}, vol.~07, p.~004, 2024.

\bibitem{13}
R.~Bluhm, N.~Gagne, R.~Potting, and A.~Vrublevskis, ``Constraints and stability
  in vector theories with spontaneous lorentz violation,'' {\em Phys. Rev. D},
  vol.~77, p.~125007, 2008.

\bibitem{11}
V.~Kostelecky and S.~Samuel, ``Phenomenological gravitational constraints on
  strings and higher dimensional theories,'' {\em Phys. Rev. Lett.}, vol.~63,
  p.~224, 1989.

\bibitem{10}
V.~Kostelecky and S.~Samuel, ``Gravitational phenomenology in higher
  dimensional theories and strings,'' {\em Phys. Rev. D}, vol.~40, p.~1886,
  1989.

\bibitem{9}
V.~Kostelecky, ``Gravity, lorentz violation, and the standard model,'' {\em
  Phys. Rev. D}, vol.~69, p.~105009, 2004.

\bibitem{amarilo2024gravitational}
K.~M. Amarilo, M.~B. Ferreira~Filho, A.~A. Ara{\'u}jo~Filho, and J.~A. A.~S.
  Reis, ``Gravitational waves effects in a lorentz--violating scenario,'' {\em
  Physics Letters B}, vol.~855, p.~138785, 2024.

\bibitem{schreck2014quantum}
M.~Schreck, ``Quantum field theoretic properties of lorentz-violating operators
  of nonrenormalizable dimension in the fermion sector,'' {\em Physical Review
  D}, vol.~90, no.~8, p.~085025, 2014.

\bibitem{schreck2014quantum2}
M.~Schreck, ``Quantum field theoretic properties of lorentz-violating operators
  of nonrenormalizable dimension in the photon sector,'' {\em Physical Review
  D}, vol.~89, no.~10, p.~105019, 2014.

\bibitem{paperrainbow}
A.~A. Araújo~Filho, J.~Furtado, H.~Hassanabadi, and J.~Reis, ``Thermal
  analysis of photon--like particles in rainbow gravity,'' {\em Physics of Dark
  Universe}, vol.~42, no.~8, p.~101310, 2023.

\bibitem{aa2021lorentz}
A.~A. Ara{\'u}jo~Filho, ``Lorentz-violating scenarios in a thermal reservoir,''
  {\em The European Physical Journal Plus}, vol.~136(4), 417 (2021).

\bibitem{araujo2021higher}
A.~A. Ara{\'u}jo~Filho and A.~Y. Petrov, ``Higher-derivative lorentz-breaking
  dispersion relations: a thermal description,'' {\em The European Physical
  Journal C}, vol.~81(9), 843 (2021).

\bibitem{araujo2021thermodynamic}
A.~A. Ara{\'u}jo~Filho and R.~V. Maluf, ``Thermodynamic properties in
  higher-derivative electrodynamics,'' {\em Brazilian Journal of Physics},
  vol.~51, no.~3, pp.~820--830, 2021.

\bibitem{araujo2022thermal}
A.~A. Ara{\'u}jo~Filho, {\em Thermal aspects of field theories}.
\newblock Amazon. com, 2022.

\bibitem{araujo2022does}
A.~A. Ara{\'u}jo~Filho and J.~Reis, ``How does geometry affect quantum
  gases?,'' {\em International Journal of Modern Physics A}, vol.~37,
  no.~11n12, p.~2250071, 2022.

\bibitem{Anacleto:2018wlj}
M.~A. Anacleto, F.~A. Brito, E.~Maciel, A.~Mohammadi, E.~Passos, W.~O. Santos,
  and J.~R.~L. Santos, ``{Lorentz-violating dimension-five operator
  contribution to the black body radiation},'' {\em Phys. Lett. B}, vol.~785,
  pp.~191--196, 2018.

\bibitem{reis2021thermal}
J.~Reis {\em et~al.}, ``Thermal aspects of interacting quantum gases in
  lorentz-violating scenarios,'' {\em The European Physical Journal Plus},
  vol.~136, no.~3, pp.~1--30, 2021.

\bibitem{Capozziello:2024ucm}
S.~Capozziello, S.~De~Bianchi, and E.~Battista, ``{Avoiding singularities in
  Lorentzian-Euclidean black holes: The role of~atemporality},'' {\em Phys.
  Rev. D}, vol.~109, no.~10, p.~104060, 2024.

\bibitem{DeBianchi:2025bgn}
S.~De~Bianchi, S.~Capozziello, and E.~Battista, ``{Atemporality from
  Conservation Laws of Physics in Lorentzian-Euclidean Black Holes},'' {\em
  Found. Phys.}, vol.~55, no.~3, p.~36, 2025.

\bibitem{Capozziello:2025wwl}
S.~Capozziello, E.~Battista, and S.~De~Bianchi, ``{Null geodesics, causal
  structure, and matter accretion in Lorentzian-Euclidean black holes},'' {\em
  Phys. Rev. D}, vol.~112, no.~4, p.~044009, 2025.

\bibitem{20}
R.~Maluf and J.~Neves, ``Black holes with a cosmological constant in bumblebee
  gravity,'' {\em Phys. Rev. D}, vol.~103, p.~044002, 2021.

\bibitem{23}
R.~Xu, D.~Liang, and L.~Shao, ``Bumblebee black holes in light of event horizon
  telescope observations,'' {\em Astrophys. J.}, vol.~945, p.~148, 2023.

\bibitem{24}
D.~Liang, R.~Xu, Z.-F. Mai, and L.~Shao, ``Probing vector hair of black holes
  with extreme-mass-ratio inspirals,'' {\em Phys. Rev. D}, vol.~107, p.~044053,
  2023.

\bibitem{22}
Z.-F. Mai, R.~Xu, D.~Liang, and L.~Shao, ``Extended thermodynamics of the
  bumblebee black holes,'' {\em Phys. Rev. D}, vol.~108, p.~024004, 2023.

\bibitem{21}
R.~Xu, D.~Liang, and L.~Shao, ``Static spherical vacuum solutions in the
  bumblebee gravity model,'' {\em Phys. Rev. D}, vol.~107, p.~024011, 2023.

\bibitem{14}
R.~Casana, A.~Cavalcante, F.~Poulis, and E.~Santos, ``Exact schwarzschild-like
  solution in a bumblebee gravity model,'' {\em Phys. Rev.}, vol.~97,
  p.~104001, 2018.

\bibitem{19}
R.~Oliveira, D.~Dantas, and C.~Almeida, ``Quasinormal frequencies for a black
  hole in a bumblebee gravity,'' {\em EPL}, vol.~135, p.~10003, 2021.

\bibitem{Liu:2022dcn}
W.~Liu, X.~Fang, J.~Jing, and J.~Wang, ``{QNMs of slowly rotating
  Einstein\textendash{}Bumblebee black hole},'' {\em Eur. Phys. J. C}, vol.~83,
  no.~1, p.~83, 2023.

\bibitem{18}
Z.~Cai and R.-J. Yang, ``Accretion of the vlasov gas onto a schwarzschild-like
  black hole,''

\bibitem{17}
R.-J. Yang, H.~Gao, Y.~Zheng, and Q.~Wu, ``Effects of lorentz breaking on the
  accretion onto a schwarzschild-like black hole,'' {\em Commun. Theor. Phys.},
  vol.~71, p.~568, 2019.

\bibitem{Lambiase:2024uzy}
G.~Lambiase, R.~C. Pantig, and A.~{\"O}vg{\"u}n, ``{Weak field deflection angle
  and analytical parameter estimation of the Lorentz-violating Bumblebee
  parameter through the black hole shadow using EHT data},'' {\em EPL},
  vol.~148, no.~4, p.~49001, 2024.

\bibitem{Lambiase:2023zeo}
G.~Lambiase, L.~Mastrototaro, R.~C. Pantig, and A.~Ovgun, ``{Probing
  Schwarzschild-like black holes in metric-affine bumblebee gravity with
  accretion disk, deflection angle, greybody bounds, and neutrino
  propagation},'' {\em JCAP}, vol.~12, p.~026, 2023.

\bibitem{Pantig:2025paj}
R.~C. Pantig and A.~{\"O}vg{\"u}n, ``{Multimodal signatures of asymptotic (A)dS
  Kalb{\textendash}Ramond black holes: Constraints through the shadow, weak
  deflection angle, and topological photon spheres},'' {\em Annals Phys.},
  vol.~480, p.~170104, 2025.

\bibitem{Ovgun:2025pwy}
A.~{\"O}vg{\"u}n and R.~C. Pantig, ``{Gravitational Aharonov-Bohm phase in
  Kalb-Ramond spacetimes},'' {\em Phys. Dark Univ.}, vol.~50, p.~102179, 2025.

\bibitem{Pantig:2024kqy}
R.~C. Pantig, ``{On the analytic generalization of particle deflection in the
  weak field regime and shadow size in light of EHT constraints for
  Schwarzschild-like black hole solutions},'' {\em Eur. Phys. J. C}, vol.~85,
  no.~1, p.~52, 2025.

\bibitem{Nandi2016}
K.~K. Nandi, A.~A. Potapov, R.~N. Izmailov, A.~Tamang, and J.~C. Evans,
  ``{Stability and instability of Ellis and phantom wormholes: Are there
  ghosts?},'' {\em Phys. Rev. D}, vol.~93, no.~10, p.~104044, 2016.

\bibitem{Filho:2023ydb}
J.~Furtado, H.~Hassanabadi, J.~A. A.~S. Reis, {\em et~al.}, ``Thermal analysis
  of photon-like particles in rainbow gravity,'' {\em arXiv preprint
  arXiv:2305.08587}, 2023.

\bibitem{araujo2023thermodynamical}
A.~A. Ara{\'u}jo~Filho, J.~Furtado, J.~A. A.~S. Reis, and G.~Silva, J.~E,
  ``Thermodynamical properties of an ideal gas in a traversable wormhole,''
  {\em Class. Quant. Grav.}, vol.~40, no.~24, p.~245001, 2023.

\bibitem{filho2025modified}
A.~A. Ara{\'u}jo~Filho, J.~A. A.~S. Reis, and A.~{\"O}vg{\"u}n, ``Modified
  particle dynamics and thermodynamics in a traversable wormhole in bumblebee
  gravity,'' {\em The European Physical Journal C}, vol.~85, no.~1, p.~83,
  2025.

\bibitem{filho2025particlemotion}
A.~A. A~Ara{\'u}jo~Filho, ``Particle motion and thermal effects around a
  kalb--ramond black hole,'' {\em The European Physical Journal C}, vol.~85,
  no.~9, p.~1002, 2025.

\bibitem{liu2025charged}
J.-Z. Liu, W.-D. Guo, S.-W. Wei, and Y.-X. Liu, ``Charged spherically symmetric
  and slowly rotating charged black hole solutions in bumblebee gravity,'' {\em
  The European Physical Journal C}, vol.~85, no.~2, p.~145, 2025.

\bibitem{Liu:2025oho}
J.-Z. Liu, S.-P. Wu, S.-W. Wei, and Y.-X. Liu, ``{Exact Black Hole Solutions in
  Bumblebee Gravity with Lightlike or Spacelike VEVS},'' 10 2025.

\bibitem{blackledge2005digital}
J.~M. Blackledge, {\em Digital image processing: mathematical and computational
  methods}.
\newblock Elsevier, 2005.

\bibitem{gradshteyn2014table}
I.~S. Gradshteyn and I.~M. Ryzhik, {\em Table of integrals, series, and
  products}.
\newblock Academic press, 2014.

\bibitem{Touati:2024tqy}
A.~Touati, ``{Elastic scattering of electron by a Yukawa potential in
  non-commutative spacetime},'' {\em Phys. Lett. B}, vol.~867, p.~139598, 2025.

\bibitem{gibbons2008applications}
G.~W. Gibbons and M.~C. Werner, ``Applications of the gauss--bonnet theorem to
  gravitational lensing,'' {\em Classical and Quantum Gravity}, vol.~25,
  no.~23, p.~235009, 2008.

\bibitem{araujo2023thermodynamics}
A.~A. Ara{\'u}jo~Filho, ``Thermodynamics of massless particles in curved
  spacetime,'' {\em International Journal of Geometric Methods in Modern
  Physics}, vol.~20, no.~13, p.~2350226, 2023.

\bibitem{AraujoFilho:2025fwd}
A.~A. Ara{\'u}jo~Filho, ``{Particle motion and thermal effects around a
  Kalb{\textendash}Ramond black hole},'' {\em Eur. Phys. J. C}, vol.~85, no.~9,
  p.~1002, 2025.

\bibitem{amelino2013quantum}
G.~Amelino-Camelia, ``Quantum-spacetime phenomenology,'' {\em Living Reviews in
  Relativity}, vol.~16, pp.~1--137, 2013.

\bibitem{Wagner:2023fmb}
F.~Wagner, G.~Var\~ao, I.~P. Lobo, and V.~B. Bezerra, ``{Quantum-spacetime
  effects on nonrelativistic Schr\"odinger evolution},'' {\em Phys. Rev. D},
  vol.~108, no.~6, p.~066008, 2023.

\bibitem{jacob2008lorentz}
U.~Jacob and T.~Piran, ``Lorentz-violation-induced arrival delays of
  cosmological particles,'' {\em Journal of Cosmology and Astroparticle
  Physics}, vol.~2008, no.~01, p.~031, 2008.

\bibitem{anchordoqui2003ultrahigh}
L.~Anchordoqui, T.~Paul, S.~Reucroft, and J.~Swain, ``Ultrahigh energy cosmic
  rays: The state of the art before the auger observatory,'' {\em International
  Journal of Modern Physics A}, vol.~18, no.~13, pp.~2229--2366, 2003.

\bibitem{greiner2012thermodynamics}
W.~Greiner, L.~Neise, and H.~St{\"o}cker, {\em Thermodynamics and statistical
  mechanics}.
\newblock Springer Science \& Business Media, 2012.

\bibitem{isihara2013statistical}
A.~Isihara, {\em Statistical physics}.
\newblock Academic Press, 2013.

\bibitem{wannier1987statistical}
G.~H. Wannier, {\em Statistical physics}.
\newblock Courier Corporation, 1987.

\bibitem{salinas1999introduccao}
S.~R. Salinas, {\em Introdu{\c{c}}{\~a}o {\`a} f{\'\i}sica estat{\'\i}stica}.
\newblock Edusp, 1999.

\bibitem{vogt2017statistical}
J.~Vogt, ``Statistical thermodynamics,'' {\em Exam survival guide: Physical
  chemistry}, pp.~175--211, 2017.

\bibitem{mandl1991statistical}
F.~Mandl, {\em Statistical physics}, vol.~14.
\newblock John Wiley \& Sons, 1991.

\bibitem{Heidari:2025oop}
N.~Heidari and A.~A. Ara{\'u}jo~Filho, ``{Quantum particle production and
  radiative properties of a new bumblebee black hole},'' 12 2025.

\bibitem{Muniz:2015jba}
C.~R. Muniz, M.~O. Tahim, M.~S. Cunha, and H.~S. Vieira, ``{On the Global
  Casimir Effect in the Schwarzschild Spacetime},'' {\em JCAP}, vol.~01,
  p.~006, 2018.

\bibitem{fiziev2009novel}
P.~P. Fiziev, ``Novel relations and new properties of confluent heun's
  functions and their derivatives of arbitrary order,'' {\em Journal of Physics
  A: Mathematical and Theoretical}, vol.~43, no.~3, p.~035203, 2009.

\bibitem{bezerra2014klein}
V.~B. Bezerra, H.~S. Vieira, and A.~A. Costa, ``The klein--gordon equation in
  the spacetime of a charged and rotating black hole,'' {\em Classical and
  Quantum Gravity}, vol.~31, no.~4, p.~045003, 2014.

\bibitem{barranco2014schwarzschild}
J.~Barranco, A.~Bernal, J.~C. Degollado, A.~Diez-Tejedor, M.~Megevand,
  M.~Alcubierre, D.~N{\'u}{\~n}ez, and O.~Sarbach, ``Schwarzschild scalar wigs:
  spectral analysis and late time behavior,'' {\em Physical Review D}, vol.~89,
  no.~8, p.~083006, 2014.

\bibitem{laptev2006electromagnetic}
Y.~P. Laptev and M.~Fil'Chenkov, ``Electromagnetic and gravitational radiation
  of graviatoms,'' {\em Astronomical \& Astrophysical Transactions}, vol.~25,
  no.~1, pp.~33--42, 2006.

\bibitem{lasenby2005bound}
A.~Lasenby, C.~Doran, J.~Pritchard, A.~Caceres, and S.~Dolan, ``Bound states
  and decay times of fermions in a schwarzschild black hole background,'' {\em
  Physical Review D—Particles, Fields, Gravitation, and Cosmology}, vol.~72,
  no.~10, p.~105014, 2005.

\bibitem{mostepanenko1988casimir}
V.~M. Mostepanenko and N.~N. Trunov, ``The casimir effect and its
  applications,'' {\em Soviet Physics Uspekhi}, vol.~31, no.~11, p.~965, 1988.

\bibitem{Zhu:2025fiy}
J.~Zhu and H.~Li, ``{Full Classification of Static Spherical Vacuum Solutions
  to Bumblebee Gravity with General VEVs},'' 11 2025.

\end{thebibliography}

\end{document}